\documentclass[review]{elsarticle}
\biboptions{sort&compress}

\usepackage[colorlinks,citecolor=blue,linktoc=all,linkcolor=cyan]{hyperref}
\usepackage{graphicx}

\usepackage[T1]{fontenc}
\usepackage{dsfont}               
\usepackage{mathrsfs}             
\usepackage{slashed}              
\usepackage{amsmath}
\usepackage{amssymb}
\usepackage{amsbsy}
\usepackage{amsfonts}

\newcommand{\be}{\begin{equation}}
\newcommand{\ee}{\end{equation}}
\newcommand{\bea}{\begin{eqnarray}}
\newcommand{\eea}{\end{eqnarray}}

\numberwithin{equation}{section}
\numberwithin{table}{section}
\numberwithin{figure}{section}

\journal{Progress in Particle and Nuclear Physics}

\usepackage{titlesec}
\usepackage{sectsty}
\titleformat{\section}{\normalfont\Large\bfseries}{\thesection}{1em}{}
\titleformat{\subsection}{\normalfont\large\bfseries}{\thesubsection}{1em}{}
\titleformat{\subsubsection}{\normalfont\normalsize\bfseries}{\thesubsubsection}{1em}{}

\begin{document}

\begin{frontmatter}

\title{High precision tests of QCD without scale or scheme ambiguities \\ \vspace{0.5cm} \small \emph{The 40$^{th}$ anniversary of the
Brodsky--Lepage--Mackenzie method}}

\author[1,2]{Leonardo Di Giustino\corref{mycorrespondingauthor}}
  \cortext[mycorrespondingauthor]{Corresponding author}
  \ead{leonardo.digiustino@uninsubria.it}
\author[3]{Stanley J. Brodsky }
\ead{sjbth@slac.stanford.edu}
\author[1,2]{Philip G. Ratcliffe }
\ead{philip.ratcliffe@uninsubria.it}
\author[4]{Xing-Gang Wu}
\ead{wuxg@cqu.edu.cn}
\author[5]{Sheng-Quan Wang}
\ead{sqwang@alu.cqu.edu.cn}

\address[1]{Department of Science and High Technology, University of Insubria, via Valleggio 11, I-22100, Como, Italy}
\address[2]{INFN, Sezione di Milano--Bicocca, 20126 Milano, Italy}

\address[3]{SLAC National Accelerator Laboratory, Stanford University, Stanford, California 94039, USA}

\address[4]{Department of Physics, Chongqing University, Chongqing 401331, P.R. China}

\address[5]{Department of Physics, Guizhou Minzu University, Guiyang 550025, P.R. China}


\begin{abstract}

A key issue in making precise predictions in QCD is the
uncertainty in setting the renormalization scale $\mu_r$ and thus
determining the correct values of the QCD running coupling
$\alpha_s(\mu_r)$ at each order in the perturbative expansion of a
QCD observable. It has often been conventional to simply set the
renormalization scale to the typical scale of the process $Q$ and
vary it in the range $\mu_r \in [Q/2,2Q]$ in order to estimate the
theoretical error. This is the practice of Conventional Scale
Setting (CSS). The resulting CSS prediction will however depend on
the theorist's choice of renormalization scheme and the resulting
pQCD series will diverge factorially.  It will also disagree with
renormalization scale setting used in QED and electroweak theory
thus precluding grand unification. A solution to the
renormalization scale-setting problem is offered by the Principle
of Maximum Conformality (PMC), which provides a systematic way to
eliminate the renormalization scale-and-scheme dependence in
perturbative calculations. The PMC method has rigorous theoretical
foundations, it satisfies Renormalization Group Invariance (RGI)
and preserves all self-consistency conditions derived from the
renormalization group. The PMC cancels the renormalon growth,
reduces to the Gell-Mann--Low scheme in the $N_c \to 0$ Abelian
limit and leads to scale- and scheme-invariant results. The PMC
has now been successfully applied to many high-energy processes.
In this article we summarize recent developments and results in
solving the renormalization scale and scheme ambiguities in
perturbative QCD. In particular, we present a recently developed
method the PMC$_\infty$ and its applications, comparing the
results with CSS. The method preserves the property of
renormalizable SU(N)/U(1) gauge theories defined as
\emph{Intrinsic Conformality} (\emph{iCF}).

This property underlies the scale invariance of physical
observables and leads to a remarkably efficient method to solve
the conventional renormalization scale ambiguity at every order in
pQCD.

This new method reflects the underlying conformal properties
displayed by pQCD at NNLO, eliminates the scheme dependence of
pQCD predictions and is consistent with the general properties of
the PMC. A new method to identify conformal and $\beta$-terms,
which can be applied either to numerical or to theoretical
calculations is also shown. We present results for the thrust and
$C$-parameter distributions in $e^+e^-$ annihilation showing errors
and comparison with the CSS. We also show results for a recent
innovative comparison between the CSS and the PMC$_\infty$ applied
to the thrust distribution investigating both the QCD conformal
window and the QED $N_c\rightarrow0$ limit. In order to
determine the thrust distribution along the entire renormalization
group flow from the highest energies to zero energy, we consider
the number of flavors near the upper boundary of the conformal
window. In this flavor-number regime the theory develops a
perturbative infrared interacting fixed point. These results show
that PMC$_\infty$ leads to higher precision and introduces new
interesting features in the PMC. In fact, this method preserves
with continuity the position of the peak, showing perfect
agreement with the experimental data already at NNLO.

We also show a detailed comparison of the PMC$_\infty$ with the
other PMC approaches: the multi-scale-setting approach (PMCm) and
the single-scale-setting approach (PMCs) by comparing their
predictions for three important fully integrated quantities
$R_{e^+e^-}$, $R_\tau$ and $\Gamma(H \to b \bar{b})$ up to the
four-loop accuracy.

\end{abstract}


\end{frontmatter}

\newpage

\thispagestyle{empty}
\tableofcontents

\newpage

\section{Introduction}\label{intro}

\subsubsection*{The renormalization scale and scheme ambiguities}

The renormalization scale and scheme ambiguities are an important
source of errors in many processes in perturbative QCD preventing
precise theoretical predictions for both standard model (SM) and
beyond standard model (BSM) physics. In principle, an infinite
perturbative series is devoid of this issue, given the scheme and
scale invariance of entire physical quantities
\cite{StueckelbergdeBreidenbach:1952pwl, GellMann:1954fq,
Peterman:1978tb, Callan:1970yg, Symanzik:1971vw}; in practice,
perturbative corrections are known up to a certain order of
accuracy and scale invariance is only approximated in truncated
series, leading to scheme and scale ambiguities
\cite{Celmaster:1979km, Buras:1979yt, Stevenson:1980du,
Stevenson:1981vj, Stevenson:1982qw, Stevenson:1982wn,
Grunberg:1980ja, Grunberg:1982fw, Grunberg:1989xf, Brodsky:1982gc,
Chishtie:2015lwk, Chishtie:2016wob, Abbott:1980hwa}.

Although perturbative calculations for theoretical predictions are
also affected by other sources of errors, e.g.\ the top- and
Higgs-mass uncertainties and the strong-coupling $\alpha_s(M_Z)$
uncertainty, the renormalization scale and scheme ambiguities
remain among of the main sources of errors. These scale and scheme
ambiguities play an important role for predictions in many
fundamental processes in perturbative QCD, also with respect to
the other sources of uncertainties. Processes such as gluon fusion
in Higgs production \cite{Anastasiou:2016cez} or bottom-quark
production \cite{Catani:2020kkl}, which are essential for the
physics investigated at the Large Hadron Collider (LHC) and at
future colliders, are all affected by these ambiguities. In the
present era, high-precision predictions are crucial for both SM
and BSM physics, in order to test the theory in all sectors and to
enhance sensitivity to possible new physics (NP) at colliders.

At the moment two basic strategies exist to deal with this
problem. On one hand, according to conventional practice or
conventional scale setting (CSS), this problem cannot be avoided
and is responsible for part of the theoretical errors. The simple
CSS procedure starts from the scale and scheme invariance of a
given observable, which translates into complete freedom for the
choice of renormalization scale. According to common practice, a
first evaluation of the physical observable is obtained by
calculating perturbative corrections in a given scheme (commonly
used are $\rm MS$ or $\overline{\rm MS}$) and at an initial
renormalization scale $\mu_{r}=\mu_{r}^{\rm init}$. Thus, the
renormalization scale $\mu_r$ is set to one of the typical scales
of a process, $Q$, and errors are estimated by varying the scale
over a range $[Q/2,2Q]$.

This method is claimed to evaluate uncalculated contributions from
higher-order terms and that, due to the perturbative nature of the
expansion, the introduction of higher-order corrections would
reduce the scheme and scale ambiguities order by order. There is
no doubt that the higher the order of the loop corrections
calculated, the greater is the precision of the theoretical
estimates in comparison with the experimental data, but we cannot
determine \emph{a priori} the level of accuracy necessary for the
CSS to achieve the desired precision. And in the majority of cases
at present only the NNLO corrections are available. Moreover, the
divergent nature of the asymptotic perturbative series and the
presence of factorially growing terms (i.e.\ renormalons
\cite{Gross:1974jv, Lautrup:1977hs, Beneke:1998ui} severely
compromise the theoretical predictions.

However, even though this procedure may give an indication as to
the level of conformality and convergence reached by the truncated
expansion, it leads to a numerical evaluation of theoretical
errors that is quite unsatisfactory and strongly dependent on the
value of the chosen scale. Moreover, this method gives predictions
with large theoretical uncertainties that are comparable with the
calculated order correction; different choices of the
renormalization scale may lead to very different results when
higher-order corrections are included. For example, the NLO
correction to $W+3\,$jets with the BlackHat code
\cite{Berger:2009ep} can range from negligible to extremely
severe, depending on the choice of the particular renormalization
scale. One may argue that the proper renormalization scale for a
fixed-order prediction may be judged by comparing theoretical
results with experimental data, but this method would be strongly
process dependent and would compromise the predictivity of the
pQCD approach.

Besides the complexity of the higher-order calculations and the
slow convergence of the perturbative series, there are many
critical points in the CSS method:

\begin{itemize}
\item In general, the proper renormalization scale value, $Q$, is
not known, nor the correct range over which the scale and scheme
parameters should be varied in order to obtain the correct error
estimate. In fact, in some processes there can be more than one
typical momentum scale that may be taken as the renormalization
scale according to the CSS procedure; for example, in processes
involving heavy quarks typical scales are either the
center-of-mass energy $\sqrt{s}$ or also the heavy-quark mass.
Moreover, the idea of the typical momentum transfer as the
renormalization scale only sets the order of magnitude of the
scale, but does not indicate the optimal scale.

\item No distinction is made
among different sources of errors and their relative
contributions; e.g.\ in addition to the errors due to scale-scheme
uncertainties there are also errors from missing uncalculated
higher-order terms. In such an approach theoretical uncertainties
can become quite arbitrary and unreliable.

\item The convergence of the perturbative series in QCD is
affected by uncancelled large logarithms as well as by
``renormalon'' terms that diverge as $\left(n! \beta_{i}^{n}
\alpha_{s}^{n+1}\right)$ at higher orders \cite{Gross:1974jv,
Lautrup:1977hs}; this is known as the \emph{renormalon problem}
\cite{Beneke:1998ui}. Such renormalon terms can give sizable
contributions to the theoretical estimates, as shown in $e^+e^-$
annihilation, $\tau$ decay, deeply inelastic scattering and hard
processes involving heavy quarks. These terms are responsible for
important corrections at higher orders also in the perturbative
region, leading to different predictions according to the
different choices of scale (as shown in
Ref.~\cite{Berger:2009ep}). Large logarithms on the other hand can
be resummed using the resummation technique \cite{Catani:1991kz,
Catani:1992ua, Catani:1996yz, Aglietti:2006wh, DiGiustino:2011jn,
Banfi:2014sua, Abbate:2010xh} and results are IR renormalon free.
This does not help for the renormalization scale and scheme
ambiguities, which still affect theoretical predictions with or
without resummed large logarithms. In fact, as recently shown in
Ref.~\cite{Catani:2020kkl} for the $b\bar{b}$-production
cross-section at NNLO-order accuracy at hadron colliders, the CSS
scale setting leads to theoretical uncertainties that are of the
same order as the NNLO corrections $\sim20-30\%$, taking as the
typical momentum scale the $b$-quark mass $m_b\sim 4.92$\,GeV.

\item In the Abelian limit $N_{c}\rightarrow0$ at fixed
$\alpha_{em}=C_{F}\alpha_{s}$ with
$C_{F}=\left(N_{c}^{2}-1\right)/2N_{c}$, a QCD case effectively approaches
the analogous QED
case \cite{Brodsky:1997jk, Kataev:2010tm}. Thus, to be
self-consistent, any QCD scale-setting method should also be
extendable to QED and results should be in agreement with
the Gell-Mann--Low (GM--L) scheme. This is an important
requirement also from the perspective of a grand unified theory
(GUT), where only a single global method for setting the renormalization scale may be applied and then it can be considered as a good criterion
for verifying if a scale setting is correct or not. CSS leads to
incorrect results when applied to QED processes. In the GM--L
scheme, the renormalization scale is set with no ambiguity to the
virtuality of the exchanged photon/photons, which naturally sums an
infinite set of vacuum polarization contributions into the running
coupling. Thus, the CSS approach of varying the scale by a factor
2 is not applicable to QED, since the scale is already optimized.

\item The large amount of forthcoming high-precision experimental
data, produced especially by the running at high collision energy
and high luminosity of the Large Hadronic Collider (LHC), will
require more accurate and refined theoretical estimates. The CSS
appears to be more a ``guess''; its results are afflicted by large
errors and the perturbative series converges poorly with or
without large-logarithm resummation or renormalon contributions.
Moreover, within such a framework, it is almost impossible to
distinguish between SM and BSM signals and in many cases, improved
higher-order calculations are not expected to be available in the
short term.

\end{itemize}

To sum up, the conventional scale-setting method assigns an
arbitrary range and an arbitrary systematic error to fixed-order
perturbative calculations that greatly affects the predictions of
pQCD. On the other hand, various strategies for optimization of
the truncated expansion have been proposed. We point out that
since its very beginning, several attempts have been performed to
improve the renormalization procedure and different schemes have
been introduced mainly to improve the convergence of the
perturbative series. An example is the introduction of the
$\overline{\rm MS}$, as suggested in
Refs.\cite{DERUJULA1977315,Bardeen:1978yd} and later the
introduction of the momentum subtraction scheme (MOM) as discussed
by Celmaster and Gonsalves in Refs.
\cite{Celmaster:1979dm,Celmaster:1979km}. The MS scheme,
introduced in Ref. \cite{tHooft:1972tcz}, is in fact quite
arbitrary and related to the particular regularization scheme
used, leading to a rather higher second-order contribution in the
DIS process, by reabsorbing the scheme term $\ln(4\pi-\gamma_E)$
into a redefinition of the QCD scale parameter $\Lambda$, Bardeen
{\it et al.} noticed a significant improvement of the convergence
of the perturbative series. Later, the introduction of the MOM
scheme, though being gauge dependent, has been shown to also
include other scheme terms into $\Lambda$, leading to a scheme
which is less dependent on the regularization procedure and
tending to a more "physical" scale and nearly to an "optimization
procedure" \cite{Celmaster:1979km}. A recent discussion on the
gauge dependence of the MOM scheme has been given in Ref.
\cite{Zeng:2020lwi}. We discuss these schemes in more detail in
Sec. \ref{rsdependence}.

In general, an optimization or scale-setting procedure is
considered reliable if it preserves important self-consistency
requirements. All Renormalization Group properties, such as
\emph{uniqueness}, \emph{reflexivity}, \emph{symmetry} and
\emph{transitivity} should also be preserved by the scale-setting
procedure in order to be generally applied \cite{Brodsky:2012ms}.
A fundamental requirement is the scheme independence; other
requirements can be suggested by known tested theories (such as
QED), by the convergence behavior of the series in particular
kinematic regions or by important phenomenological results.

The Principle of Minimal Sensitivity (PMS) is an optimization
procedure proposed by Stevenson \cite{Stevenson:1982wn,
Stevenson:1982qw, Stevenson:1981vj, Stevenson:1980du} based on the
assumption that, since observables should be independent of the
particular RS and scale, their optimal perturbative approximations
should be stable under small RS variations. The RS-scheme
parameters $\beta_2,\beta_3,\dots$ and the scale parameter
$\Lambda$ (or the subtraction point $\mu_r$) are considered
``unphysical'' and independent variables; their values are thus
set in order to minimize the sensitivity of the estimate to their
small variations. This is essentially the core of Optimized
Perturbation Theory (OPT) \cite{Stevenson:1981vj}, based on the
PMS procedure. The convergence of the perturbative expansion is
improved by requiring its independence from the choice of RS
and~$\mu$. Optimization is implemented by identifying the
RS-dependent parameters in the truncated series ($\beta_{i}$ for
$2 \leq i \leq n$ and $\Lambda$), with the request that the
partial derivative of the perturbative expansion of the observable
with respect to the RS-dependent and scale parameters vanishes. In
practice, the PMS scale setting is designed to eliminate the
remaining renormalization and scheme dependence in the truncated
expansions of the perturbative series.

We would argue that this approach is based more on convergence
rather than physical criteria. In particular, the PMS is a
procedure that can be extended to higher order and it can be
generally applied to calculations obtained in arbitrary initial
renormalization schemes. Although this procedure leads to results
that are suggested to be unique and scheme independent,
it unfortunately violates important properties of the
renormalization group, as shown in Ref.~\cite{Wu:2013ei}, such as
reflexivity, symmetry, transitivity and also the \emph{existence}
and \emph{uniqueness} of the optimal PMS renormalization scheme are not
guaranteed since they are strictly related to the presence of
maxima and minima.

Another optimization procedure, namely the Fastest Apparent
Convergence (FAC) criterion was introduced by Grunberg and is
based on the idea of \emph{effective charges}. As pointed out by
Grunberg \cite{Grunberg:1980ja, Grunberg:1982fw, Grunberg:1989xf},
any perturbatively calculable physical quantity can be used to
define an effective coupling, or ``effective charge'', by entirely
incorporating the radiative corrections into its definition.
Effective charges can be defined from an observable starting from
the assumption that the infinite series of a given quantity is
scheme and scale invariant. The effective charge satisfies the
same renormalization group equations as the usual coupling. Thus,
the running behavior for both the effective coupling and the usual
coupling are the same if their RG equations are calculated in the
same renormalization scheme. This idea has been discussed in more
detail in Refs.~\cite{Dhar:1983py, Gupta:1990jq}.

An important suggestion is that all effective couplings defined in
the same scheme satisfy the same RG equations. While different
schemes or effective couplings would lead to different
renormalization group equations. Hence, any effective coupling can
be used as a reference for the particular renormalization
procedure.
In general, this method can be applied to any observable calculated
in any RS, also in processes with large higher-order corrections.
The FAC scale setting, as has been shown in Ref.~\cite{Wu:2013ei},
preserves the RG self-consistency requirements, although the FAC
method can be considered more an optimization approach rather than
a proper scale-setting procedure to extend order by order. FAC
results depend sensitively on the quantity to which the method is
applied. In general, when the NLO correction is large, the FAC
proves to be a resummation of the most important higher-order
corrections and thus a RG-improved perturbation theory is
achieved.

PMS and FAC are procedures commonly in use for scale setting in
perturbative QCD together with CSS and an introduction to these
methods can be found in Refs.~\cite{Wu:2013ei, Deur:2016tte}.
However, as shown in Refs.~\cite{Kramer:1990zt}, these
optimization methods not only have the same difficulties of CSS,
but they also lead to incorrect and unphysical results in
particular kinematic regions.

A solution to the renormalization scale setting problem is offered
by the Principle of Maximum Conformality (PMC)
\cite{Brodsky:2011ig,Brodsky:2011ta,Brodsky:2012rj,DiGiustino:2022ggl}.
This method is the generalization and extension of the original
Brodsky-Lepage-Mackenzie (BLM) method \cite{Brodsky:1982gc}  to
all theories, to all orders and to all observables and it
satisfies all the theoretical requirements of a reliable
scale-setting procedure at once, leading to accurate and
consistent results. The primary purpose of the PMC method is to
solve the scale-setting ambiguity; it has been extended to all
orders \cite{Mojaza:2012mf,Brodsky:2013vpa} and it determines the
correct running coupling and the correct momentum flow according
to RGE invariance \cite{Wu:2014iba, Wu:2019mky}. This leads to
results that are invariant with respect to the initial
renormalization scale and in agreement with the requirement of
scale invariance of observables in pQCD \cite{Wu:2013ei}. The
approach provides a systematic method to eliminate renormalization
scheme and scale ambiguities from first principles by absorbing
the $\beta$ terms, which govern the behavior of the running
coupling via the renormalization group equation. Thus, the
divergent renormalon terms cancel,  improving convergence of the
perturbative QCD series. Furthermore, the resulting PMC
predictions do not depend on the particular scheme used, thereby
preserving the principles of renormalization group invariance
\cite{Brodsky:2012ms,Wu:2014iba}. The PMC procedure is also
consistent with the standard Gell-Mann--Low method in the Abelian
limit, $N_c\rightarrow0$ \cite{Brodsky:1997jk}. Moreover, in a
theory unifying all forces (electromagnetic, weak and strong
interactions), such as Grand Unified Theories, one cannot
trivially apply a different scale-setting or analytic procedure to
different sectors of the theory. The PMC offers the possibility to
apply the same method to all sectors of a theory, starting from
first principles, eliminating renormalon growth, scheme
dependence, scale ambiguity and satisfying the QED Gell-Mann--Low
scheme in the zero-color limit $N_c\to0$.


\subsubsection*{PMC and schemes}

We remark that the fundamental task of the PMC is to solve the
renormalization scale and scheme ambiguities and in order to
achieve this, it makes use of the RG-equations to reabsorb the
$n_f$-terms that are related to the UV-divergent diagrams. The
procedure of the PMC works with \emph{any} initial definition of
the scheme or scale for the running coupling, MS or
 $\overline{\rm MS}$ or the 't Hooft scheme are all equivalent, since the entire scheme
dependence of the observable is reabsorbed into the running
coupling and into the PMC scale in the final series. In
particular, the lower order terms, $\beta_0,\beta_1$, of the
$\beta$-function are scheme independent; thus a scheme
transformation cannot lead to a prescription for determining the
scale. In fact, also in the case of the 't Hooft scheme (as shown
in Refs.\cite{Shrock:2013uaa,Shrock:2014qea}) one can eliminate
all scheme dependent $\beta_i$ coefficients, but there is no
prescription for the renormalization scale, which can be
considered as the first scheme parameter. Moreover, what seems a
good prescription for the running coupling is not necessarily a
good prescription for the entire fixed-order calculated quantity.
In fact, still considering the 't Hooft case, the entire series of
$\beta_i, i\geq 2$ is removed from the definition of the coupling,
but in a fixed-order calculation these $\beta_i$ coefficients
would have an impact on the coefficients of the series at each
order of calculation and on the convergence of the series. On the
other hand, the PMC gives a prescription for fixing the scale and
reabsorbing all scheme-dependent terms of a cross-section into the
running coupling and into the PMC scale, exposing the perturbative
series to a minimization/cancellation of the effects of the scale
and scheme uncertainties. An important consequence of the PMC
procedure is the RS-invariance of the resulting series. We refer
to RS-scheme invariance as the invariance under the
extended-renormalization group and its equations
(xRGE)\cite{Stevenson:1981vj,Ryttov:2012ur,Ryttov:2012nt,Chishtie:2016wob}.
It can be shown that the PMC procedure may be performed either way
using the same xRGE\footnote{We will discuss this procedure in
detail in a future work soon.}. The xRGE show that different
scheme definitions can be related at the lowest order by a scale
transformation for the case of the minimal subtraction schemes (MS
and $\overline{\rm MS}$) and the momentum space subtraction (MOM)
scheme, but using the PMC, results would be scheme independent. A
recent argument on the scheme dependence of the first conformal
coefficient by Stevenson \cite{Stevenson:2023xvx} is incorrect.
The reason is that he assumes the $v_1$ coefficient of the scheme
to be totally free with respect to the structure of the color
factors and that it can be reabsorbed into the $\Lambda$
parameter. There are several reasons why this certainly leads to
wrong results. First, two different couplings in two different
schemes at NLO can only be related by a scale transformation
according to the RG group, i.e. by a shift of the type: $\beta_0
\log(\frac{\mu_1^2}{\mu_2^2})$ as shown also in Ref.
\cite{Chishtie:2016wob}. In fact, given the scheme invariance of
the $\beta_0, \beta_1$, the only parameter that can be varied at
NLO is the renormalization scale $\mu_r$, which can have any real
value or at most any value in the complex plane. Thus, the only
scheme terms that can be reabsorbed into the scale (either the
$\Lambda$ or the renormalization scale $\mu_r$) are those related
to a shift of the scale using the standard RG group equations.
This corresponds to the freedom of subtracting out any finite
value together with the pole, in the renormalization procedure
(e.g. MS and $\overline{\rm MS}$). If one takes the freedom to
vary the scheme according to any other relation that does not
correspond to a "proper" RG equation at NLO, this consequently
modifies the structure of the color factors for the coupling and
thus would obtain a wrong result. If one follows this misleading
assumption, one can obtain any result outside of a given initial
theory and this is certainly not the aim of the xRGE, which should
preserve the invariance of the final result. Moreover, if one
changes the color structure for the coupling, one should be aware
that, to be consistent, also the structure of the entire
fixed-order calculation should be varied accordingly; though we do
not agree with this procedure since a change of the color
structure correspond to a change of the initial SU(N) theory (e.g.
from QCD to QED Ref.\cite{Brodsky:1997jk}). Thus, any relation
among couplings in different schemes, which can have perturbative
validity in QCD but which cannot be considered as "proper" xRG
transformations, should be considered as matching relations among
quantities defined in different approximations or obtained using
different approaches. Also in this case, results can be improved
using the PMC and the residual dependence on the particular
implicit definition of the "scheme" can be suppressed
perturbatively by adding higher-order calculations. It can be
shown that also in this case the results obtained are
scheme-independent (e.g. see Ref. \cite{Wu:2018cmb}). One may
point out that, though the LO and the NLO conformal coefficient
are scheme invariant, the higher-order conformal coefficients can
be scheme dependent, we can answer that in the worst case this
scheme dependence for any of the PMCs procedures is highly
suppressed. Another argument in Ref. \cite{Stevenson:2023xvx} by
Stevenson, which is based on the principle of minimum sensitivity
(PMS), we hold to be incorrect. Since the PMS is based on the
assumption that all the unknown higher-order terms give zero
contribution to the pQCD series \cite{Stevenson:1981vj}, its
prediction directly breaks the standard renormalization group
invariance \cite{Brodsky:2012ms,Wu:2014iba}, its pQCD series does
not have normal perturbative features \cite{Ma:2014oba} and can be
treated as an effective prediction only when we know the series up
to high enough orders and the conventional series has already
shown good perturbative features \cite{Ma:2017xef}. On the other
hand, the PMC respects all features of the renormalization group,
and its prediction satisfies all the requirements of standard
renormalization group invariance
\cite{Brodsky:2012ms,Wu:2014iba,Wu:2018cmb,Wu:2019mky}. Moreover,
given that the PMC preserves the RG invariance, it is possible to
define CSR - Commensurate Scale Relations\cite{Brodsky:2013vpa}
among the effective charges relating observables in different
"schemes" preserving all the group properties. Applying the PMC
and the CSRs, one can relate effective couplings, as also
conformal coefficients, leading to scheme-independent results for
the observables. We remark that in order to apply the PMC
correctly, one should be able to distinguish among the nature of
the different $n_f$-terms: whether they are related to the running
of the coupling, to the running of masses or to UV-finite diagrams
and in a deeper analysis also to the particular UV-divergent
diagram (as discussed in Refs.
\cite{DiGiustino:2023jiq,DiGiustino:2022ggl}). Once all
$n_f$-terms have been associated with the correct diagram or
parameter, conformal coefficients are RG invariant and match the
coefficients of a conformal theory. Applications of the PMC to
different quantities (see Ref. \cite{Huang:2021hzr,Wang:2020ckr})
have recently shown a direct improvement of theoretical
predictions. Moreover, a deeper insight into the QCD coupling
$\alpha_s(Q)$ at all scales, including $Q^2=0$, has been recently
achieved (e.g. see Refs. \cite{Deur:2023dzc, Deur:2017cvd}),
showing results consistent with the PMC.


\subsubsection*{PMC applications}

 So far, the PMC approach
has been successfully applied to many high-energy processes,
including Higgs boson production at the LHC \cite{Wang:2016wgw},
Higgs boson decays to $\gamma\gamma$ \cite{Wang:2013akk,
Yu:2018hgw}, $gg$ and $b\bar{b}$ \cite{Wang:2013bla, Zeng:2015gha,
Zeng:2018jzf, Gao:2021wjn, Yan:2022foz, Wang:2020ckr} processes,
top-quark pair production at the LHC and Tevatron
\cite{Brodsky:2012sz, Brodsky:2012rj, Brodsky:2012ik,
Wang:2014sua, Wang:2015lna, Wang:2017kyd, Wang:2020mel}, decay
process \cite{Meng:2022htg, Wang:2023ttk}, semihard processes
based on the BFKL approach \cite{Hentschinski:2012kr,
Zheng:2013uja, Caporale:2015uva}, electron--positron annihilation
to hadrons \cite{Mojaza:2012mf, Brodsky:2013vpa, Wu:2014iba},
hadronic $Z^0$ boson decays \cite{Wang:2014aqa, Huang:2020skl},
the event shapes in electron--positron annihilation
\cite{Wang:2019ljl, Wang:2019isi, DiGiustino:2020fbk,
DiGiustino:2021nep, Wang:2021tak}, the electroweak parameter
$\rho$ \cite{Wang:2014wua, Yu:2021ujk}, $\Upsilon(1S)$ leptonic
decay \cite{Shen:2015cta, Huang:2019frb}, and charmonium
production \cite{Wang:2013vn, Sun:2018rgx, Yu:2020tri} and decay
\cite{Zhang:2014qqa, Du:2017lmz, Yu:2019mce, Zhou:2021zvx}. As
shown in Ref.~\cite{Yu:2021yvw}, by using the PMC, one can obtain
a smooth transition for the running behavior between Bjorken sum
rule effective coupling $\alpha_s^{g_1}(Q)$ in the V-scheme and
the Light-Front Holographic QCD (LFHQCD) coupling, going from the
perturbative to the non nonperturbative domain respectively. In
addition, the PMC provides a possible solution to the $B\to\pi\pi$
puzzle \cite{Qiao:2014lwa} and to the
$\gamma\gamma^*\rightarrow\eta_c$ puzzle \cite{Wang:2018lry}.

In particular, with regard to the PMC application to top-quark
pair hadroproduction, the resulting production cross-sections
agree with precise experimental data, and the large discrepancies
of the top-quark forward-backward asymmetries between SM
estimations and experimental measurements are greatly
reduced~\cite{Brodsky:2012rj,Brodsky:2012sz,Brodsky:2012ik,Wang:2014sua,Wang:2015lna}.
Recently, an improved QCD prediction for the top-quark decay
process obtained by using the PMC has also been presented in
Ref.\cite{Meng:2022htg}.
\begin{figure} [htb]
\centering
\includegraphics[width=0.60\textwidth]{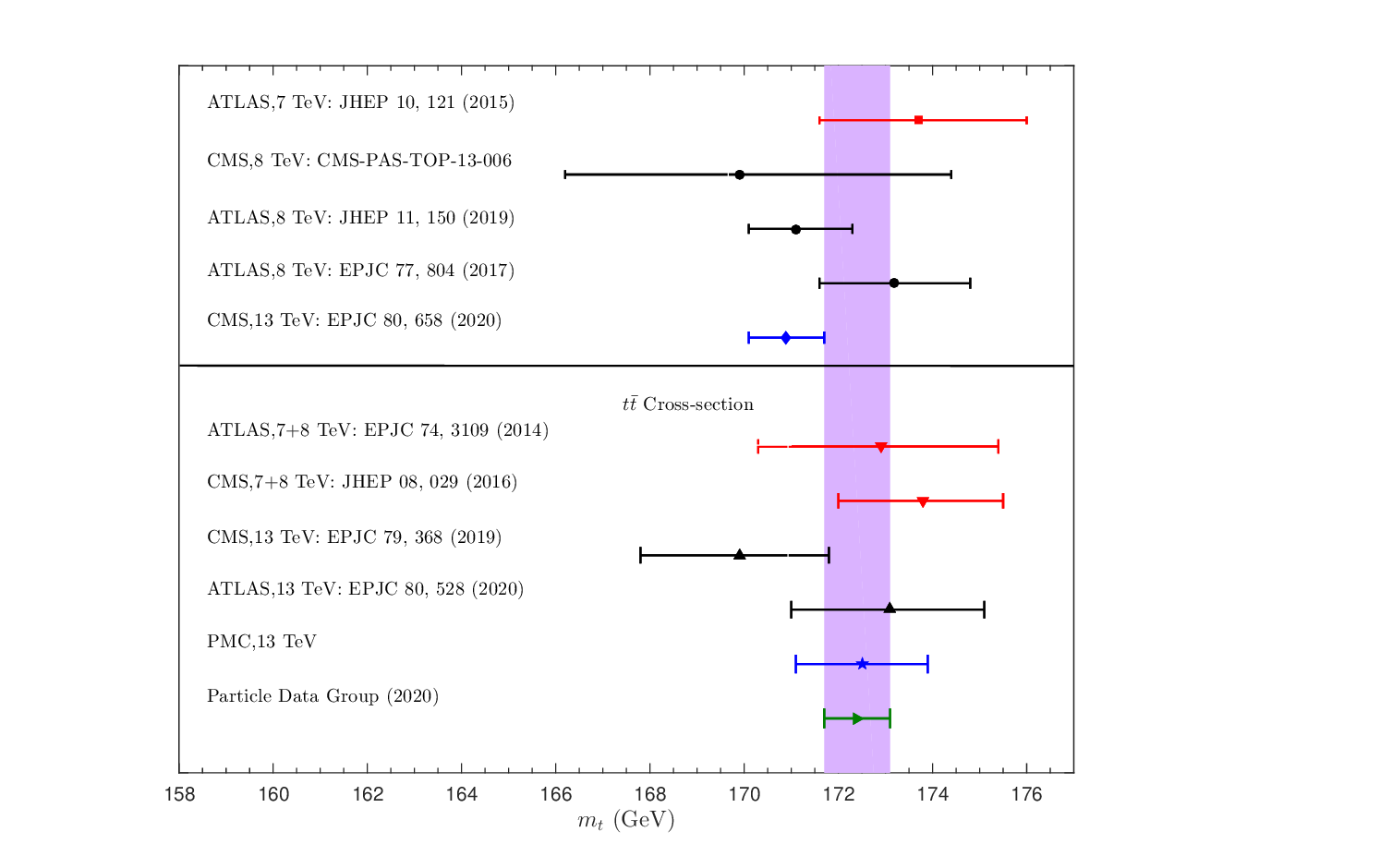}
\caption{ Summary of the top-quark pole masses,
 where the PMC result and previous determinations from collider measurements at different energies and different techniques are presented.
 The top-quark pole mass from the PDG~\cite{ParticleDataGroup:2020ssz} is also presented as the shaded band for reference.}
\label{masscompa}
\end{figure}
A precise top-quark mass can be extracted by the comparison of
precise PMC predictions of production cross-sections with
experimental data measured at
LHC~\cite{Wang:2017kyd,Wang:2020mel}. The determined top-quark
pole mass $m_t=172.5\pm1.4$ GeV, from the LHC measurement at
$\sqrt{s}=13$ TeV~\cite{Wang:2020mel} agrees with the world
average cited by the Particle Data Group
(PDG)~\cite{ParticleDataGroup:2020ssz}. More explicitly, we
present a summary of the top-quark pole masses in
Fig.(\ref{masscompa}),
 where our PMC result and previous determinations from collider
 measurements at different energies and different techniques are presented.
Owing to unknown higher-order contributions, this leads to two
kinds of residual scale dependence for PMC predictions. However,
these two residual scale dependencies are distinct from the
conventional scale ambiguities, they are given by the unknown
higher-order uncalculated contributions, while the scale
dependence of the CSS is related more to the calculated orders.
Thus, the residual scale dependence of the PMC results from the
unknown higher-order terms and is an unavoidable intrinsic feature
of the perturbative approach. It is to be noted that two of the
three PMC ambiguities identified in Ref.\cite{Chawdhry:2019uuv}
correspond exactly to two kinds of residual scale dependence.
These two residual scale dependencies are highly suppressed;
however, if the pQCD convergence of the perturbative series of
either the PMC scale or the pQCD approximant is weak, such
residual scale dependence could be significant. A recent
discussion of the residual scale dependence of PMC predictions can
also be found in the review~\cite{Wu:2019mky}.


\subsubsection*{Outline}

In this review we display recent developments in solving the
renormalization-scale and -scheme ambiguities based on the PMC and
present some fundamental applications and results.

With respect to the previous reviews on the PMC (Refs.
\cite{Wu:2013ei,Wu:2019mky}), in this review we focus on the
recently developed method, namely the {\emph{Infinite-Order
Scale-Setting using the Principle of Maximum Conformality}}
(PMC$_\infty$), showing first some of its applications and new
features, and then a comparison of this method with the CSS and
the other PMC approaches, such as the multi-scale (PMCm), and the
single-scale PMC (PMCs), on the Event Shape Variables and on some
crucial fully integrated quantities: $R_{e^+e^-}$, $R_\tau$ and
$\Gamma(H \to b \bar{b})$.

More in detail: in \textbf{Section~I} we give an introduction to
the renormalization scale setting problem in QCD; in
\textbf{Section~II} we recall fundamental equations of the
renormalization group and their extended version, starting from
the renormalization procedure of the strong coupling
$\alpha_s(\mu)$ and its renormalization scale dependence; in
\textbf{Section~III} we summarize formulas and basic concepts of
the PMC approach with its features and methods (PMCm, PMCs); in
\textbf{Section~IV} we introduce the newly developed method
PMC$_\infty$ and its new features; in \textbf{Section~V} we show
results for the application of the PMC$_\infty$ to the event-shape
variables: thrust and $C$-parameter, we show particularly new
features of the PMC performing interesting limits for thrust (IR
conformal and QED limit) and comparing the results under the CSS
and the PMC$_\infty$ methods; in this section we also show a novel
method to determine the strong coupling and its behavior
$\alpha_s(Q)$ over a wide range of scales, from a single
experiment at a single scale, using the event-shape variable
results; in \textbf{Section~VI} we present a detailed comparison
of the CSS, PMCs, PMCm and PMC$_\infty$ methods tested on
fundamental fully integrated quantities at 4~loops: $R_{e^+e^-}$,
$R_\tau$ and $\Gamma(H \to b \bar{b})$; finally, in
\textbf{Section~VII} we summarize and discuss the results.


  \newpage

\section{Renormalization Scale and Scheme invariance}
\label{rginvariance}
The focus of this section is the strong coupling constant and its
renormalization-scale and -scheme dependence. We summarize
fundamental basic theoretical results and updated formulas
regarding the renormalization group.

\subsection{The Renormalization Group \label{RGroup}}

QCD is a renormalizable theory, which means that the infinite
number of ultraviolet (UV) singularities that arise in loop
integration may be reabsorbed into a finite number of parameters
entering the Lagrangian: the masses, coupling constant and fields.

The procedure starts from the assumption that the variables
entering the Lagrangian are not the effective quantities measured
in experiments, but are unknown functions affected by
singularities. The origin of the ultraviolet singularities is
often interpreted as a manifestation that a QFT is a low-energy
effective theory of a more fundamental yet unknown theory. The use
of regularization UV cut-offs shields the very short-distance
domain, where the perturbative approach to QFT ceases to be valid.

Once the coupling has been renormalized to a measured value and at
a given energy scale, the effective coupling is no longer
sensitive to the ultraviolet (UV) cut-off, nor to any unknown
phenomena arising beyond this scale. Thus, the scale dependence of
the coupling can be well understood formally and
phenomenologically. Actually, gauge theories are affected not only
by UV, but also by infrared (IR) divergencies. The cancellation of
the latter is guaranteed by the Kinoshita-Lee-Nauenberg (KLN)
theorem \cite{Kinoshita:1962ur, Lee:1964is}.

Considering first the Lagrangian of a massless theory, which is
free of any particular scale parameter, in order to deal with
these divergences a regularization procedure is introduced.
Referring to the dimensional regularization procedure
\cite{tHooft:1972tcz, Cicuta:1972jf, Bollini:1972ui}, one varies
the dimension of the loop integration, $D=4-2\varepsilon$ and
introduces a scale $\mu$ in order to restore the correct dimension
of the coupling.

In order to determine the renormalized gauge coupling, we consider
the quark-quark-gluon vertex and its loop corrections.
UV-divergences arise from loop integration for higher-order
contributions for both the external fields and the vertex. The
renormalization constants of the vertices are related by
Slavnov-Taylor identities, in particular: \be
Z^{-1}_{\alpha_s}=(\sqrt{Z_3} Z_2 /Z_1)^2, \label{eqn:zalphas} \ee
where $Z_{\alpha_s}$ is the coupling renormalization constant,
$Z_1\left(Q \right)$ is the vertex renormalization constant,
$Z_3\left(Q \right)$ and $Z_{2}\left(Q \right)$ are:
$$
A_{a, \mu}=Z_3^{1 / 2}\left(Q \right) A_{a, \mu}^{R}\left(Q
\right) \quad \text{and} \quad \psi=Z_{2}^{1 / 2}\left(Q \right)
\psi^{R}\left(Q \right),
$$
the renormalization constants for gluon and quark fields
respectively. The superscript $R$ indicates renormalized fields.
The renormalization constants in dimensional regularization are
given by: \bea Z_1(Q)&=&
1-\frac{\alpha_s(Q)}{4\pi} \big(N_c+C_F\big) \frac{1}{\varepsilon} \\
Z_2(Q)&=& 1-\frac{\alpha_s(Q)}{4\pi} C_F  \frac{1}{\varepsilon} \\
Z_3(Q) &=& 1+\frac{\alpha_s(Q)}{4\pi}
\left(\frac{5}{3}N_c-\frac{2}{3}N_f\right) \frac{1}{\varepsilon}
 \eea
where ${\rm \varepsilon}$ is the regulator parameter for the
UV-ultraviolet singularities. We have labelled $N_f$ the number of
flavors related to the UV singular diagrams.\footnote{Throughout
the paper we use the following notation: $n_f$ for the general
number of active flavors, $N_f$ to indicate the number of flavors
related to the UV-divergent and $N_F$ for the UV-finite diagrams.}
By substitution, we have that the UV divergence: \be
Z_{\alpha}\left(Q \right)=1-\frac{\alpha_{s}\left(Q \right)}{4
\pi} \beta_{0} \frac{1}{\varepsilon}, \ee with \be
\beta_{0}=11-\frac{2 N_f}{3}. \ee

The singularities related to the UV poles are subtracted out by a
redefinition of the coupling. In the $\rm MS$ scheme, the
renormalized strong coupling $\alpha_s(Q)$ is related to the bare
coupling $\overline{\alpha_{s}}$ by: \be
\overline{\alpha_{s}}=Q^{2 \varepsilon} Z_{\alpha}\left(Q \right)
\alpha_{s}\left(Q \right). \label{eqn:alphasren}\ee In the minimal
subtraction scheme ($\rm MS$) only the pole $1/\varepsilon$
related to the UV singularity is subtracted out. A more suitable
scheme is $\overline{\rm MS}$ \cite{Bardeen:1978yd,
tHooft:1973mfk, Weinberg:1973xwm}, where the constant term $\ln(4
\pi)-\gamma_E$ is also subtracted out. Different schemes can also
be related by scale redefinition, e.g.\ $\mu^2 \rightarrow 4 \pi
\mu^2 e^{-\gamma_E}$. We remark that the renormalization procedure
leads to a unique renormalization constant $Z_{\alpha_s}\equiv
Z^2_g $ for the strong coupling. In fact, the other
renormalization constants, such as $Z_1^V$, $V\in(3g, 4g, ccg,
qqg)$, i.e.\ the renormalization constants for 3-gluon, 4-gluon,
ghost-ghost-gluon, quark-quark-gluon vertex respectively, are
related to it via the Slavnov--Taylor identities
\cite{Chetyrkin:2004mf}: \bea
Z_g & = & Z_1^{3 g}\left(Z_3\right)^{-3/2}, \\
Z_g & = & \sqrt{Z_1^{4 g}}\left(Z_3\right)^{-1}, \\
Z_g & = & Z_1^{ccg}\left(Z_3\right)^{-1/2}\left(Z_3^c\right)^{-1}, \\
Z_g & = & Z_1^{\psi\psi g}\left(Z_3\right)^{-1/2}\left(Z_2\right)^{-1}.
\eea

Thus, the renormalization procedure depends both on the particular
choice of scheme and on the subtraction point~$\mu$. Hence,
even though there are no dimensionful parameters in the initial
bare Lagrangian, a mass scale $\mu$ is acquired during the
renormalization procedure. The emergence of $\mu$ from a
Lagrangian with no explicit scale is called dimensional
transmutation \cite{Coleman:1973sx}. The value of $\mu$ is
arbitrary and is the momentum at which the UV divergences are
subtracted out. Hence $\mu$ is called the subtraction point or
renormalization scale. Thus, the definition of the renormalized
coupling $\alpha_s^{\overline{\rm MS}}(\mu)$ depends at the same
time on both the chosen scheme, $\overline{\rm MS}$ and the
renormalization scale~$\mu$.

Renormalization scale invariance is recovered by introducing
the Renormalization Group Equations.
The scale dependence of the coupling can be determined by considering
that the bare coupling $\overline{\alpha_{s}}$ and renormalized
couplings, $\alpha_s$, at different scales are related by:
 \be
\overline{\alpha_{s}}=Q^{2 \varepsilon} Z_{\alpha}\left(Q \right)
\alpha_{s}\left(Q \right)=\mu^{2 \varepsilon} Z_{\alpha}\left(\mu
\right) \alpha_{s}\left(\mu \right),
 \label{eqn:alphascales}\ee
where $\varepsilon$ is the regularization parameter, integrals are
carried out in $4-2 \varepsilon$ dimensions and the UV divergences
are regularized to $1 / \varepsilon$ poles. The $Z_{i}$ are
constructed as functions of $1 / \varepsilon$, such that they
cancel all $1 / \varepsilon$ poles. From
Eq.~\eqref{eqn:alphascales} we obtain the relation from two
different couplings at two different scales: \be \alpha_{s}\left(Q
\right)=\mathcal{Z}_{\alpha}\left(Q, \mu\right)
\alpha_{s}\left(\mu \right), \ee with $\mathcal{Z}_{\alpha}\left(Q
, \mu \right) \equiv\left(\mu^{2 \varepsilon} / Q^{2
\varepsilon}\right)\left[Z_{\alpha}\left(Q \right) /
Z_{\alpha}\left(\mu \right)\right]$. The $\mathcal{Z}_{\alpha}$
then clearly form a group with a composition law \be
\mathcal{Z}_{\alpha}\left(Q , \mu
\right)=\mathcal{Z}_{\alpha}\left(Q , \mu_{0} \right)
\mathcal{Z}_{\alpha}\left(\mu_{0}, \mu \right),\ee a unity element
$\mathcal{Z}_{\alpha}\left(Q , Q \right)=1$ and an inverse:
$\mathcal{Z}_{\alpha}\left(Q , \mu
\right)=\mathcal{Z}_{\alpha}^{-1}\left(\mu , Q \right)$. The
fundamental properties of the Renormalization Group are:
\emph{reflexivity}, \emph{symmetry} and \emph{transitivity}. Thus,
the scale invariance of a given perturbatively calculated quantity
is recovered by the invariance of the theory under the
Renormalization Group Equations (RGE).


\subsection{The evolution of $\alpha_s(\mu)$ in perturbative QCD}

As shown in the previous section the renormalization procedure is
not void of ambiguities. The subtraction of the singularities
depends on the subtraction point or renormalization scale $\mu$
and on the renormalization scheme (RS). Physical observables
cannot depend on the particular scheme or scale, given that the
theory stems from a conformal Lagrangian. This implies that scale
invariance must be recovered imposing the invariance of the
renormalized theory under the renormalization group equation
(RGE). In this section we discuss the dependence of the
renormalized coupling $\alpha_s(Q)$ on the scale~$Q$. As shown in
QED by Gell-Mann and Low, this dependence can be described
introducing the $\beta$-function given by: \be \frac{1}{4
\pi}\frac{d\alpha_s(Q)}{d \log Q^2}=\beta(\alpha_s)
\label{betafun1}\ee and \be
\beta\left(\alpha_{s}\right)=-\left(\frac{\alpha_{s}}{4
\pi}\right)^{2} \sum_{n=0} \left(\frac{\alpha_{s}}{4
\pi}\right)^{n} \beta_{n}. \label{betafun10}\ee

Neglecting quark masses, the first two $\beta$-terms are RS
independent and have been calculated in
Refs.~\cite{Gross:1973id, Politzer:1973fx, Caswell:1974gg, Jones:1974mm, Egorian:1978zx}
for the $\overline{\rm MS}$ scheme:
\begin{eqnarray}
\beta_{0}&=& \frac{11}{3}C_{A}\!-\!\frac{4}{3}T_{R}N_{f},\\
\beta_{1}&=&
\frac{34}{3}C_{A}^{2}\!-\!4\left(\frac{5}{3}C_{A}\!+\!C_{F}\right)T_{R}N_{f}
\end{eqnarray}
where $C_F=\frac{\left(N_{c}^{2}-1\right)}{2 N_{c}}$, $C_A=N_c$
and $T_R=1/2$ are the color factors for the ${\rm SU(3)}$ gauge group \cite{Mojaza:2010cm}.
At higher loops $\beta_i,$ $i\geq 2$ are scheme
dependent; results for $\overline{\rm MS}$ are given in
Ref.~\cite{Larin:1993tp}
\begin{eqnarray}
 \beta_{2}&=&\frac{2857}{2}-\frac{5033}{18}
N_{f}+\frac{325}{54} N_{f}^{2},
\end{eqnarray}
in Ref.~\cite{vanRitbergen:1997va}
\begin{eqnarray}
\beta_3&=& \left(\frac{149753}{6}+3564  \zeta_3 \right)- \left(\frac{1078361}{162}+\frac{6508}{27}  \zeta_3 \right) N_{f} \nonumber \\
&&\quad+ \left(\frac{50065}{162}+\frac{6472}{81} \zeta_3 \right)
N_{f}^{2}+\frac{1093}{729} N_{f}^{3},
\end{eqnarray}
and in Ref.~\cite{Baikov:2016tgj}
\begin{eqnarray}
\beta_{4}&=& \left\{\frac{8157455}{16}+\frac{621885}{2} \zeta_3-\frac{88209}{2} \zeta_{4}-288090 \zeta_{5}\right. \nonumber \\
&&\quad+N_{f}\left[-\frac{336460813}{1944}-\frac{4811164}{81} \zeta_3+\frac{33935}{6}\zeta_{4}+\frac{1358995}{27}\zeta_5 \right]
\nonumber \\
&&\quad+N_{f}^{2}\left[\frac{25960913}{1944} +\frac{698531}{81} \zeta_3-\frac{10526}{9} \zeta_{4}-\frac{381760}{81}\zeta_{5}\right]
\nonumber \\
&&\quad+N_{f}^{3}\left[-\frac{630559}{5832}-\frac{48722}{243}
\zeta_3+\frac{1618}{27} \zeta_{4}+\frac{460}{9} \zeta_{5}\right]
+ \left.
N_{f}^{4}\left[\frac{1205}{2916}-\frac{152}{81}\zeta_3\right]\right\},
\end{eqnarray}
with $\zeta_3 \simeq 1.20206$, $ \zeta_{4}\simeq
1.08232 $ and $ \zeta_{5} \simeq 1.03693$, the Riemann zeta
function. Given the renormalizability of QCD, new UV singularities
arising at higher orders can be cancelled by redefinition of the
same parameter, i.e.\ the strong coupling. This procedure leads to
the renormalization constant:
\begin{eqnarray}
Z_{a}(\mu)&=& 1-\frac{\beta_{0}}{\epsilon} a
+\left(
  \frac{\beta_{0}^{2}}{\epsilon^{2}}-\frac{\beta_{1}}{2\epsilon}
\right)a^{2}
-\left(\frac{\beta_{0}^{3}}{\epsilon^{3}}-\frac{7}{6} \frac{\beta_{0} \beta_{1}}{\epsilon^{2}}+\frac{\beta_{2}}{3 \epsilon}\right) a^{3}
\nonumber \\
& &\qquad +\left(\frac{\beta_{0}^{4}}{\epsilon^{4}}-\frac{23 \beta_{1}
\beta_{0}^{2}}{12 \epsilon^{3}}+\frac{5 \beta_{2} \beta_{0}}{6
\epsilon^{2}}+\frac{3 \beta_{1}^{2}}{8
\epsilon^{2}}-\frac{\beta_3}{4 \epsilon}\right) a^{4}+\cdots,
\label{eqn:zexp}
\end{eqnarray}
where $a=\alpha_s(\mu)/(4 \pi)$ in the $\rm MS$ scheme. Given the
arbitrariness of the subtraction procedure of also including part
of the finite contributions (e.g.\ the constant $[\ln(4
\pi)-\gamma_E]$ in $\overline{\rm MS}$), there is an inherent
ambiguity for these terms that translates into the RS dependence.
In order to solve any truncated version of Eq.~\eqref{betafun1}, this being a
first order differential equation, we need an initial value of
$\alpha_s$ at a given energy scale~$\alpha_s(\mu_0)$. For this
purpose, we set the initial scale $\mu_0=M_{Z}$ the $Z^0$ mass, with
the value $\alpha_s(M_Z)$ being determined phenomenologically. In QCD
the number of colors $N_c$ is set to 3, while $N_f$, i.e.\ the
number of active flavors, varies with energy scale across quark
thresholds.


\subsubsection{One-loop result and asymptotic freedom\label{oneloop}}

When all quark masses are set to zero, two physical parameters
dictate the dynamics of the theory and these are the numbers of
flavors $N_f$ and colors~$N_c$. In this section we determine the
exact analytical solution to the truncated Eq.~\eqref{betafun1}.
Considering the formula: \be
\int_{\alpha_s(\mu_0)}^{\alpha_s(\mu)} \frac{1}{4\pi}
\frac{d\alpha_{s}}{\beta(\alpha_{s}) }=-\int_{\mu_0^2}^{\mu^2}
\frac{d Q^{2}}{Q^{2}}, \ee and retaining only the first term: \be
\frac{Q^{2}}{\alpha_{s}^{2}} \frac{\partial \alpha_{s}}{\partial
Q^{2}}=-\frac{1}{4 \pi} \beta_{0} \ee we achieve the solution for
the coupling: \be \frac{4
\pi}{\alpha_{s}\left(\mu_{0}\right)}-\frac{4
\pi}{\alpha_{s}\left(Q\right)}=\beta_{0} \ln
\left(\frac{\mu_{0}^{2}}{Q^{2}}\right). \label{1loopalphas} \ee
This solution can be given in the explicit form: \be \alpha_{s}(Q)
= \frac{\alpha_s(\mu_0)}{1+\beta_0 \frac{\alpha_s(\mu_0)}{4 \pi}
\ln(Q^2/\mu_0^2)}. \label{1loopalphas2} \ee This solution relates
one known (measured value) of the coupling at a given scale
$\mu_0$ with an unknown value~$\alpha_s(Q)$. More conveniently,
the solution can be given introducing the QCD scale
parameter~$\Lambda$. At $\beta_{0}$ order, this is defined as: \be
\Lambda^{2} \equiv \mu^{2} e^{-\frac{4 \pi}{\beta_{0}
\alpha_{s}\left(\mu\right)}}, \label{lambda1loop}\ee which yields
the familiar one-loop solution:
$$
\alpha_{s}\left(Q\right)=\frac{4 \pi}{\beta_{0} \ln \left(Q^{2} /
\Lambda^{2}\right)}.
$$

Already at the one-loop level one can distinguish two regimes of
the theory. For the number of flavors larger than $11N_c/2$ (i.e.\ the zero of the $\beta_0$ coefficient) the theory possesses an
infrared noninteracting fixed point and at low energies the
theory is known as non-Abelian quantum electrodynamics
(non-Abelian QED). The high-energy behavior of the theory is
uncertain, it depends on the number of active flavors and there is
the possibility that it could develop a critical number of flavors
above which the theory reaches an UV fixed
point \cite{Antipin:2017ebo} and therefore becomes safe. When the
number of flavors is less than $11N_c/2$, the noninteracting fixed
point becomes UV in nature and then we say that the theory is \emph{asymptotically free}.

It is straightforward to check the asymptotic limit of the
coupling in the deep-UV region: \be \lim_{s \rightarrow \infty}
\alpha_s(s)=0. \ee This result is known as \emph{asymptotic
freedom} and it is the outstanding result that has justified QCD
as the most accredited candidate for the theory of strong
interactions. On the other hand, we have that the perturbative
coupling diverges at the scale $\Lambda\sim(200-300)\,{\rm MeV}$.
This is sometimes referred to as the \emph{Landau ghost pole} to
indicate the presence of a singularity in the coupling that is
actually unphysical and implies the breakdown of the perturbative
regime. This itself is not an explanation for confinement, though
it might indicate its presence. When the coupling becomes too
large, the use of a nonperturbative approach to QCD is mandatory
in order to obtain reliable results. We remark that the scale
parameter $\Lambda$ is RS dependent and its definition depends on
the order of accuracy of the coupling~$\alpha_s(Q)$. Considering
that the solution $\alpha_{s}$ at order $\beta_{0}$ or $\beta_{1}$
is universal, the definition of $\Lambda$ at the first two orders
is usually preferred, i.e.\ $\Lambda$ given at 1-loop by
Eq.~\eqref{lambda1loop} or at 2-loops (see later) by
Eq.~\eqref{landaupole}.

\subsubsection{Two-loop solution and the perturbative conformal
window\label{twoloops}}


In order to determine the solution for the strong coupling
$\alpha_s$ at NNLO, it is convenient to introduce the following
notation: $x(\mu)\equiv \frac{\alpha_s(\mu)}{2 \pi}$,
$t=\log(\mu^2/\mu_0^2)$, $B=\frac{1}{2}\beta_0$ and
$C=\frac{1}{2}\frac{\beta_1}{\beta_0}$, $x^*\equiv-\frac{1}{C}$.
The truncated NNLO approximation of the Eq.~\eqref{betafun1} leads
to the differential equation:
\begin{equation}
\frac{dx}{dt}=-B x^2(1+C x) \label{lambert1}
\end{equation}
An implicit solution of Eq.~\eqref{lambert1} is given by the Lambert
$W(z)$ function:
\begin{equation}
W e^W = z \label{W}
\end{equation}
with: $ W=\left(\frac{x^*}{x}-1\right)$. The general solution for
the coupling is:
\begin{eqnarray}
x &=& \frac{x^*}{1+W}, \label{xz1} \\
 z &=& e^{\frac{x^*}{x_0}-1} \left(\frac{x^*}{x_0}-1 \right) \left( \frac{\mu^2}{\mu_0^2}
\right)^{x^* B}. \label{xz}
\end{eqnarray}
Here we shall discuss the solutions to Eq.~\eqref{lambert1} with
respect to the particular initial phenomenological value
$x_0\equiv \alpha_s(M_Z) /(2\pi)= 0.01877 \pm 0.00014$ given by
the coupling determined at the $Z^0$ mass scale
\cite{Workman:2022ynf}.

The signs of $\beta_0,\beta_1$ and consequently of $B,x^*$,
depend on the values of the $N_c,N_f$, since the number $N_c$ is
set by the SU$(N_c)$ theory, we discuss the possible regions
varying only the number of flavors~$N_f$. We point out that
different regions are defined by the signs of the
$\beta_0,\beta_1$, which have zeros in
$\bar{N_f^0}=\frac{11}{2}N_c$, $\bar{N_f^1}=\frac{34 N_c^3}{13
N_c^2-3}$ respectively with $\bar{N_f^0}> \bar{N_f^1}$.
In the range $N_f<\bar{N_f^1}$ and $N_f>\bar{N_f^0}$ we have
$B>0$, $C>0$ and the physical solution is given by the $W_{-1}$
branch, while for $\bar{N_f^1}< N_f < \bar{N_f^0}$ the solution
for the strong coupling is given by the $W_{0}$ branch. By
introducing the phenomenological value $x_0$, we define a
restricted range for the IR fixed point discussed by Banks and
Zaks \cite{Banks:1981nn}. Given the value $\bar{N}_f =x^{*-1}(x_0)
= 15.222 \pm 0.009$, we have that in the range $\frac{34 N_c^3}{13
N_c^2-3}< N_f<\bar{N}_f$ the $\beta$-function has both a UV and an
IR fixed point, while for $N_f> \bar{N}_f$ we no longer have the
asymptotically free UV behavior. The two-dimensional region in the
number of flavors and colors where asymptotically free QCD
develops an IR interacting fixed point is colloquially known as
the \emph{conformal window of pQCD}.

\begin{figure}[htb]
\centering
\includegraphics[width=8cm]{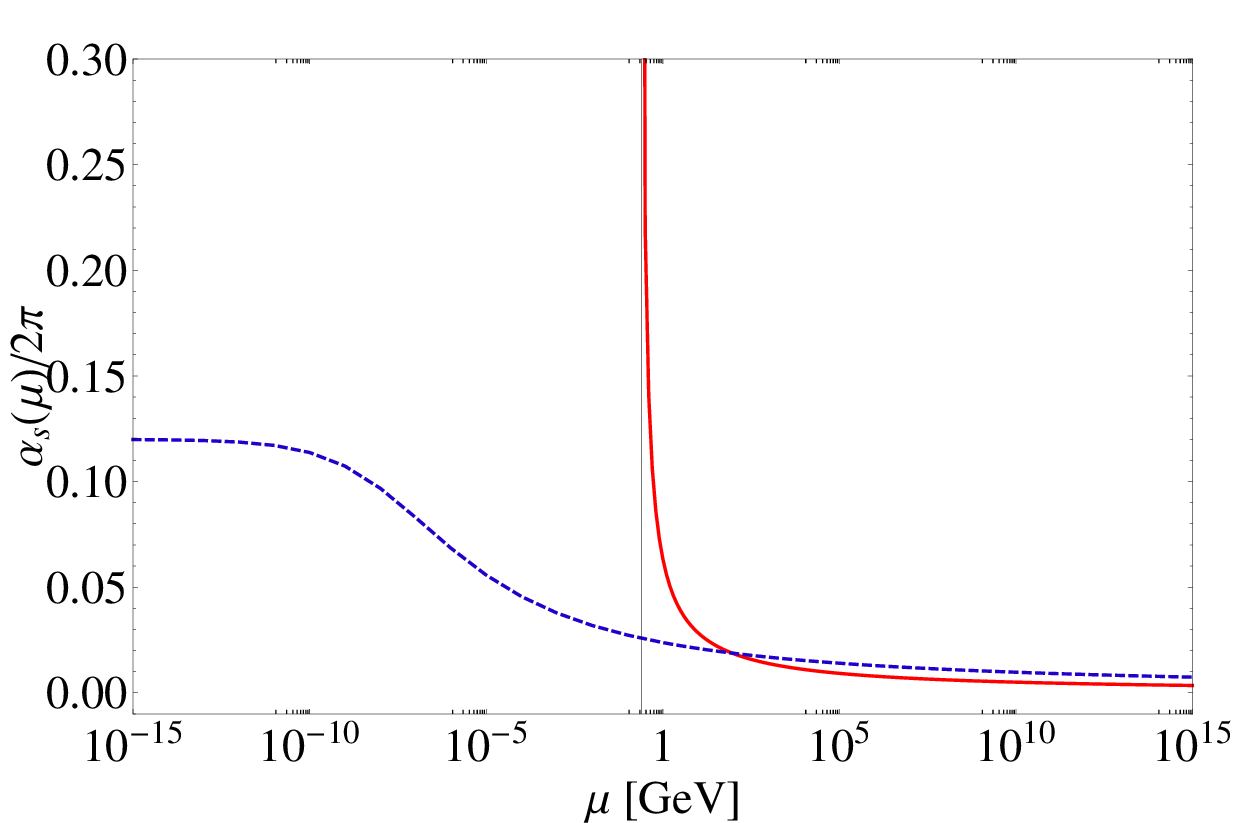}
\caption{The strong running coupling $\alpha_s(\mu)$ for $N_f=12$
(blue dashed) and for $N_f=5$ (solid red).
\cite{DiGiustino:2021nep} } \label{Lambert}
\end{figure}

Thus, the actual physical range of a conformal window for pQCD is
given by $\frac{34 N_c^3}{13 N_c^2-3}< N_f<\bar{N}_f$. The
behavior of the coupling is shown in Fig.~\ref{Lambert}. In the IR
region the strong coupling approaches the IR finite limit, $x^*$,
in the case of values of $N_f$ within the conformal window (e.g.\ black dashed curve of Fig.~\ref{Lambert}), while it diverges at
\begin{equation} \Lambda= \mu_0 \left(1+ \frac{|x^*|}{x_0}
\right)^{\frac{1}{2 B |x^*|}} e^{-\frac{1}{2 B x_0}}
\label{landaupole}\end{equation} outside the conformal window
given the solution for the coupling with $W_{-1}$ (e.g.\ the solid
red curve of Fig.~\ref{Lambert}). The solution of the NNLO
equation for the case $B>0, C>0$, i.e.\ $N_f<\frac{34 N_c^3}{13
N_c^2-3}$, can also be given using the standard QCD scale
parameter $\Lambda$ of Eq.~\eqref{landaupole},
\begin{eqnarray}
x &=& \frac{x^*}{1+W_{-1}}, \\
 z &=& -\frac{1}{e}  \left( \frac{\mu^2}{\Lambda^2}
\right)^{x^* B}. \label{xz2}
\end{eqnarray}
Different solutions can be achieved using different schemes, i.e.\ different definitions of the scale
parameter $\Lambda$ \cite{Gardi:1998qr}. We stress that the presence of a
Landau ``ghost'' pole in the strong coupling is only an effect of
the breaking of the perturbative regime, including
nonperturbative contributions, or using nonperturbative QCD, a
finite limit is expected at any $N_f$ \cite{Deur:2016tte}. Both
solutions have the correct UV asymptotically free behavior. In
particular, for the case $\bar{N}_f<N_f<\frac{11}{2}N_c$, we have
a negative $z$, a negative $C$ and a multi-valued solution, one
real and the other imaginary, actually only one (the real) is
acceptable given the initial conditions, but this solution is not
asymptotically free. We thus restrict our analysis to the range
$N_f<\bar{N}_f$, where we have the correct UV behavior. In general,
IR and UV fixed points of the $\beta$-function can also be
determined at different values of the number of colors $N_c$
(different gauge group $SU(N)$) and $N_f$ also extending this analysis
to other gauge theories \cite{Ryttov:2017khg}.

\subsubsection{$\alpha_s$ at higher loops}

The 3-loop truncated RG equation~\ref{betafun1}, written using the
same normalization as Eq.~\eqref{lambert1} is given by \be
\beta(x)=\frac{d x}{d t}=-B x^{2}\left(1+C x+C_{2} x^{2}\right),
\label{3loopbeta}\ee
with  $C_2=\frac{\beta_2}{4\beta_0}$.

A straightforward integration of this equation would be hard to
invert, as shown in Ref.~\cite{Gardi:1998qr}, it is more convenient
to extend the approach of the previous section
by using the Pad\'e Approximant Approach (PAA) \cite{Basdevant:1972fe, Samuel:1992qg, Samuel:1995jc}.
The Pad\'e Approximant of a given quantity calculated
perturbatively in QCD up to order $n$, i.e.\ of the series: \be
S(x)=x\left(1+r_{1} x+r_{2} x^{2}+\cdots+r_{n} x^{n}\right)
\label{pade0} \ee
is defined as the rational function
\be x[N / M]=x \, \frac{1+a_{1}
x+\ldots+a_{N} x^{N}}{1+b_{1} x+\ldots+b_{M} x^{M}}, \quad
x[N/M]=S+x \mathcal{O}\left(x^{N+M+1}\right), \label{padeapp} \ee
whose Taylor expansion up to order $N+M=n$ is identical to the
original truncated series. The use of the PA makes the integration
of Eq.~\eqref{3loopbeta} straightforward. PA's may also be used
either to predict the next term of a given perturbative expansion,
called a Pad\'e Approximant prediction (PAP), or to estimate the
sum of the entire series, called Pad\'e Summation. Features of the
PA are described in Ref.~\cite{Gardi:1996iq}.


The Pad\'e Approximant ($x^{2}[1 / 1]$) of the 3-loop $\beta$-function is
given by
\be \beta_{\rm PA}(x)=-B x^{2} \,
\frac{1+\left[C-\left(C_{2} / C\right)\right] x}{1-\left(C_{2} /
C\right) x}, \label{padebeta} \ee
which leads to the solution: \be B
\ln \left(Q^{2} / \Lambda^{2}\right)=\frac{1}{x}-C \ln
\left[\frac{1}{x}+C-\frac{C_{2}}{C}\right] \ee and finally,
\bea
x\left(Q^{2}\right)&=&-\frac{1}{C} \frac{1}{1-\left(C_{2} / C^{2}\right)+W(z)}, \\
z&=&-\frac{1}{C} \exp \left[-1+\left(C_{2} / C^{2}\right)-B t /
C\right], \eea the sign of $C$ determines the sign of $z$ and also
the physically relevant branches of the Lambert function $W(z)$:
for $C>0, z<0$ and the physical branch is $W_{-1}(z)$, taking real
values in the range $(-\infty,-1)$, while for $C<0, z>0$ and the
physical branch is given by the $W_{0}(z)$, taking real values in
the range $(0,\infty)$.

We notice that the only significant difference between the 3-loop
solution and the 2-loop solution \eqref{xz} is in the solution
 $x\left(Q^{2}\right)$. This is because the
difference in the definition of $z$ can be reabsorbed into an
appropriate redefinition of the scale parameter:
$$
\Lambda^{2} \longrightarrow \tilde{\Lambda}^{2}=\Lambda^{2}
e^{C_{2} /\left(B C\right)}.$$

For orders up to $\beta_{4}$, an approximate analytical solution
is obtained integrating Eq.~\eqref{betafun1}:
\begin{eqnarray}
\ln \frac{\mu^{2}}{\Lambda^{2}} &=& \int \frac{d a}{\beta(a)}~~=~~\frac{1}{\beta_{0}}\left[\frac{1}{a}+b_{1} \ln a+a\left(-b_{1}^{2}+b_{2}\right)+a^{2}\left(\frac{b_{1}^{3}}{2}-b_{1} b_{2}+\frac{b_3}{2}\right)\right.
\nonumber \\
&&\hspace{8em}\null
+a^3\left(-\frac{b_{1}^{4}}{3}+b_{1}^{2} b_{2}-\frac{b_{2}^{2}}{3}
-\left.\frac23 b_1b_3+\frac{b_{4}}{3}\right)+O\left(a^{4}\right)\right]+C
\end{eqnarray}
where $a=\alpha_s(\mu)/(4 \pi)$ and $b_N\equiv \beta_N/\beta_0$,
$(N=1,\dots,4)$ and performing the inversion of the last formula by
iteration as shown in Ref.~\cite{Kniehl:2006bg}, achieving the
final result of the coupling at five-loop accuracy:
\begin{eqnarray}
a &=& \frac{1}{\beta_{0} L}-\frac{b_{1} \ln L}{\left(\beta_{0} L\right)^{2}}+\frac{1}{\left(\beta_{0} L\right)^{3}}\left[b_{1}^{2}\left(\ln^{2} L-\ln L-1\right)+b_{2}\right]
\nonumber \\
&&+\frac{1}{\left(\beta_{0} L\right)^{4}}\left[b_{1}^{3}\left(-\ln^{3} L+\frac{5}{2} \ln^{2} L+2 \ln L-\frac{1}{2}\right)
-3 b_{1} b_{2} \ln L+\frac{b_3}{2}\right]
\nonumber\\
&&+\frac{1}{\left(\beta_{0} L\right)^{5}}\left[b_{1}^{4} \left(\ln^{4} L-\frac{13}{3} \ln^{3} L\right.
-\frac{3}{2} \ln^{2} L+4 \ln L+\frac{7}{6}\right)+3 b_{1}^{2} b_{2}\left(2 \ln^{2} L-\ln L-1\right)
\nonumber \\
&&\hspace{6em}-\left. b_{1} b_3\left(2 \ln L+\frac{1}{6}\right)+\frac{5}{3}
b_{2}^{2}+\frac{b_{4}}{3}\right]+O\left(\frac{1}{L^{6}}\right).
\end{eqnarray}
where $L=\ln(\mu^2/\Lambda^2)$. The same definition of the $\Lambda$
scale given in Eq.~\eqref{landaupole} has been used for the
$\overline{\rm MS}$ scheme, which leads to setting the constant $C=
\left(b_{1} / \beta_{0}\right) \ln (\beta_{0})$.

Alternatively, we can relate the values of the coupling at two
different scales by the 5-loop perturbative solution:
\bea
a_{Q_2} & = & a_{Q_1}+\beta_0 \ln \left(\frac{Q_1^2}{Q_2^2}\right) a_{Q_1}^2+\left[\beta_0^2 \ln^2 \left(\frac{Q_1^2}{Q_2^2}\right)+\beta_1 \ln\left(\frac{Q_1^2}{Q_2^2}\right) \right] a_{Q_1}^3
\nonumber \\
& & +\left[\beta_0^3 \ln^3 \left(\frac{Q_1^2}{Q_2^2}\right)+\frac{5}{2} \beta_0 \beta_1 \ln^2\left(\frac{Q_1^2}{Q_2^2}\right)+\beta_2 \ln \left(\frac{Q_1^2}{Q_2^2}\right)\right] a_{Q_1}^4
\nonumber \\
& & +\left[\beta_0^4 \ln^4\left(\frac{Q_1^2}{Q_2^2}\right)
+\frac{13}{3} \beta_0^2 \beta_1
\ln^3\left(\frac{Q_1^2}{Q_2^2}\right)+\frac{3}{2} \beta_1^2
\ln^2\left(\frac{Q_1^2}{Q_2^2}\right)+3 \beta_2 \beta_0
\ln^2\left(\frac{Q_1^2}{Q_2^2}\right)+\beta_3 \ln
\left(\frac{Q_1^2}{Q_2^2}\right)\right] a_{Q_1}^5
\nonumber \\
&  & +\left[  \beta_0^5 \ln^5 \left(\frac{Q_1^2}{Q_2^2}\right)
+\frac{77}{12}\beta_1\beta_0^3
\ln^4\left(\frac{Q_1^2}{Q_2^2}\right)+ \left(6 \beta_2\beta_0^2+
\frac{35}{6}\beta_1^2\beta_0 \right)
\ln^3\left(\frac{Q_1^2}{Q_2^2}\right) \right.
\nonumber \\
& & \hspace{15em}
\left.+\frac{7}{2} \left(\beta_3\beta_0 +\beta_2\beta_1\right)
\ln^2\left(\frac{Q_1^2}{Q_2^2}\right) +\beta_4 \ln
\left(\frac{Q_1^2}{Q_2^2}\right)  \right] a_{Q_1}^6 +\cdots.
\label{5loopalphaspert}
\eea


\subsection{Renormalization scheme dependence}\label{rsdependence}

The $\beta_{i}$ are the coefficients of the $\beta$-function
arising in the loop expansion, i.e.\ in powers of~$\hbar$.
Although the first two coefficients $\beta_0,\beta_1$ are
universally scheme-independent coefficients, depending only on the
number of colors $N_c$ and flavors $N_f$, the higher-order terms
are, in contrast, scheme dependent. In particular, for the 't
Hooft scheme \cite{thooftscheme} the higher $\beta_i, ~i\geq 2$
terms are set to zero, leading to the solution of
Eq.~\eqref{twoloops} for the $\beta$-function valid at all orders.
Moreover, in all $\rm MS$-like schemes all the $\beta_{i}$
coefficients are gauge independent, while other schemes, such as
the momentum space subtraction (MOM) scheme
\cite{Celmaster:1979dm,Celmaster:1979km}, depend on the particular
gauge. Using the Landau gauge, the $\beta$ terms for the MOM
scheme are given by \cite{Boucaud:2005rm}
$$
\beta_{2}=3040.48-625.387 N_f +19.3833 N_f^{2}
$$
and
$$
\beta_3=100541-24423.3 N_f+1625.4 N_f^{2}-27.493 N_f^{3}.
$$
Results for the minimal MOM scheme and Landau gauge are given in
Ref.~\cite{Chetyrkin:2000dq}. The renormalization condition for
the MOM scheme sets the virtual quark propagator to the same form
as a free massless propagator. Different MOM schemes exist and the
above values of $\beta_{2}$ and $\beta_3$ are determined with the
MOM scheme defined by subtracting the 3-gluon vertex to a point
with one null external momentum. This leads to a coupling that is
not only RS dependent but also gauge dependent. The values of
$\beta_{2}$ and $\beta_3$ given here are only valid in the Landau
gauge. Values in the $\mathrm{V}$-scheme defined by the static
heavy-quark potential \cite{Appelquist:1977tw, Fischler:1977yf,
Peter:1996ig, Schroder:1998vy, Smirnov:2008pn, Smirnov:2009fh,
Anzai:2009tm} can be found in Ref.~\cite{Kataev:2015yha}. They
result in $\beta_{2}=4224.18-746.01 N_f+20.88 N_f^{2}$ and
$\beta_3=43175-12952 N_f+707.0 N_f^{2}$ respectively. We recall
that the signs of the $\beta_{i}$ control the running
of~$\alpha_{s}$. We have $\beta_{0}>0$ for $N_f \leq 16,
\beta_{1}>0$ for $N_f \leq 8, \beta_{2}>0$ for $N_f \leq 5$ and
$\beta_3$ is always positive. Consequently, $\alpha_{s}$ decreases
at high momentum transfer, leading to the asymptotic freedom of
pQCD. Note that, $\beta_{i}$ are sometimes defined with an
additional multiplying factor $1 /(4 \pi)^{i+1}$.
 Different schemes are characterized by different $\beta_i, i\geq 2
 $ and lead to different definitions for the effective coupling.


The $\Lambda$ parameter represents the Landau ghost pole in the
perturbative coupling in QCD. We recall that the Landau pole was
initially identified in the context of Abelian QED. However, the
presence of this pole does not affect QED. Given its value,
$\Lambda\sim10^{30-40}\,{\rm GeV}$, above the Planck scale
\cite{Gockeler:1997dn}, at which new physics is expected to occur
in order to suppress the unphysical divergence. The QCD $\Lambda$
parameter in contrast is at low energies, its value depends on the
RS, on the order of the $\beta$-series, $\beta_{i}$, on the
approximation of the coupling $\alpha_s(\mu)$ at orders higher
than $\beta_{1}$ and on the number of flavors~$N_f$. Although mass
corrections due to light quarks at higher order in perturbative
calculations introduce negligible terms, they actually indirectly
affect $\alpha_s$ through~$N_f$. In fact, the number of active
quark flavors runs with the scale $Q$ and a quark $q$ is
considered active in loop integration if the scale $Q\geq m_q$.
Thus, in general, light quarks can be considered massless
regardless of whether they are active or not, while $\alpha_s$
varies smoothly when passing a quark threshold, rather than in
discrete steps. The matching of the values of $\alpha_s$ below and
above a quark threshold makes $\Lambda$ depend on~$N_f$. Matching
requirements at leading order $\beta_0$, imply that:
$$
\alpha_{s}^{N_f-1}\left(Q{=}m_{q}\right)=
\alpha_{s}^{N_f}\left(Q{=}m_{q}\right)
$$
and therefore that:
$$
\Lambda^{N_f}=\Lambda^{N_f-1}\left(\frac{\Lambda^{N_f-1}}{m_{q}}\right)^{2
/\left(33-2 N_f\right)}
$$
The formula with $\beta_{1}$, can be found in \cite{Larin:1994va}
and the four-loop matching in the $\overline{\rm MS}$ RS is given
in \cite{Chetyrkin:1997sg}.

As shown in the previous section at the lowest order $\beta_{0}$,
the Landau singularity is a simple pole on the positive real axis
of the $Q^2$-plane, whereas at higher order it acquires a more
complicated structure. This pole is unphysical and is located on
the positive real axis of the complex $Q^{2}$-plane. This
singularity of the coupling indicates that the perturbative regime
of QCD breaks down and it may also suggest that a new mechanism
takes over, such as the confinement. Thus, the value of $\Lambda$
is often associated with the confinement scale, or equivalently to
the hadronic mass scale. An explicit relation between hadron
masses and the $\Lambda$ scale has been obtained in the framework
of holographic QCD \cite{Brodsky:2014yha}. Landau poles on the
other hand, usually do not appear in nonperturbative approaches,
such as AdS/QCD.

Different schemes are related perturbatively by: \be
\alpha_{s}^{(2)}\left(Q\right)=\alpha_{s}^{(1)}\left(Q\right)\left[1+v_{1}
\alpha_{s}^{(1)}\left(Q\right) /(4 \pi)\right]
+\mathcal{O}(\alpha_s^2)\ee where $v_{1}$ is the leading order
difference between $\alpha_{s}\left(Q \right)$ in the two schemes.
In the case of the V-scheme and $\overline{\rm MS}$ scheme we
have: $v_1^{\overline{\mathrm{MS}}}=\frac{31}{9}
C_{A}-\frac{20}{9} T_{F} n_{l}$. Thus, the relation between
$\Lambda_{1}$ in a scheme 1 and $\Lambda_{2}$ in a scheme 2 is, at
the one-loop order, given by:
$$
\Lambda_{2}=\Lambda_{1} e^{\frac{v_{1}}{2 \beta_{0}}}.
$$
For example, the $\overline{\rm MS}$ and V-scheme scale parameters
are related by:
$$
\Lambda_{\rm V}=\Lambda_{\overline{\rm MS}} e^{\frac{93-10 N_f}{2(
99-6 N_f)}}
$$
The relation is valid at each threshold, translating all values for
the scale from one scheme to the other.

\subsection{The Extended Renormalization Group Equations}

Given that physical predictions cannot depend on the choice of the
renormalization scale nor on the scheme, the same approach used
for the renormalization scale based on the invariance under RGE is
extended to scheme transformations. This approach leads to the
Extended Renormalization Group Equations, which were introduced
first by St\"uckelberg and
Peterman \cite{StueckelbergdeBreidenbach:1952pwl}, then discussed
by Stevenson \cite{Stevenson:1980du, Stevenson:1981vj, Stevenson:1982wn, Stevenson:1982qw}
and also improved by Lu and Brodsky \cite{Lu:1992nt}. A physical
quantity, $R$, calculated at the $N$-th order of accuracy is
expressed as a truncated expansion in terms of a coupling constant
$\alpha_{S}(\mu)$ defined in the scheme $\rm S$ and at the scale
$\mu$, such as
\be R_{N}=r_{0} \alpha_{S}^{p}(\mu)+r_{1}(\mu)
\alpha_{S}^{p+1}(\mu)+\cdots+r_{N}(\mu) \alpha_{S}^{p+N}(\mu).
\label{eqn:truncated}\ee
At any finite order, the scale and scheme
dependences of the coupling constant $\alpha_{S}(\mu)$ and of the
coefficient functions $r_{i}(\mu)$ do not totally cancel, this
leads to a residual dependence in the finite series and to the
scale and scheme ambiguities.

In order to generalize the RGE approach, it is convenient to
improve the notation by introducing the universal coupling
function as the extension of an ordinary coupling constant to
include the dependence on the scheme parameters
$\left\{c_{i}\right\}$:
\be \alpha=\alpha\left(\mu / \Lambda,\left\{c_{i}\right\}\right).
\ee where $\Lambda$ is the standard two-loop $\overline{\rm MS}$
scale parameter. The subtraction prescription is now characterized
by an infinite set of continuous \emph{scheme parameters}
$\left\{c_{i}\right\}$ and by the renormalization scale~$\mu$.
Stevenson \cite{Stevenson:1981vj} has shown that one can identify
the beta-function coefficients of a given renormalization scheme
with the scheme parameters.  Considering that the first two
coefficients of the $\beta$-function are scheme independent, each
scheme is identified by its $\left\{\beta_{i}, ~i=2,3,
\ldots\right\}$ parameters.

More conveniently, let us define the rescaled coupling constant
and the rescaled scale parameter as \be
a=\frac{\beta_{1}}{\beta_{0}} \frac{\alpha}{4 \pi}, \quad
\tau=\frac{2 \beta_{0}^{2}}{\beta_{1}} \log (\mu / \Lambda). \ee
Then, the rescaled $\beta$-function takes the canonical form: \be
\beta(a)=\frac{d a}{d \tau}=-a^{2}\left(1+a+c_{2} a^{2}+c_3
a^{3}+\cdots\right) \ee with $c_{n}=\beta_{n} \beta_{0}^{n-1} /
\beta_{1}^{n}$ for $n=2,3,\cdots$.


The scheme and scale invariance of a given observable $R$, can be
 expressed as: \bea \frac{\delta R}{\delta \tau} &=& \beta
\frac{\partial
R}{\partial a}+\frac{\partial R}{\partial \tau}=0 \nonumber \\
\frac{\delta R}{\delta c_{n}}&=&\beta_{(n)} \frac{\partial
R}{\partial a}+\frac{\partial R}{\partial c_{n}}=0.
\label{extendedrge} \eea

The fundamental beta function that appears in
Eqs.~\eqref{extendedrge} reads: \be
\beta\left(a,\left\{c_{i}\right\}\right) \equiv \frac{\delta
a}{\delta \tau}=-a^{2}\left(1+a+c_{2} a^{2}+c_3
a^{3}+\cdots\right) \ee and the extended or scheme-parameter beta
functions are defined as: \be
\beta_{(n)}\left(a,\left\{c_{i}\right\}\right) \equiv \frac{\delta
a}{\delta c_{n}}. \ee The extended beta functions can be
expressed in terms of the fundamental beta function. Since the
$(\tau,\{c_i\})$ are independent variables, second partial
derivatives respect the commutativity relation: \be
\frac{\delta^{2} a}{\delta \tau \delta c_{n}}=\frac{\delta^{2}
a}{\delta c_{n} \delta \tau}, \ee which implies \be \frac{\delta
\beta_{(n)}}{\delta \tau}=\frac{\delta \beta}{\delta c_{n}},\ee
\be \beta \beta_{(n)}^{\prime}=\beta_{(n)} \beta^{\prime}-a^{n+2},
\ee where $\beta_{(n)}^{\prime}=\partial \beta_{(n)} / \partial a$
and $\beta^{\prime}=\partial \beta / \partial a$. From here \be
\beta^{-2}\left(\frac{\beta_{(n)}}{\beta}\right)^{\prime}=-a^{n+2},
\ee \be
\beta_{(n)}\left(a,\left\{c_{i}\right\}\right)=
-\beta\left(a,\left\{c_{i}\right\}\right)
\int_{0}^{a} d x \,
\frac{x^{n+2}}{\beta^{2}\left(x,\left\{c_{i}\right\}\right)}, \ee
where the lower limit of the integral has been set to satisfy the
boundary condition
$$
\beta_{(n)} \sim O\left(a^{n+1}\right).
$$
That is, a change in the scheme parameter $c_{n}$ can only affect
terms of order $a^{n+1}$ or higher in the evolution of the
universal coupling function.

The extended renormalization group equations
Eqs.~\eqref{extendedrge} can be written in the form: \bea
\frac{\partial R}{\partial \tau}&=& -\beta \frac{\partial R}{\partial a} \nonumber \\
\frac{\partial R}{\partial c_{n}}&=&-\beta_{(n)} \frac{\partial
R}{\partial a}. \label{eqn:pms}\eea Thus, provided we know the
extended beta functions, we can determine any variation of the
expansion coefficients of $R$ under scale-scheme transformations.
In particular, we can evolve a given perturbative series into
another, determining the expansion coefficients of the latter and
vice versa. Thus, different schemes and scales can be related
according to the extended renormalization group equations and the
fundamental requirement of ``renormalization scale and scheme
invariance'' is recovered via the extended renormalization group
invariance of perturbative QCD.
Unfortunately, these relations and in general all perturbative
calculations are known only up to a certain level of accuracy and the
truncated formulas are responsible for an important source of
uncertainties: the \emph{scheme} and \emph{scale ambiguities}.

\subsection{The running coupling constant $\alpha_s(\mu)$\label{sec:alphas}}

The strong coupling $\alpha_s$, is a fundamental parameter of the
SM theory and determines the strength of the interactions among
quarks and gluons in quantum chromodynamics~(QCD).

In order to understand hadronic interactions, it is necessary to
determine the magnitude of the coupling and its behavior over a
wide range of values, from low- to high-energy scales. Long and
short distances are related to low and high energies respectively.
In the high-energy region the strong coupling has an
\emph{asymptotic behavior} and QCD becomes perturbative, while in
the region of low energies, e.g.\ at the proton-mass scale, the
dynamics of QCD is affected by processes such as quark
confinement, soft radiation and hadronization. In the first case
experimental results can be matched with theoretical calculations
and a precise determination of $\alpha_s$ depends on both
experimental accuracy and theoretical errors. In the latter case
experimental results are difficult to achieve and theoretical
predictions are affected by the confinement and hadronization
mechanisms, which are rather model dependent. Various processes
also involve a precise knowledge of the coupling in both the high
and low momentum transfer regions and in some cases calculations
must be improved with electroweak (EW) corrections. Thus, the
determination of the QCD coupling over a wide range of energy
scales is a crucial task in order to achieve results and to test
QCD to the highest precision. Theoretical uncertainties in the
value of $\alpha_s(Q)$ contribute to the total theoretical
uncertainty in the physics investigated at the Large Hadron
Collider (LHC), such as the Higgs sector, e.g.\ Higgs production
via gluon fusion \cite{Anastasiou:2016cez}. The behavior of the
perturbative coupling at low-momentum transfer is also fundamental
for the scale of the proton mass, in order to understand hadronic
structure, quark confinement and hadronization processes. IR
effects, such as soft radiation and renormalon factorial growth,
spoil the perturbative nature of QCD in the low-energy domain.
Higher-twist effects can also play an important role. Processes
involving the production of heavy quarks near threshold require
knowledge of the QCD coupling at very low momentum scales. Even
reactions at high energies may involve the integration of the
behavior of the strong coupling over a large domain of momentum
scales, including IR regions. Precision tests of the coupling are
crucial also for other aspects of QCD that are still under
continuous investigation, such as the hadron masses and their
internal structure. In fact, the strong interaction is responsible
for the mass of hadrons in the zero-mass limit of the $u$, $d$
quarks.

The origin and phenomenology of the behavior of $\alpha_s(\mu)$ at
small distances, where asymptotic freedom appears, is well
understood and explained in many textbooks on Quantum Field Theory
and Particle Physics. Numerous reviews exist; see e.g.\
Refs.~\cite{Prosperi:2006hx, Altarelli:2013bpa}. However, standard
explanations often create an apparent puzzle. Other questions
remain even in this well understood regime: a significant issue is
how to identify the scale $Q$ that controls a given hadronic
process, especially if it depends on many physical scales.

 \newpage

\section{The Principle of Maximum Conformality -- PMC scale
setting}\label{sec:blm}

The Principle of Maximum Conformality
(PMC) \cite{Brodsky:2011ig, Brodsky:2011ta, Brodsky:2012rj, Mojaza:2012mf, Brodsky:2013vpa}
is the principle underlying BLM and it generalizes the BLM method
to all possible applications and to all orders.

BLM scale setting is greatly inspired by QED. The standard
Gell-Mann--Low scheme determines the correct renormalization scale
identifying the scale with the virtuality of the exchanged photon
\cite{GellMann:1954fq}. For example, in electron--muon elastic
scattering, the renormalization scale is given by the virtuality
of the photon exchanged, i.e.\ the spacelike momentum transfer
squared $\mu_r^2 = q^2 = t$. Thus,
\begin{equation} \alpha(t) = {\alpha(t_0) \over 1 - \Pi(t,t_0)},
\label{qed1}
\end{equation} where
$$ \Pi(t,t_0) = {\Pi(t) -\Pi(t_0)\over 1-\Pi(t_0) } $$ is the
vacuum polarization (VP) function. From Eq.~\eqref{qed1} it follows
that the renormalization scale $\mu^2_R=t$ can be determined by
the $\beta_0$-term at the lowest order. This scale is sufficient
to sum all the vacuum polarization contributions into the dressed
photon propagator, both proper and improper to all orders.
 Starting from a first evaluation of the physical
observable that is obtained by calculating perturbative
corrections in a given scheme (commonly used are $\rm MS$ or
$\overline{\rm MS}$) and at an initial renormalization scale
$\mu_{r}=\mu_{r}^{\rm init }$, one obtains the truncated
expansion:
\begin{equation}
\rho_{n}=\mathcal{C}_{0}
\alpha_{s}^{p}\left(\mu_{r}\right)+\sum_{i=1}^{n}
\mathcal{C}_{i}\left(\mu_{r}\right)
\alpha_{s}^{p+i}\left(\mu_{r}\right), \quad(p \geq 0),
\label{observable-init0}
\end{equation}
where $\mathcal{C}_{0}$ is the tree-level term, while
$\mathcal{C}_{1}, \mathcal{C}_{2}, \dots, \mathcal{C}_{n}$ are the
one-loop, two-loop and $n$-loop corrections respectively and $p$ is
the power of the coupling at tree-level. In order to improve the
pQCD estimate of the observable, after the initial renormalization
a change of scale using the RGE is performed according to the BLM
scale setting.

Following the GM--L scheme in QED, the BLM scales can be determined
at LO in perturbation theory by writing explicit
contributions coming from the different $N_f$ terms of the NLO
coefficient in a physical observable as \cite{Brodsky:1982gc}:
\bea
\rho &=& C_{0} \alpha_{s,\overline{\rm MS}}^{p}\left(\mu_{r}\right)\left[1+\left(A N_{f}+B\right) \frac{\alpha_{s,\overline{\rm MS}}\left(\mu_{r}\right)}{\pi}\right] \nonumber \\
&=& C_{0} \alpha_{s, \overline{\rm
MS}}^{p}\left(\mu_{r}\right)\left[1+\left(-\frac{3}{2} A
\beta_{0}+\frac{33}{2} A+B\right) \frac{\alpha_{s, \overline{\rm
MS}}\left(\mu_{r}\right)}{\pi}\right], \eea
where $\mu_{r}=\mu_{r}^{\rm  init }$ stands for an initial
renormalization scale, which practically can be taken as the
typical momentum transfer of the process. The $N_{f}$ term is due
to the quark vacuum polarization. Calculations are in the
$\overline{\rm MS}$-scheme.

At the NLO level, all $N_{f}$ terms should be resummed into the
coupling. Using the NLO $\alpha_{s}$-running formula: \be
\alpha_{s, \overline{\rm
MS}}\left(\mu_{r}^{*}\right)=\frac{\alpha_{s,\overline{\rm
MS}}\left(\mu_{r}\right)}{1+\frac{\beta_{0}}{4 \pi} \alpha_{s,
\overline{\rm MS}}\left(\mu_{r}\right) \ln\left(
\frac{\mu_r^*}{\mu_r}\right)}, \ee we obtain \be \rho=C_{0}
\alpha_{s,\overline{\rm
MS}}^{p}\left(\mu_{r}^{*}\right)\left[1+C_{1}^{*} \frac{\alpha_{s,
\overline{\rm
MS}}\left(\mu_{r}^{*}\right)}{\pi}\right],\label{eqn:conformalform}
\ee where
$$
\mu_{r}^{*}=\mu_{r} \exp \left(\frac{3 A}{p}\right) $$ is the BLM
scale and
$$ C_{1}^{*}=\frac{33}{2} A+B
$$
is the \emph{conformal} coefficient, i.e.\ the NLO coefficient not
depending on the RS and scale~$\mu$. Both the effective BLM scale
$\mu_{r}^{*}$ and the coefficient $C_{1}^{*}$ are $N_{f}$
independent and conformal at LO.  By including the term $33 A / 2$
into the scale we eliminate the $\beta_0$ term of the NLO
coefficient $C_{1}$, which is responsible for the running of the
coupling constant, and the observable in the final results can be
written in its \emph{maximal conformal form},
Eq.~\eqref{eqn:conformalform}.

The BLM method can be extended to higher orders in a systematic
way by including the $n_f$ terms arising at higher order into the
BLM scales consistently. In order to extend the BLM beyond the
NLO, the following points are considered essential:
\begin{enumerate}
\item All $n_f$-terms associated with the $\beta$-function (i.e.\
$N_f$ terms) and then with the renormalization of the coupling
constant, must be absorbed into the effective coupling, while
those $n_f$-terms that have no relation with UV-divergent diagrams
(i.e.\ $N_F$-terms) should be identified and considered as part of
the conformal coefficients. After BLM scale setting, the
perturbative series for the physical observable becomes a
conformal series, all nonconformal terms should be absorbed into
the effective coupling in a consistent manner. \item New
$N_f$-terms (corresponding to new $\beta_0$ coefficients) arise at
each perturbative order, thus a new BLM scale that sums these
terms consistently into the running coupling, should be introduced
at each calculated perturbative order. In fact, there is no reason
to use a unified effective scale for the entire perturbative
series as shown in Refs.~\cite{Grunberg:1991ac, Grunberg:1992mp}.
\item The BLM scales themselves should be a RG-improved
perturbative series \cite{Brodsky:2011ta}. The length of the
perturbative series for each BLM scale depends on how many new
$N_f$-terms (or $\beta_i$-terms) we have from the higher-order
calculation and to what perturbative order we have performed.
\end{enumerate}
Actually, the last point is not mandatory and needs clarification.
In order to apply the BLM/PMC using perturbative scales, the
argument of the coupling in the expansion of the BLM/PMC scale
should be the physical scale of the process $Q$, that may be
either the center-of-mass energy $\sqrt{s}$ or even another
variable, such as $\sqrt{t},\sqrt{u},M_H,\dots$, depending on the
process. Setting the initial scale to the physical scale would
greatly simplify the BLM/PMC procedure, preserving the original
scale invariance of the observable and eliminating the initial
scale dependence from the BLM/PMC scales. In case the BLM/PMC
scales are not perturbatively calculated, as will be shown in
Section~\ref{sec:icf}, the initial scale can be treated as an
arbitrary parameter.

In agreement with these indications, it is possible to achieve a
scale setting method extendible iteratively to all orders, which
leads to the correct coefficients ${\mathcal{C}_i}(\mu^*_{BLM})$
for the final ``maximally conformal'' series:

\begin{equation}
\rho_{n}=\mathcal{C}_{0} \alpha_{s}^{p}\left(\mu_{\rm
BLM}^{*}\right)+{\mathcal{C}}_{1}\left(\mu_{\rm BLM}^{* *}\right)
\alpha_{s}^{p+1}\left(\mu_{\rm BLM}^{*
*}\right)+{\mathcal{C}}_{2}\left(\mu_{\rm BLM}^{* * *}\right)
\alpha_{s}^{p+2}\left(\mu_{\rm BLM}^{* * *}\right)+\cdots
\label{eqn:conf}
\end{equation}
where the BLM scales $\mu^*_{\rm BLM},\mu^{**}_{\rm BLM},\dots$ are
set by a recursive use of the RG equations in order to cancel all
the $N_f$ terms from the series. We remark that since the
coefficients ${\mathcal{C}_i}(\mu^*_{BLM})$ have been obtained
cancelling all $\beta$ terms related to running of the coupling
they actually are free of any scale and scheme dependence left.
In other words, the ${\mathcal{C}_i}(\mu^*_{BLM}) \equiv
\tilde{\mathcal{C}_i}$, where the $\tilde{\mathcal{C}_i}$ are
conformal coefficients not depending on the renormalization scale.
Hence, the BLM approach leads to an maximally conformal observable,
i.e.\ where all the renormalization scale and scheme dependence has
been confined to the effective coupling and to its renormalization
scale $\alpha_s(\mu_{\rm BLM})$.

Fundamental features of the BLM method:
\begin{enumerate}
\item [A)] BLM scales at LO, are set simply by identifying the
coefficient $A$ of the $N_f$ term. \item [B)] Since all
$n_f$-terms related to the running of the coupling are reabsorbed,
scheme differences do not affect the results and the perturbative
expansions in $\alpha_s(\mu_r^*)$ in two different schemes, e.g.\
$\rm MS$ and $\overline{\rm MS}$, are identical. We notice that
$n_f$-terms related to the UV finite diagrams, may arise at every
order in perturbation theory. These terms might be related either
to the particular kinematics of the initial state or even to
finite loop diagrams arising at higher orders, thus in both cases
are insensitive to the UV cutoff or to the RS and cannot be
considered as $\beta$-terms. We label these terms as $N_F$-terms
and they do not give contributions to the BLM scales. \item [C)]
Using BLM scale setting, the perturbative expansion does not
change across quark threshold, given that all vacuum-polarization
effects due to a new quark are automatically absorbed into the
effective coupling. This implies that in a process with fixed
kinematic variables (e.g.\ a total cross-section), we can use a
naive LO/NLO $\alpha_s(\mu_r^{\rm BLM})$-running, with the number
of active flavor $N_f$ fixed to the value determined by the BLM
scale, to perform the calculation \cite{Brodsky:1998mf}. \item
[D)] The BLM method preserves all the RG properties of
\emph{existence} and \emph{uniqueness}, \emph{reflexivity},
\emph{symmetry} and \emph{transitivity}. As shown in
Refs.~\cite{Lu:1991yu, Lu:1991qr, Brodsky:1995ds, Brodsky:1994eh},
the RG invariance of the BLM leads to scheme-independent
transformations that relate couplings in different schemes. These
are known as \emph{commensurate scale relations} (CSRs) and it has
been shown that, even though the expansion coefficients under
different renormalization schemes can be different, after a proper
scale setting, one can determine a relation between the effective
couplings leading to an invariant result for the calculated
quantity. Using this approach it is also possible to extend
conformal properties to renormalizable gauge theories, such as the
generalized Crewther relation \cite{Crewther:1972kn,
Broadhurst:1993ru, Baikov:2010je, Brodsky:1995tb}. \item [E)] The
BLM approach reduces to the GM--L scheme for QED in the Abelian
limit $N_c \rightarrow 0$ \cite{Brodsky:1997jk}; the results are
in perfect agreement. \item [F)] The elimination of the $N_f$ term
related to the $\beta_0$ coefficient from the perturbative series
eliminates the renormalon terms $n!\beta^n_0\alpha^{n+1}_s$ over
the entire range of the accessible physical energies and not only
in the low-energy domain. The convergence of the resulting series
is then greatly improved.
\end{enumerate}

Several extended versions of the BLM approach beyond the NLO have
been proposed in the literature, such as the dressed-skeleton
expansion, the large-$\beta_0$ expansion, the BLM expansion with
an overall renormalization scale and the sequential BLM (seBLM), an
extension to the sequential BLM (xBLM) in
Refs.~\cite{Beneke:1994qe, Ball:1995ni, Brodsky:1994eh, Brodsky:1995tb, Grunberg:1991ac, Neubert:1994vb, Mikhailov:2004iq, Kataev:2014jba}.
These different extensions of the BLM are mostly partial or \emph{ad hoc} improvements of the first LO-BLM \cite{Brodsky:1982gc} in
some cases up to NNLO, in other cases using a rather effective
approach, i.e.\ by introducing an overall effective BLM scale for
the entire perturbative expansion. Results obtained with these
approaches also did not respect the basic points (1--3). Most
importantly, these methods lead to results that are still dependent
on the initial renormalization scale. The fundamental feature of
the BLM is to obtain results free of scale ambiguities and thus
independent of the choice of initial renormalization scale. The
first aim of the BLM scale is to eliminate the renormalization
scale and scheme uncertainties; thus any extension of the BLM not
respecting this basic requirement does not represent a real
improvement of the standard conventional scale setting CSS method.

The reasons for the different extensions of the BLM method to
higher orders were mainly two: firstly, it was not clear how to
generalize this approach to all possible quantities, which
translates into the question: what is the principle underlying the
BLM method? And secondly, what is the correct procedure to identify
and reabsorb the $N_f$-terms unambiguously order-by-order? A
practical reason that renders the extension to higher orders not
straightforward is the presence of finite UV corrections given by
the three- and four-gluon vertices of the additional $N_{F}$-terms
that are unrelated to the running of~$\alpha_{s}$.

In its first formulation in Ref.~\cite{Brodsky:2011ig} it was
suggested to use a unique PMC scale at LO to reabsorb all
$\beta$ contributions related to different skeleton-graph scales
by properly weighting the two contributions, such as that of the
$t$-channel and $s$-channel. This approach was more oriented towards a
single PMC scale that reabsorbs all $\beta$ terms related to the
running coupling. A multi-scale approach was later developed
considering different scales arising at each order of accuracy
including different $\beta$ coefficients according to the
perturbative expansion. We remark that the PMC method preserves
all the properties (A--F) of the BLM procedure and it
extends these properties to all orders, eliminating the
renormalization scale and scheme ambiguities. The PMC also
generalizes this approach to all gauge theories. Firstly, this is
crucial in order to apply the same method to all SM sectors.
Secondly, in the perspective of a \emph{grand unified theory}
(GUT), only one scale-setting method may be applied for
consistency, and this method must agree with the GM--L scheme and
with the QED results.

In order to apply the PMC, is convenient to follow the flowchart
shown in Fig.~\ref{flowchart} and to write the observable of
Eq.~\eqref{observable-init0} with the explicit contributions of the
$n_f$ terms in the coefficients calculated at each order of
accuracy:
\begin{figure}[htb]
\begin{center}
  \includegraphics[width=10cm]{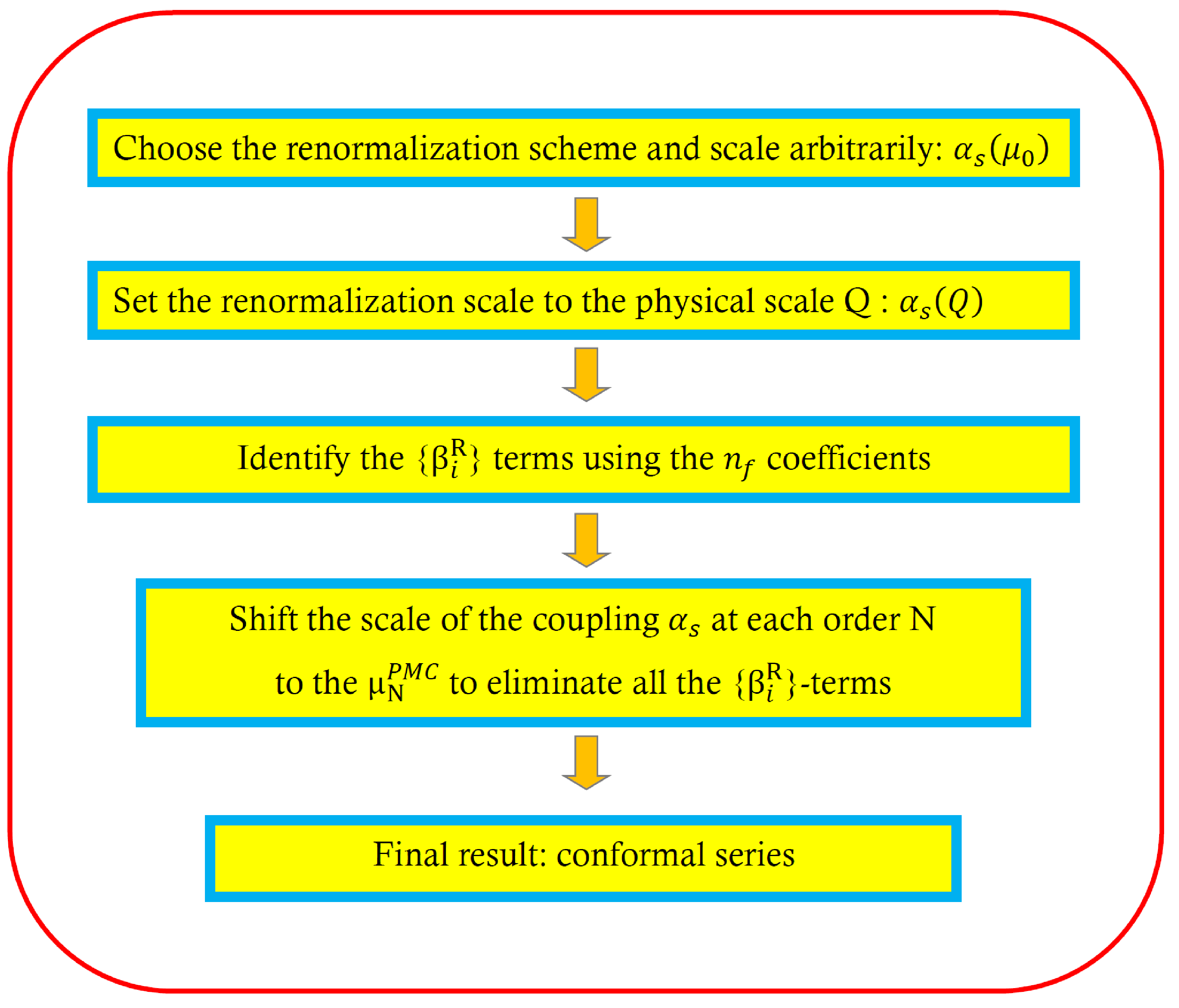}\\
  \caption{Flowchart for the PMC procedure.}\label{flowchart}
\end{center}
\end{figure}
\begin{equation}
\rho(Q)= \sum^{n}_{i=1} \left(\sum^{i-1}_{j=0} c_{i,j}(\mu_r, Q)
n_f^{j}\right) a_s^{p+i-1}(\mu_r), \label{nf}
\end{equation}
where $Q$ represents the kinematic scale or the physical scale of
the measured observable and $p$ is the power of $\alpha_s$
associated with the tree-level terms. In general, this procedure
is always possible either for analytic or numerical (e.g.\ Monte Carlo) calculations, given that both strategies keep track of
terms related to different color factors. The core of the PMC
method, as was for BLM, is that all $N_f$ terms related to the
$\beta$-function arising in a perturbative calculation must be
summed, by proper definition of the renormalization scale
$\mu_{\rm PMC}$, into the effective coupling $\alpha_s$ by
recursive use of the RGE. Essentially the difference between the
two procedures is that while in BLM the scales are set iteratively
order-by-order to remove all $n_f$ terms, in the PMC the $n_f$ terms
are written first as $\beta$ terms and then reabsorbed into the
effective coupling. The two procedures are related by the \emph{correspondence principle} \cite{Brodsky:2011ta}.

\subsection{The multi-scale Principle of Maximum Conformality: PMCm
\label{sec:pmcm}}

The PMCm method is based on a multi-scale application of the PMC.
In this section we show how to implement this method to any order
of accuracy. First, as shown in the flowchart in
Fig.~\ref{flowchart}, for the pQCD approximant \eqref{nf}, it is
convenient to transform the $\{n_f\}$ series at each order into
the $\{\beta_i\}$ series. The QCD degeneracy
relations \cite{Bi:2015wea} ensure the realizability of such a
transformation. For example, Eq.~\eqref{nf} may be rewritten
as \cite{Mojaza:2012mf, Brodsky:2013vpa}
\begin{eqnarray}
\rho(Q)&=&r_{1,0}\,a_s(\mu_r) +
\bigg(r_{2,0}+\beta_{0}r_{2,1}\bigg)a_{s}^{2}(\mu_r)\nonumber\\
&&+\bigg(r_{3,0}+\beta_{1}r_{2,1}+ 2\beta_{0}r_{3,1}+
\beta_{0}^{2}r_{3,2}\bigg)a_{s}^{3}(\mu_r)\nonumber\\
&& +\bigg(r_{4,0}+\beta_{2}r_{2,1}+ 2\beta_{1}r_{3,1} +
\frac{5}{2}\beta_{1}\beta_{0}r_{3,2}
+3\beta_{0}r_{4,1}+3\beta_{0}^{2}r_{4,2}+\beta_{0}^{3}r_{4,3}\bigg)
a_{s}^{4}(\mu_r)+\cdots, ~\label{rij1}
\end{eqnarray}
where $r_{i,j}$ can be derived from $c_{i,j}$, $r_{i,0}$ are
conformal coefficients and $r_{i,j}$ $(j\neq0)$ are nonconformal.
For definiteness and without loss of generality, we have set $p=1$
and $n=4$ to illustrate the PMC procedures. Different types of
$\{\beta_i\}$-terms can be absorbed into $\alpha_s$ in an
order-by-order manner by using the RGE, which leads to distinct
PMC scales at each order:
\begin{eqnarray}
a^k_s(Q_k) &\leftarrow& a^k_s(\mu_r)\bigg\{1 + k \beta_0 \frac{r_{k+1,1}}{r_{k,0}} a_s(\mu_r) \nonumber \\
&&\hspace{4em}
+k\left(\beta_1 \frac{r_{k+1,1}}{r_{k,0}} +\frac{k+1}{2} \beta_0^2 \frac{r_{k+2,2}}{r_{k,0}} \right) a^{2}_s(\mu_r)\nonumber \\
&&\hspace{4em}
+k\bigg(\beta_2 \frac{r_{k+1,1}}{r_{k,0}}
+\frac{2k+3}{2}\beta_0\beta_1 \frac{r_{k+2,2}}{r_{k,0}}
+\frac{(k+1)(k+2)}{3!}\beta_0^3 \frac{r_{k+3,3}}{r_{k,0}}\bigg)
a^{3}_s(\mu_r)+\cdots \bigg\}. \label{scaledis1}
\end{eqnarray}
The coefficients $r_{i,j}$ are generally functions of $\mu_r$,
which can be redefined as
\begin{eqnarray}
r_{i,j}=\sum^j_{k=0}C^k_j{\hat r}_{i-k,j-k}{\rm
ln}^k(\mu_r^2/Q^2),\label{rijrelation}
\end{eqnarray}
where the reduced coefficients ${\hat r}_{i,j}=r_{i,j}|_{\mu_r=Q}$
(specifically, we have ${\hat r}_{i,0}=r_{i,0}$) and the
combinatorial coefficients $C^k_j=j!/(k!(j-k)!)$. As discussed in
the previous section, we set the renormalization scale $\mu_r$ to
the physical scale of the process $Q$:
\begin{eqnarray}
\ln\frac{Q^2}{Q_1^2}&=& \frac{{\hat r}_{2,1}}{{\hat r}_{1,0}}+\beta_0\left(\frac{{\hat r}_{1,0}{\hat r}_{3,2}-{\hat r}_{2,1}^2}{{\hat r}_{1,0}^2}\right)a_s(Q)\nonumber \\
&&+\left[\beta_1\left(\frac{3{\hat r}_{3,2}}{2r_{1,0}}-\frac{3{\hat r}_{2,1}^2}{2{\hat r}_{1,0}^2}\right)+\beta_0^2\left(\frac{{\hat r}_{4,3}}{{\hat r}_{1,0}}-\frac{2{\hat r}_{3,2}{\hat r}_{2,1}}{{\hat r}_{1,0}^2}+\frac{{\hat r}_{2,1}^3}{{\hat r}_{1,0}^3} \right)\right]a^2_s(Q)+\cdots, \label{Q1} \\
\ln\frac{Q^2}{Q_2^2}&=& \frac{{\hat r}_{3,1}}{{\hat r}_{2,0}}+3\beta_0\frac{{\hat r}_{2,0}{\hat r}_{4,2}-{\hat r}_{3,1}^2}{2{\hat r}_{2,0}^2}a_s(Q)+\cdots, \label{Q2}\\
\ln\frac{Q^2}{Q_3^2}&=& \frac{{\hat r}_{4,1}}{{\hat
r}_{3,0}}+\cdots. \label{Q3}
\end{eqnarray}
Note that the PMC scales are of a perturbative nature, which is
also a sort of resummation and we need to know more loop terms to
achieve more accurate predictions. The PMC resums all the known
same type of $\{\beta_i\}$-terms to form precise PMC scales for
each order. Thus, the precision of the PMC scale for the
high-order terms decreases at higher and higher orders due to
the less known $\{\beta_i\}$-terms in those higher-order terms.
For example, $Q_1$ is determined up to next-to-next-to-leading-logarithm
(N$^2$LL) accuracy, $Q_2$ is determined up to NLL accuracy and
$Q_3$ is determined at LL accuracy. Thus, the PMC scales at
higher orders are of less accuracy due to more of the perturbative
terms being unknown. This perturbative property of the PMC scale
causes the \emph{first kind of residual scale dependence}.

After fixing the magnitude of $a_s(Q_k)$, we achieve a conformal
series
\begin{eqnarray}
\rho(Q) &=& \sum_{i=1}^{4}{\hat r}_{i,0} a^i_s(Q_i) +\cdots.
\label{PMCmseries}
\end{eqnarray}
The PMC scale for the highest-order term, e.g.\ $Q_4$ for the
present case, is unfixed, since there is no $\{\beta_i\}$-term to
determine its magnitude. This renders the last perturbative term
unfixed and causes the \emph{second kind of residual scale
dependence}. Usually, the PMCm suggests setting $Q_4$ as the last
determined scale $Q_3$, which ensures the scheme independence of
the prediction due to commensurate scale relations among the
predictions under different renormalization
schemes \cite{Brodsky:1994eh, Huang:2020gic}. The pQCD series
\eqref{PMCmseries} is renormalization scheme and scale
independent and becomes more convergent due to the elimination of
the $\beta$ terms including those related to the renormalon
divergence. Thus, a more accurate pQCD prediction can be achieved
by applying the PMCm. Two residual scale dependences are due to
perturbative nature of either the pQCD approximant or the PMC
scale, which is in principal different from the conventional
arbitrary scale dependence. In practice, we have found that these
two residual scale dependences are quite small even at low orders.
This is due to a generally faster pQCD convergence after applying
the PMCm. Some examples can be found in Ref.~\cite{Wu:2015rga}.

\subsection{The single-scale Principle of Maximum Conformality: PMCs\label{sec:pmcs}}

In some cases, the perturbative series might have a weak
convergence and the PMC scales might retain a comparatively larger
residual scale dependence. To overcome this, a single-scale
approach has been proposed, namely the PMCs, in order to
suppress the residual scale dependence by directly fixing a single
effective~$\alpha_s$. Following the standard procedures of
PMCs \cite{Shen:2017pdu}, the pQCD approximant \eqref{rij1}
changes to the following conformal series,
\begin{eqnarray}
\rho(Q)&=& \sum_{i=1}^{4} {\hat r}_{i,0}a^i_s(Q_*) + \cdots.
\label{conformal}
\end{eqnarray}
As in the previous section, we have set $p=1$ and $n=4$ for
illustrating the procedure. The PMC scale $Q_*$ can be determined
by requiring all the nonconformal terms to vanish, which can be fixed
up to N$^2$LL accuracy for $p=1$ and $n=4$, i.e.\ $\ln Q^2_* / Q^2$
can be expanded as a power series over $a_s(Q)$,
\begin{eqnarray}
\ln\frac{Q^2_*}{Q^2}=T_0+T_1 a_s(Q)+T_2 a^2_s(Q)+ \cdots,
\label{qstar}
\end{eqnarray}
where the coefficients $T_i~(i=0, 1, 2)$ are
\begin{align}
T_0&=-\frac{{\hat r}_{2,1}}{{\hat r}_{1,0}}, \\
T_1&=\frac{\beta_0 ({\hat r}_{2,1}^2-{\hat r}_{1,0} {\hat
r}_{3,2})}{{\hat r}_{1,0}^2}+\frac{2 ({\hat r}_{2,0} {\hat
r}_{2,1}-{\hat r}_{1,0} {\hat r}_{3,1})}{{\hat r}_{1,0}^2},
\nonumber\\[-2ex]\intertext{and}\nonumber\\[-9ex]
T_2&=\frac{3 \beta_1 ({\hat r}_{2,1}^2-{\hat r}_{1,0} {\hat r}_{3,2})}{2 {\hat r}_{1,0}^2}\nonumber\\
&\quad+\frac{4({\hat r}_{1,0} {\hat r}_{2,0} {\hat r}_{3,1}-{\hat r}_{2,0}^2 {\hat r}_{2,1})+3({\hat r}_{1,0} {\hat r}_{2,1} {\hat r}_{3,0}-{\hat r}_{1,0}^2 {\hat r}_{4,1})}{{\hat r}_{1,0}^3} \nonumber \\
&\quad+\frac{\beta_0 (4 {\hat r}_{2,1} {\hat r}_{3,1} {\hat r}_{1,0}-3 {\hat r}_{4,2} {\hat r}_{1,0}^2+2 {\hat r}_{2,0} {\hat r}_{3,2} {\hat r}_{1,0}-3 {\hat r}_{2,0} {\hat r}_{2,1}^2)}{{\hat r}_{1,0}^3}\nonumber\\
&\quad+\frac{\beta_0^2 (2 {\hat r}_{1,0} {\hat r}_{3,2} {\hat
r}_{2,1}- {\hat r}_{2,1}^3- {\hat r}_{1,0}^2
{\hat r}_{4,3})}{{\hat r}_{1,0}^3}.
\end{align}

Eq.~\eqref{qstar} shows that the PMC scale $Q_*$ is also a power
series over $\alpha_s$, which resums all the known
$\{\beta_i\}$-terms and is explicitly independent of $\mu_r$ at
any fixed order, but depends only on the physical scale Q. It
represents the correct momentum flow of the process and determines
an overall effective $\alpha_s$ value. Together with the
$\mu_r$-independent conformal coefficients, the resultant PMC pQCD
series is scheme and scale independent \cite{Wu:2018cmb}. By using
a single PMC scale determined with the highest accuracy from the
known pQCD series, both the \emph{first} and the \emph{second kind
of residual scale dependence} are suppressed.
\newpage
\section{Infinite-Order Scale Setting via the Principle of
Maximum Conformality: PMC$_\infty$} \label{pmcinfty}

In this section we introduce a parametrization of the observables
that stems directly from the analysis of the perturbative QCD
corrections and which reveals interesting properties, such as
scale invariance, independently of the process or of the
kinematics. We point out that this parametrization can be an
intrinsic general property of gauge theories and we define this
property \emph{intrinsic conformality} (iCF\footnote{Here the
conformality must be understood as RG invariance only.}). We also
show how this property directly indicates the correct
renormalization scale $\mu_r$ at each order of calculation and we
define this new method PMC$_\infty$: \emph{Infinite-Order Scale
Setting using the Principle of Maximum Conformality}. We apply the
iCF property and the PMC$_\infty$ to the case of the thrust and
$C$-parameter distributions in $e^+ e^-\rightarrow 3\,$jets and we
display the results.

\subsection{Intrinsic conformality (iCF)\label{sec:icf}}

In order to introduce intrinsic conformality (iCF), we consider
the case of a normalized IR-safe single-variable distribution and
write the explicit sum of pQCD contributions calculated up to NNLO
at the initial renormalization scale $\mu_0$:
\begin{eqnarray}
\frac{1}{\sigma_{0}} \! \frac{O d \sigma(\mu_{0})}{d O}\! &\!\! =
\!\! & \! \left\{\! \frac{\alpha_{s}(\mu_0)}{2 \pi} \frac{O d
A_{\mathit{O}}(\mu_0)}{d O} + \!
\left(\!\frac{\alpha_{s}(\mu_{0})}{2 \pi}\!\right)^{\!\!2} \!\!
\frac{O d B_{\mathit{O}}(\mu_0)}{d O}
+ \left(\frac{\alpha_{s}(\mu_0)}{2
\pi}\right)^{\!\!3} \! \frac{O dC_{\mathit{O}}(\mu_0)}{d O}\!+
\!{\cal O}(\alpha_{s}^4) \right\},
 \label{observable1}
\end{eqnarray}
where the $\sigma_0$ is a tree-level hadronic cross-section, $A_O,
B_O, C_O$ are respectively the LO, NLO and NNLO coefficients, $O$
is the selected unintegrated variable. For the sake of simplicity,
we shall refer to the perturbatively calculated differential
coefficients as \emph{implicit coefficients} and drop the
derivative symbol, i.e.\
\begin{eqnarray}
A_O(\mu_0) &\equiv& \frac{O d A_{O}(\mu_0)}{d O}, \qquad
B_O(\mu_0)~~\equiv~~\frac{O d B_{O}(\mu_0)}{d O}, \nonumber \\
C_O(\mu_0) &\equiv& \frac{O d C_{O}(\mu_0)}{d O}.
\label{implicitcoefficients}
\end{eqnarray}
We define here the \emph{intrinsic conformality} as the property
of a renormalizable SU(N)/U(1) gauge theory, such as QCD, which
yields a particular structure of the perturbative corrections that
can be made explicit by representing the perturbative coefficients
using the following parametrization:\,\footnote{We are neglecting
here other running parameters, such as the mass terms.}
\begin{eqnarray}
 A_{O}(\mu_0)\!\!\! &=& \!\!\! A_{\mathit{Conf}}, \nonumber \\
B_{O}(\mu_0) \!\!\! &=& \!\!\! B_{\mathit{Conf}}+\frac{1}{2} \beta_{0} \ln \left(\frac{\mu_0^{2}}{\mu_{\rm I}^{2}}\right) A_{\mathit{Conf}},  \nonumber \\
C_{O}(\mu_0)\!\!\! &=& \!\!\! C_{\mathit{Conf}} +\beta_{0}
\ln\left(\frac{\mu_{0}^{2}}{\mu_{\rm II}^{2}}\right)B_{\mathit{Conf}}
+ \frac{1}{4}\left[\beta_{1}+\beta_{0}^{2}
\ln \left(\frac{\mu_0^{2}}{\mu_{\rm I}^{2}}\right)\right]
\ln\left(\frac{\mu_0^{2}}{\mu_{\rm I}^{2}}\right)
A_{\mathit{Conf}},
\label{newevolution}
\end{eqnarray}
where the $A_{\mathit{Conf}}, B_{\mathit{Conf}},
C_{\mathit{Conf}}$ are the scale-invariant \emph{Conformal
Coefficients} (i.e.\ the coefficients of each perturbative order
not depending on the scale $\mu_0$), while we define the $\mu_{\rm
N}$ as \emph{Intrinsic Conformal Scales} and $\beta_0,\beta_1$
are the first two coefficients of the $\beta$-function. We recall
that the implicit coefficients are defined at the scale $\mu_0$
and that they change according to the standard RG equations under
a change of the renormalization scale according to:
\begin{eqnarray}
 A_{O}(\mu_r)\!\!\! &=& \!\!\! A_{O}(\mu_0), \nonumber \\
B_{O}(\mu_r) \!\!\! &=& \!\!\! B_{O}(\mu_0)+\frac{1}{2} \beta_{0}
\ln \left(\frac{\mu_r^{2}}{\mu_{0}^{2}}\right) A_{O}(\mu_0),  \nonumber \\
C_{O}(\mu_r)\!\!\! &=& \!\!\! C_{O}(\mu_0) +\beta_{0} \ln
\left(\frac{\mu_{r}^{2}}{\mu_{0}^{2}}\right)B_{O}(\mu_0)
+\frac{1}{4}\left[\beta_{1}+\beta_{0}^{2} \ln
\left(\frac{\mu_r^{2}}{\mu_{0}^{2}}\right)\right] \ln
\left(\frac{\mu_r^{2}}{\mu_{0}^{2}}\right) A_{O}(\mu_0).
\label{standardevolution}
\end{eqnarray}
It can be shown that the form of Eq.~\eqref{newevolution} is scale
invariant and it is preserved under a change of the
renormalization scale from $\mu_0$ to $\mu_r$ by standard RG
equations Eq.~\eqref{standardevolution}, i.e.:
\begin{eqnarray}
A_{O}(\mu_r)\!\!\! &=& \!\!\! A_{\mathit{Conf}}, \nonumber \\
B_{O}(\mu_r) \!\!\! &=& \!\!\! B_{\mathit{Conf}} +\frac{1}{2}
\beta_{0} \ln \left(\frac{\mu_r^{2}}{\mu_{\rm I}^{2}}\right)
A_{\mathit{Conf}},
\nonumber \\
C_{O}(\mu_r)\!\!\! &=& \!\!\! C_{\mathit{Conf}} +\beta_{0} \ln
\left(\frac{\mu_{r}^{2}}{\mu_{\rm II}^{2}}\right)B_{\mathit{Conf}}
+\frac{1}{4}\left[\beta_{1}+\beta_{0}^{2} \ln
\left(\frac{\mu_r^{2}}{\mu_{\rm I}^{2}}\right)\right] \ln
\left(\frac{\mu_r^{2}}{\mu_{\rm I}^{2}}\right) A_{\mathit{Conf}}.
\label{evolved}
\end{eqnarray}
We note that the form of Eq.~\eqref{newevolution} is invariant and
that the initial scale dependence is exactly removed by~$\mu_r$.
Extending this parametrization to all orders we achieve a
scale-invariant quantity: \emph{the iCF parametrization is a
sufficient condition in order to obtain a scale-invariant
observable}.

In order to show this property we collect together the terms
identified by the same \emph{conformal coefficient}, we call
each set a \emph{conformal subset} and extend the property
to order~$n$:
\begin{eqnarray}
& \sigma_{\rm I} &  =\left\{\left(\frac{\alpha_{s}(\mu_{0})}{2
\pi}\right)+\frac{1}{2}\beta_{0} \ln \left(
\frac{\mu_0^{2}}{\mu_{\rm
I}^{2}}\right)\left(\frac{\alpha_{s}(\mu_0)}{2 \pi}\right)^2
\right.  \nonumber \\
&& \qquad \qquad +\left. \frac{1}{4}\! \left[ \beta_{1} + \beta_{0}^{2}
\ln \left(\frac{\mu_0^{2}}{\mu_{\rm I}^{2}}\right) \right] \ln
\left(\frac{\mu_0^{2}}{\mu_{\rm I}^{2}}\right)
\left(\frac{\alpha_{s}(\mu_{0})}{2
\pi}\right)^3 +\ldots \right\} A_{\mathit{Conf}}   \nonumber \\
&  \sigma_{\rm II} & =  \left\{\left(\frac{\alpha_{s}(\mu_{0})}{2
\pi}\right)^2  + \beta_0 \ln  \left(  \frac{\mu_0^2}{\mu_{\rm
II}^2}\right) \left(\frac{\alpha_{s}(\mu_{0})}{2 \pi}\right)^3
+\ldots \right\}
B_{\mathit{Conf}}   \nonumber \\
& \sigma_{\rm III} & = \left\{\left(\frac{\alpha_{s}(\mu_{0})}{2
\pi}\right)^3 +\ldots \right\}C_{\mathit{Conf}}, \nonumber \\
& \vdots & \qquad\qquad
\boldsymbol{\cdot^{\textstyle\cdot^{\textstyle\cdot}}} \nonumber \\
&  \sigma_{\rm n} & = \left\{\left(\frac{\alpha_{s}(\mu_{0})}{2
\pi}\right)^n \right\}\mathcal{L}_{n \mathit{Conf}}.
\label{confsubsets}
\end{eqnarray}
In each subset we have only one intrinsic scale and only one
conformal coefficient and the subsets are disjoint; thus, no mixing
terms among the scales or the coefficients are introduced in this
parametrization. Moreover, the structure of the subsets remains
invariant under a global change of the renormalization scale, as
shown from Eq.~\eqref{evolved}. The structure of each conformal set
$\sigma_{\rm I}, \sigma_{\rm II}, \sigma_{\rm III},\dots$ and
consequently the iCF are preserved, also if we fix a different
renormalization scale for each conformal subset, i.e.
\begin{eqnarray}
\left(\mu^2 \frac{\partial}{\partial \mu^2} +\beta
(\alpha_s)\frac{\partial}{\partial \alpha_s}\right) \sigma_{\rm
n}=0. \label{sigmainvariance}
\end{eqnarray}
We define here this property of Eq.~\ref{confsubsets} of
separating an observable into the union of ordered scale-invariant
disjoint subsets $\sigma_{\rm I}, \sigma_{\rm II}, \sigma_{\rm
III},\dots$ an \emph{ordered scale invariance}.

In order to extend the iCF to all orders, we perform the
$n\rightarrow \infty$ limit using the following strategy: we first
perform a partial limit $J_{/n}\rightarrow \infty$ including the
higher-order corrections relative only to those
$\beta_0,\beta_1,\beta_2,\dots,\beta_{n-2}$ terms that have been
determined already at order $n$ for each subset and we then
perform the complementary $\bar{n}$ limit, which consists in
including all the remaining higher-order terms. For the $J_{/n}$
limit we have:

\begin{eqnarray}
 \lim_{J_{/n}\rightarrow\infty}
\sigma_{\rm I} & \rightarrow &
\left(\frac{\left. \alpha_{s}(\mu_{\rm I})\right|_{n-2}}{2 \pi}\right) A_{\mathit{Conf}} \nonumber \\
\lim_{J_{/n}\rightarrow\infty} \sigma_{\rm II} & \rightarrow &
\left(\frac{\left.\alpha_{s}(\mu_{\rm II})\right|_{n-3}}{2
\pi}\right)^2  B_{\mathit{Conf}}  \nonumber \\
 \lim_{J_{/n}\rightarrow\infty} \sigma_{\rm III} & \rightarrow &
\left(\frac{\left.\alpha_{s}(\mu_{\rm III})\right|_{n-4}}{2
\pi}\right)^3
C_{\mathit{Conf}} \nonumber \\
 & \vdots & \qquad \vdots \nonumber \\
\lim_{J_{/n}\rightarrow\infty} \sigma_{\rm n} & \equiv &
 \left(\frac{\alpha_{s}(\mu_{0})}{2
\pi}\right)^n \mathcal{L}_{n \mathit{Conf}}, \label{jnlimit}
\end{eqnarray}
where $\left.\alpha_{s}(\mu_{\rm I})\right|_{n-2}$ is the
coupling calculated up to $\beta_{n-2}$ at the intrinsic
scale~$\mu_{\rm I}$. Given the particular ordering of the powers of the
coupling, in each conformal subset we have the coefficients of the
$\beta_0,\dots,\beta_{n-k-1}$ terms, where $k$ is the order of the
conformal subset and the $n$ is the order of the highest subset
with no $\beta$ terms. We note that the limit of each conformal
subset is finite and scale invariant up to~$\sigma_{n-1}$. The
remaining scale dependence is confined to the coupling of the
$n^{th}$ term. Any combination of the $\sigma_{\rm
I},\dots,\sigma_{n-1}$ subsets is finite and scale invariant. We can
now extend the iCF to all orders performing the $\bar{n}$ limit.
In this limit we include all the remaining higher-order
corrections. For the calculated conformal subsets this leads to
defining the coupling at the same scales but including all the
missing $\beta$ terms. Thus, each conformal subset remains scale
invariant. We point out that we are not making any assumption on
the convergence of the series for this limit. We thus have:
\begin{eqnarray}
\lim_{\bar{n}\rightarrow\infty} \sigma_{\rm I} & \rightarrow &
\left(\frac{\alpha_{s}(\mu_{\rm I})}{2 \pi}\right) A_{\mathit{Conf}} \nonumber \\
\lim_{\bar{n}\rightarrow\infty} \sigma_{\rm II} & \rightarrow &
\left(\frac{\alpha_{s}(\mu_{\rm II})}{2
\pi}\right)^2  B_{\mathit{Conf}}  \nonumber \\
\lim_{\bar{n}\rightarrow\infty} \sigma_{\rm III} & \rightarrow &
\left(\frac{\alpha_{s}(\mu_{\rm III})}{2 \pi}\right)^3
C_{\mathit{Conf}} \nonumber \\
& \vdots & \qquad
\qquad \vdots \nonumber \\ \hphantom{Conformal Limit}
\lim_{\bar{n}\rightarrow\infty} \sigma_{\rm n} & \equiv &
\lim_{n \rightarrow\infty}
\left(\frac{\alpha_{s}(\mu_{0})}{2
\pi}\right)^n \!\! \mathcal{L}_{n \mathit{Conf}}
\rightarrow \hbox{Conformal Limit},\label{nlimit}
\end{eqnarray}
where here now $\alpha_{s}(\mu_{\rm I})$ is the complete coupling
determined at the same scale~$\mu_{\rm I}$. Equation~\eqref{nlimit} shows
that the entire renormalization scale dependence has been
completely removed. In fact, neither the intrinsic scales
$\mu_{\rm N}$ nor the conformal coefficients
$A_{\mathit{Conf}},B_{\mathit{Conf}},C_{\mathit{Conf}},\dots,\mathcal{L}_{n
\mathit{Conf}},\dots$ depend on the particular choice of the initial
scale. The only term with a residual $\mu_0$ dependence is the
$n$-th term, but this dependence cancels in the limit $n\rightarrow\infty$.
The scale dependence is totally confined to the coupling
$\alpha_s (\mu_0)$ and its behavior does not depend on the
particular choice of any scale $\mu_0$ in the perturbative region,
i.e.\ $\lim_{n\rightarrow\infty} \alpha_s(\mu_0)^n \sim a^n$ with
$a<1$. Hence, the limit of $\lim_{n\rightarrow\infty} \sigma_{\rm
n}$ depends only on the properties of the theory and not on the
scale of the coupling in the perturbative regime.
The proof given here shows that the iCF is \emph{sufficient} to have a
scale-invariant observable and it does not depend on the particular
convergence of the series.

In order to show the \emph{necessary}
condition, we separate the two cases of a convergent series and an
asymptotic expansion. For the first case the \emph{necessary}
condition stems directly from the uniqueness of the iCF form,
since given a finite limit and the scale invariance, any other
parametrization can be reduced to the iCF by means of appropriate
transformations in agreement with the RG equations. For the second
case, we have that an asymptotic expansion though not convergent,
can be truncated at a certain order $n$, which is the case of
Eq.~\eqref{confsubsets}. Given the particular structure of the iCF
we can perform the first partial limit $J_{/n}$ and we would
achieve a finite and scale-invariant prediction,
$\sigma_{N-1}=\Sigma_{i=1}^{n-1} \sigma_i$, for a truncated
asymptotic expansion, as shown in Eq.~\eqref{jnlimit}. Given the
truncation of the series in the region of maximum of convergence
the $n$-th term would be reduced to the lowest value and so the
scale dependence of the observable would reach its minimum. Given
the finite and scale-invariant limit $\sigma_{N-1}$ we conclude
that the iCF is unique and thus \emph{necessary} for an \emph{ordered} scale-invariant truncated asymptotic expansion up to the
$n$-th order.

We point out that in general the iCF form is the
most general and irreducible parametrization that leads to
scale invariance; other parametrization are forbidden, since if we were to
introduce more scales\,\footnote{Here we refer to the form of
Eq.~\eqref{newevolution}. In principle, it is possible to write other
parametrizations preserving the scale invariance, but these can be
reduced to the iCF in agreement with the RG equations.} into the
logarithms of one subset, we would spoil the invariance under the
RG transformation and we could not achieve Eq.~\eqref{evolved},
while on the other hand no scale dependence can be introduced into
the intrinsic scales since it would remain in the observable
already in the first partial limit $J_{/n}$ and it could not be
eliminated. The conformal coefficients are conformal
at each order by definition; thus, they do not depend on the renormalization
scale and they do not have a perturbative expansion. Hence \emph{the iCF is a necessary and sufficient condition for scale invariance.}

\subsection{Comments on the iCF and ordered scale invariance\label{sec:osi}}

The iCF parametrization can stem either from an inner property of
the theory, the iCF, or from direct parametrization of the
scale-invariant observable. In both cases the iCF parametrization
makes the scale dependence of the observable explicit and it
exactly preserves the scale invariance. The iCF parametrization is
invariant with respect to the choice of initial scale $\mu_0$,
this implies that the same calculation performed choosing
different arbitrary initial scales, $\mu_0,\,\mu_0'$ leads to the
same result in the limit $J_{/n}$, a limit that is scale and
scheme independent. The iCF is also strongly motivated by the
renormalizability of QCD and by the uniqueness of the $\beta$-function
in a given scheme; i.e.\ two different $\beta_i,\,\beta_i'$
do not occur in a perturbative calculation at any order in one RS
and the UV divergencies are cancelled by redefinition of the same
parameters at lowest and higher orders. We remark that the
conservation of the iCF form in one observable is strongly related
to the validity of the RG transformations; we thus expect the iCF
to be well preserved in the deep Euclidean region.

Once we have defined an observable in the iCF-form, we have not
only the scale invariance of the entire observable, but also the
\emph{ordered scale invariance} (i.e.\ the scale invariance of
each subset $\sigma_{\rm n}$ or $\sigma_{\rm N-1}$). The latter
property is crucial in order to obtain scale-invariant observables
independently from the particular kinematic region and
independently from the starting order of the observable or the
order of the truncation of the series. Since in general, a theory
is blind with respect to the particular observable/process that we
might investigate, the theory should preserve the \emph{ordered}
scale invariance in order to always define scale-invariant
observables. Hence if the iCF is an inner property of the theory,
it leads to implicit coefficients that are neither independent nor
conformal. This is made explicit in Eq.~\eqref{newevolution}, but it
is hidden in the perturbative calculations in the case of the
implicit coefficients. For instance, the presence of the iCF
clearly reveals itself when a particular kinematic region is
approached and the $A_O$ becomes null. This would cause a breaking
of the scale invariance since a residual initial scale dependence
would remain in the observable in the higher-order coefficients.
The presence of the iCF solves this issue by leading to the
correct redefinition of all the coefficients at each order
preserving the correct scale invariance exactly. Thus, in the case
of a scale-invariant observable $O$, defined according to the
implicit form (Eq.~\eqref{observable1}), by the coefficients $
\{A_O, B_O, C_O,..,O_O,\dots\},$ it cannot simply undergo the change
$ \rightarrow \{0, B_O, C_O,..,O_O,\dots\}$, since this would
break the scale invariance. In order to preserve the scale
invariance, we must redefine the coefficients $\{\tilde{A}_O=0,
\tilde{B}_O, \tilde{C}_O,..,\tilde{O}_O,\dots\}$ cancelling out all
the initial scale dependence originating from the LO coefficient
$A_O$ at all orders. This is equivalent to subtracting out an entire
invariant conformal subset $\sigma_{\rm I}$ related to the
coefficient $A_{\mathit{Conf}}$ from the scale-invariant
observable~$O$. This mechanism is clear in the case of the
explicit form of the iCF, Eq.~\eqref{newevolution}, where, if
$A_{\mathit{Conf}}=0$, then the entire conformal subset is null and
the scale invariance is preserved.

We stress that the conformal coefficients may acquire all possible
values without breaking scale invariance, they contain the
essential information on the physics of the process, while all the
correlation factors can be reabsorbed into the renormalization
scales as shown by the PMC method
\cite{Brodsky:2011ta,Brodsky:2012rj,Mojaza:2012mf,Brodsky:2013vpa}.
Hence, if a theory has the property of \emph{ordered scale
invariance}, it exactly preserves the scale invariance of
observables independently of the process, the kinematics and the
starting order of the observable. We stress that if a theory has
intrinsic conformality, all renormalized quantities, such as cross
sections, can be parametrized with the iCF-form. This property
should be preserved by the renormalization scheme or by the
definition of IR safe quantities and it should also be preserved
in observables defined in effective theories. The iCF shows that
point (3) of the BLM/PMC approach (Section~\ref{sec:blm}) can be
improved by eliminating the perturbative expansion of the BLM/PMC
scales, leading to a scale- and scheme-invariant result. We remark
though that the perturbative corrections in the BLM or PMCm scales
are suppressed in the perturbative region.

\subsection{The PMC$_\infty$\label{sec:pmcinf}}

We introduce here a new method for eliminating the scale-setting
ambiguity in single variable scale-invariant distributions, which
we call PMC$_\infty$. This method is based on the original PMC
principle and agrees with all the different PMC formulations for
the PMC scales at lowest order. The core of the
PMC$_\infty$ is essentially the same for all BLM-PMC prescriptions, i.e.\ the effective running-coupling value and hence its renormalization
scale at each order is determined by the $\beta_0$-term of the
next-higher order, or equivalently by the \emph{intrinsic
conformal scale}~$\mu_{\rm N}$. The PMC$_\infty$ preserves the
iCF and thus the scale and scheme invariance, absorbing an
infinite set of $\beta$-terms to all orders.

This method differs
from the other PMC prescriptions since, due to the presence of the
intrinsic conformality, no perturbative correction in $\alpha_s$
needs to be introduced at higher orders in the PMC scales. Given
that all the $\beta$-terms of a single conformal subset are
included in the renormalization scale already with the definition
at lowest order, no initial scale or scheme dependences are left
due to the unknown $\beta$-terms in each subset. The
PMC$_\infty$ scale of each subset can be unambiguously determined
by $\beta_0$-term of each order, we stress that all logarithms
of each subset have the same argument and all the differences
arising at higher orders have to be included only in the conformal
coefficients. Reabsorbing all the $\beta$-terms into the scale
also eliminates the $n!\beta_0^n\alpha_s^n$ terms (related to
renormalons \cite{Beneke:1998ui}); thus, the
precision is improved and the perturbative QCD predictions can be
extended to a wider range of values. The initial scale dependence
is totally confined in the unknown PMC$_\infty$ scale of the last
order of accuracy (i.e.\ up to NNLO case in the
$\alpha_s(\mu_0)^3$).
Thus, if we fix the renormalization scale independently to
the proper intrinsic scale for each subset $\mu_{\rm N}$, we end
up with a perturbative sum of totally conformal contributions up
to the order of accuracy:
\begin{eqnarray}
\frac{1}{\sigma_{0}} \frac{O d \sigma(\mu_{\rm I},\mu_{\rm
II},\mu_{\rm III})}{d O}&=&\left\{\frac{\alpha_{s}(\mu_{\rm
I})}{2 \pi} \frac{O dA_{\mathit{Conf}}}{d
O}+\left(\frac{\alpha_{s}(\mu_{\rm II})}{2 \pi}\right)^{2} \frac{O
d B_{\mathit{Conf}}}{d O} \right. \nonumber
\\ & & \hspace{10em} + \left.
\left(\frac{\alpha_{s}(\mu_{\rm III})}{2 \pi}\right)^{3}\frac{O d
C_{\mathit{Conf}}}{d O} \right\}+{\cal O}(\alpha_{s}^4).
 \label{observable2}
\end{eqnarray}
At this order, the last scale is set to the physical scale $Q$, i.e.\ $\mu_{\rm III}=\mu_0=Q$.

\subsection{The iCF coefficients and scales: a new ``How-To'' method}
\label{sec:icfcoeffandscales}

We describe here how all the coefficients of
Eq.~\eqref{newevolution} can be identified from either a numerical
or analytical perturbative calculation. This method applies in
general to any perturbative calculation once results for the
different color factors are kept separate; however, we refer to the
particular case of the NNLO thrust distribution results calculated
in Refs.~\cite{Weinzierl:2008iv, Weinzierl:2009ms} for the purpose.
Since the leading order is already ($A_{\mathit{Conf}}$) void of
$\beta$-terms, we start with NLO coefficients. A general
numerical/analytical calculation keeps tracks of all the color
factors and the respective coefficients:
\begin{eqnarray}
B_{O}(N_f)=C_F \left[ C_A B_{O}^{N_c}+C_F B_{O}^{C_F}+ T_F N_f
B_{O}^{N_f}\right] \label{Bcoeff}
\end{eqnarray}
where $C_F=\frac{\left(N_{c}^{2}-1\right)}{2 N_{c}}$, $C_A=N_c$
and $T_F=1/2.$ The dependence on $N_f$ is made explicit here for
sake of clarity. We can determine the conformal coefficient
$B_{\mathit{Conf}}$ of the NLO order straightforwardly, by fixing
the number of flavors $N_f$ in order to kill the $\beta_0$ term:
\begin{eqnarray}
B_{\mathit{Conf}}&=& B_{O} \left( N_f \equiv \frac{33}{2} \right),\nonumber \\
B_{\beta_0} \equiv  \log  \frac{\mu_0^2}{\mu_{\rm I}^2}  & = & 2
\frac{B_O-B_{\mathit{Conf}}}{\beta_0 A_{\mathit{Conf}}}.
\label{Bconf}
\end{eqnarray}
We would achieve the same results in the usual PMC
way; i.e.\ by identifying the $N_f$ coefficient with the $\beta_0$
term and then determining the conformal coefficient. Both methods
are consistent and results for the intrinsic scales and the
coefficients are in perfect agreement. At NNLO a general
coefficient is composed of the contribution of six different color
factors:
\begin{eqnarray}
C_{O}(N_f)&=& \frac{C_F}{4} \left\{N_{c}^{2} C_{O}^{N_c^2
}+C_{O}^{N_c^0}+\frac{1}{N_{c}^{2}} C_{O}^{\frac{1}{N_c^2}}
\right. \nonumber \\ & & \qquad \qquad
+\left.  N_{f} N_{c}\cdot C_{O}^{N_f N_c}+\frac{N_{f}}{N_{c}}
C_{O}^{N_f/N_c}+N_{f}^{2} C_{O}^{N_f^2}\right\}. \label{Ccoeff}
\end{eqnarray}
In order to identify all the terms of Eq.~\eqref{newevolution}, we
notice first that the coefficients of the terms $\beta_0^2$ and
$\beta_1$ are already given by the NLO coefficient $B_{\beta_0}$;
we thus need to determine only the $\beta_0$ and the conformal
$C_{\mathit{Conf}}$ terms. In order to determine the latter
coefficients, we use the same procedure used for the NLO; i.e.\ we set the number of flavors $N_f \equiv 33/2$ in order to remove all the $\beta_0$ terms. We then have
\begin{eqnarray}
C_{\mathit{Conf}}&=& C_{O} \left( N_f \equiv \frac{33}{2} \right)- \frac{1}{4}\overline{\beta}_1 B_{\beta_0} A_{\mathit{Conf}},\nonumber \\
C_{\beta_0} \equiv \log\left(\frac{\mu_0^2}{\mu_{\rm II}^2}\right)
& = & \frac{1}{\beta_0 B_{\mathit{Conf}}} \left(
C_O-C_{\mathit{Conf}}
-\frac{1}{4} \beta_0^2 B_{\beta_0}^2
A_{\mathit{Conf}}- \frac{1}{4} \beta_1 B_{\beta_0}
A_{\mathit{Conf}} \!\right)\!,
 \label{Cconf}
\end{eqnarray}
with $\overline{\beta}_1\equiv\beta_1(N_f=33/2)=-107$.
Up to accuracy $\mathcal{O}(\alpha_s^5)$, we have:
\begin{eqnarray}
D_{\mathit{Conf}}&=& D_{O} \left( N_f \equiv \frac{33}{2} \right)- \frac{1}{8}\overline{\beta}_2 B_{\beta_0} A_{\mathit{Conf}}-  \frac{1}{2} \overline{\beta}_1 C_{\beta_0} B_{\mathit{Conf}},\nonumber \\
D_{\beta_0} \equiv \log\left(\frac{\mu_0^2}{\mu_{\rm
III}^2}\right) & = & \frac{2}{3 \beta_0 C_{\mathit{Conf}}}
 \left[ D_O-D_{\mathit{Conf}} -\frac{1}{8} \left(\beta_0^3
B_{\beta_0}^3 +\frac{5}{2}\beta_0 \beta_1 B_{\beta_0}^2+ \beta_2
B_{\beta_0}\right) A_{\mathit{Conf}}\right. \nonumber \\ & &\hspace{14em}
- \left.\frac{1}{4} \left( 3 \beta_0^2 C_{\beta_0}^2 + 2 \beta_1
C_{\beta_0}\right)
B_{\mathit{Conf}} \right],
 \label{Dconf}
\end{eqnarray}
with $\overline{\beta}_2\equiv \beta_2(N_f=33/2).$

This procedure may be extended to all orders and one may decide
whether to cancel $\beta_0$, $\beta_1$ or $\beta_2$ by fixing the
appropriate number of flavors. The results can be compared
to exactly determine all the coefficients. We point out that
extending the intrinsic conformality to all orders, we can at this stage
predict the coefficients of all the color factors of the
higher orders related to the $\beta$-terms, except those related to
the higher-order conformal coefficients and $\beta_0$-terms (e.g.\ at $N^4$LO, $E_{\mathit{Conf}}$ and $E_{\beta_0}$).

\subsection{Comment on the PMC/PMC$_\infty$ scales}\label{PMCscales}

PMC scales stem directly from the renormalization of the UV-divergent
diagrams. As pointed out in Sec.~\ref{RGroup}, the finite part of
the divergent integrals contribute to the $\beta$-terms.
In fact, these coefficients derive from the UV-divergent
diagrams connected with the running of the coupling constant and
not from UV-finite diagrams. UV-finite $N_F$ terms may arise but
would not contribute to the $\beta$-terms. These terms can be
easily identified by the kinematic constraint at lowest order or
by checking deviations of the $n_f$ coefficients from the iCF
form. In fact, only the $N_f$ terms coming from UV-divergent
diagrams, depending dynamically on the virtuality of the
underlying quark and gluon subprocesses have to be considered as
$\beta$-terms and they would determine the intrinsic conformal
scales.  In general, each $\mu_{\rm N}$ is an independent function
of the physical scale of the process $\sqrt{s}$ (or
$\sqrt{t},\sqrt{u},\dots$), of the selected variable $O$ and it
varies with the number of colors $N_c$ mainly due to the $ggg$ and
$gggg$ vertices. The latter terms arise at higher orders only in a
non-Abelian theory, but they are not expected to spoil the
iCF-form. We stress that iCF applies to scale-invariant
single-variable differential distributions, in case one is
interested in the renormalization of a particular diagram, e.g.\ the $ggg$
vertex, contributions from different $\beta$-terms
should be singled out in order to identify the respective
intrinsic conformal scale consistently with the renormalization of
the non-Abelian $ggg$ vertex, as shown in \cite{Binger:2006sj}.

In the renormalization procedure of gauge theories, one first identifies the
UV singularities of a scattering amplitude, which appear as poles
using dimensional regularization. The UV-divergent contributions
are absorbed into renormalization constants $Z_i$ adopting a
particular scheme (e.g.\ the $\overline{MS}$ scheme). This cancels
the UV divergences and at the same time defines the finite part of
the loop integral.

The finite parts, such as the finite $\beta_i$ and $\gamma_i$
contributions, are associated with the renormalization of the
running coupling and running mass, respectively. The $\beta_i$
terms can then be summed into the running coupling using the
standard renormalization group equations; this is basically the
core of the BLM-PMC scale-setting procedure that is analogous to
the Gell-Mann--Low scheme in QED. This procedure eliminates the
scale ambiguity and reabsorbs the scheme dependence at once into
the effective running coupling up to the computed order. In
addition, the factorial renormalon divergence is eliminated. One
thus can use the same renormalization procedure for QED, QCD and
EW in a grand unified theory,

Given the renormalizability of QCD, once the coupling is
renormalized all the vertices are finite, but this does not cancel
the contributions of the finite parts of the integrals, i.e.\ the
$\beta_n$ terms, which define the PMC scale for each vertex
($3g$, $4g$, $ccg$, $qqg$).

\newpage

\section{PMC$_\infty$ results for thrust and the $C$-parameter\label{sec:thrustandc}}

The thrust distribution and event-shape variables are fundamental tools
for probing the geometrical structure of a
given process at colliders. Being observables that are exclusive
enough with respect to the final state, they allow for a deeper
geometrical analysis of the process and are also particularly
suited for measurement of the strong coupling $\alpha_s$ \cite{Kluth:2006bw}.

Given the high-precision data collected at LEP and SLAC
\cite{ALEPH:2003obs, DELPHI:2003yqh, OPAL:2004wof, L3:2004cdh, SLD:1994idb},
refined calculations are crucial in order to extract information
to the highest possible precision. Although extensive studies on
these observables have been produced during the last few decades,
including higher-order corrections from next-to-leading order (NLO)
calculations \cite{Ellis:1980wv, Kunszt:1980vt, Vermaseren:1980qz, Fabricius:1981sx, Giele:1991vf, Catani:1996jh}
to the next-to-next-to-leading
order(NNLO) \cite{Gehrmann-DeRidder:2014hxk, Gehrmann-DeRidder:2007nzq, GehrmannDeRidder:2007hr, Weinzierl:2008iv, Weinzierl:2009ms}
and including resummation of the large
logarithms \cite{Abbate:2010xh, Banfi:2014sua}, the theoretical
predictions are still affected by significant theoretical
uncertainties that are related to large renormalization scale
ambiguities.
In the particular case of the three-jet event-shape distributions
the conventional practice of CSS leads to results that do not
match the experimental data and the extracted values of $\alpha_s$
deviate from the world average \cite{Workman:2022ynf}.


The thrust ($T$) and $C$-parameter ($C$) are defined by
\begin{eqnarray}
T=\max\limits_{\vec{n}}
\left(\frac{\sum_{i}|\vec{p}_i\cdot\vec{n}|}{\sum_{i}|\vec{p}_i|}\right),
\end{eqnarray}
\begin{eqnarray}
C=\frac{3}{2}\frac{\sum_{i,j}
|\vec{p_i}||\vec{p_j}|\sin^2\theta_{ij}}{\left(\sum_i|\vec{p_i}|\right)^2},
\end{eqnarray}
where the sum runs over all particles in the hadronic final state
and $\vec{p}_i$ denotes the three-momentum of particle~$i$. The
unit vector $\vec{n}$ is varied to maximize thrust $T$; the
corresponding $\vec{n}$ is called the thrust axis and denoted
by~$\vec{n}_T$. The variable $(1-T)$ is often used, which for the LO
of 3-jet production is restricted to the range $(0<1-T<1/3)$. We
have a back-to-back or a spherically symmetric event
at $T=1$ and at $T=2/3$ respectively.
For the $C$-parameter, $\theta_{ij}$ is the angle between
$\vec{p_i}$ and~$\vec{p_j}$. At LO for 3-jet production the
$C$-parameter is restricted by
kinematics to the range $0\leq C\leq0.75$.

In general, a normalized IR-safe single-variable observable, such
as the thrust distribution for $e^+ e^-\rightarrow3\,$jets
\cite{DelDuca:2016ily, DelDuca:2016csb}, is the sum of pQCD
contributions calculated up to NNLO at the initial renormalization
scale $\mu_0=\sqrt{s}=M_{Z}$:
\begin{eqnarray}
\frac{1}{\sigma_{tot}} \! \frac{O d \sigma(\mu_{0})}{d O}\! & = &
\left\{x_0 \cdot \frac{O d \bar{A}_{\mathit{O}}(\mu_0)}{d O} +
x_0^2 \cdot \frac{O d \bar{B}_{\mathit{O}}(\mu_0)}{d O}
 + x_0^{3} \cdot \frac{O d\bar{C}_{\mathit{O}}(\mu_0)}{d O}
 + {\cal O}(\alpha_{s}^4) \right\},
 \label{observable1-thrust}
\end{eqnarray}
where $x(\mu)\equiv \alpha_s(\mu)/(2\pi)$, $O$ is the selected
event-shape variable, $\sigma$ the cross-section of the process,
$$\sigma_{tot} = \sigma_{0}
\Big( 1 + x_0 A_{t o t}+ x_0^{2} B_{t o t}
  + {\cal O}\big(\alpha_{s}^{3}\big) \Big)$$
is the total hadronic cross-section and $\bar{A}_O, \bar{B}_O,
\bar{C}_O$ are respectively the normalized LO, NLO and NNLO
coefficients:
\begin{eqnarray}
\bar{A}_{O} &=&A_{O} \nonumber \\
\bar{B}_{O} &=&B_{O}-A_{t o t} A_{O} \\
\bar{C}_{O} &=&C_{O}-A_{t o t} B_{O}-\left(B_{t o t}-A_{t o
t}^{2}\right) A_{O}, \nonumber
\end{eqnarray}
where $A_O, B_O, C_O$ are the coefficients normalized to the
tree-level cross-section $\sigma_0$ calculated by Monte Carlo (see
e.g.\ the EERAD and Event2 codes \cite{Gehrmann-DeRidder:2014hxk,
Gehrmann-DeRidder:2007nzq, GehrmannDeRidder:2007hr,
Weinzierl:2008iv, Weinzierl:2009ms}) and $A_{\mathit{tot}},
B_{\mathit{tot}}$ are
\begin{eqnarray}
A_{\mathit{tot}} &= & \frac{3}{2} C_F ; \nonumber \\
B_{\mathit{tot}} &= & \frac{C_F}{4}N_c +\frac{3}{4}C_F
\frac{\beta_0}{2} \big(11-8\zeta(3)\big) -\frac{3}{8} C_F^2,
\label{norm}
\end{eqnarray}
where $\zeta$ is the Riemann zeta function.

In general, according to CSS the renormalization scale is set to
$\mu_0=\sqrt{s}=M_Z$ and theoretical uncertainties are evaluated
using standard criteria. In this case, we have used the definition
of the parameter $\delta$ given in Ref.~\cite{Gehrmann-DeRidder:2007nzq};
we define the average error for the event-shape variable
distributions as:
\begin{equation}
\bar{\delta}=\frac{1}{N} \sum_i^N \frac{{\rm
max}_{\mu}\big(\sigma_i(\mu)\big)- {\rm min}_{\mu} \big(\sigma_i(\mu)\big)}{2
\sigma_i(\mu{=}M_Z)}, \label{delta}
\end{equation}
where $i$ is the index of the bin and $N$ is the
total number of bins, the renormalization scale is varied in the
range $\mu\in[M_Z/2,2M_Z]$.


\subsection{The PMC$_\infty$ scales at LO and NLO\label{sec:lonloscales}}

According to the PMC$_\infty$ prescription, we fix the
renormalization scale to $\mu_{\rm N}$ at each order absorbing all
the $\beta$ terms into the coupling. We notice a small mismatch
between the zeroes of the conformal coefficient
$B_{\mathit{Conf}}$ and those of the remaining $\beta_0$ term in
the numerator (the formula is shown in Eq.~\eqref{Cconf}). Due to
our limited knowledge of the strong coupling at low energies, in
order to avoid singularities in the NLO scale $\mu_{\rm II}$, we
introduce a regularization that leads to a finite scale
$\tilde{\mu}_{\rm II}$ over the entire range of values of the
variable $(1-T)$. These singularities might be due either to the
presence of UV finite $N_F$ terms or to the logarithmic behavior
of the conformal coefficients when low values of the variable
$1-T$ are approached. Large logarithms arise from the
IR-divergence cancellation procedure and they can be resummed in
order to restore a predictive perturbative regime
\cite{Catani:1991kz, Catani:1992ua, Catani:1996yz, Banfi:2014sua, Abbate:2010xh}.
We point out that IR cancellation should not spoil the iCF
property. Whether this is an actual deviation from the iCF-form
must be investigated further. However, since the discrepancies
between the coefficients are rather small, we introduce a
regularization method based on redefinition of the norm of the
coefficient $B_{\mathit{Conf}}$ in order to cancel out these
singularities in the $\mu_{\rm II}$ scale. This regularization is
consistent with the PMC principle and up to the accuracy of the
calculation it does not introduce any bias effect in the results
or any ambiguity in the NLO-PMC$_\infty$ scale. All
differences introduced by the regularization would enter at
$\rm N^3LO$ accuracy and they may be reabsorbed later in
the higher-order PMC$_\infty$ scales. For the
PMC$_\infty$ scales, $\mu_{\rm N}$ we thus obtain
\begin{eqnarray}
\mu_{\rm I} & = & \sqrt{s} \cdot e^{f_{sc}-\frac{1}{2}
B_{\beta_0}}
\hspace{4.35cm}(1-T)<0.33,  \label{icfscale1} \\
\tilde\mu_{\rm II} & = &
\left\{
\begin{array}{lr}
\sqrt{s} \cdot e^{f_{sc}-\frac{1}{2} C_{\beta_0} \cdot
\frac{B_{\mathit{Conf}}}{B_{\mathit{Conf}}+\eta \cdot
A_{\mathit{tot}} A_{\mathit{Conf}} }} \hspace{1.5cm} (1-T)<0.33,
\\[0.75ex]
\sqrt{s}\cdot e^{f_{sc}-\frac{1}{2} C_{\beta_0}} \hspace{3.9cm}
(1-T)>0.33, \label{icfscale2}
\end{array}
\right.
\label{PMC12}
\end{eqnarray}
where $\sqrt{s}=M_{Z}$ and the third scale is set to $\mu_{\rm
III}=\mu_0=\sqrt{s}$. The renormalization scheme factor for the
QCD results is set to $f_{sc}\equiv0$. This scheme factor
also reabsorbs the scheme difference into the renormalization
scale and is related to the particular choice of the scale
parameter $\Lambda$ as discussed in Section~\ref{rsdependence}.
The coefficients $B_{\beta_0},C_{\beta_0}$ are the coefficients
related to the $\beta_0$-terms of the NL and NNL perturbative
order of the thrust distribution respectively. They are determined
from the calculated $A_O, B_O, C_O$ coefficients.

The $\eta$ parameter is a regularization term to cancel the
singularities of the NLO scale, $\mu_{\rm II}$, in the range
$(1-T)<0.33$, depending on non-matching zeroes between numerator
and denominator in~$C_{\beta_0}$. In general, this term is not
mandatory for applying the PMC$_{\infty}$, it is necessary only in
case one is interested in applying the method over the entire
range covered by thrust, or any other observable. Its value
has been determined as $\eta=3.51$ for the thrust distribution and
it introduces no bias effects up to the accuracy of the
calculations and the related errors are totally negligible up to
this stage.

\begin{figure}[htb]
\centering
\includegraphics[width=12cm]{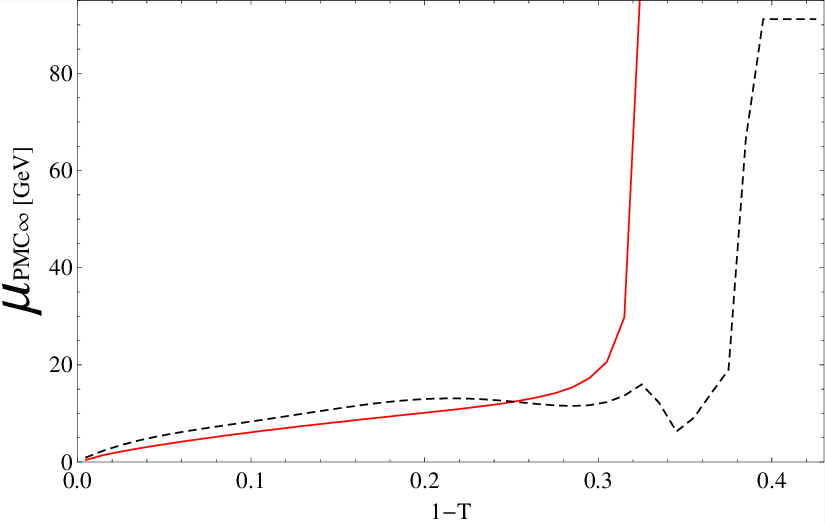}
\caption{The LO-PMC$_\infty$ (solid red) and the NLO-PMC$_\infty$
(dashed black) scales for thrust. \cite{DiGiustino:2020fbk}}
\label{Tscales}
\end{figure}

We point out that in the region $(1-T)>0.33$ we
have a clear example of intrinsic conformality-iCF where the
kinematic constraints set the $A_{\mathit{Conf}}=0$. According to
Eq.~\eqref{evolved} setting the $A_{\mathit{Conf}}=0$ the entire
conformal subset $\sigma_{\rm I}$ becomes null. In this case all
the $\beta$ terms at NLO and NNLO disappear except the
$\beta_0$-term at NNLO, which determines the $\mu_{\rm II}$ scale.
The surviving $n_f$ terms at NLO or the $n_f^2$ at NNLO are
related to the finite $N_F$-term at NLO and to the mixed $N_f
\cdot N_F$ term arising from $B_O \cdot \beta_0$ at NNLO. Using
the parametrization with explicit $n_f$ terms, we have for
$(1-T)>0.33$:
\begin{eqnarray}
A_{O} &=&  0, \nonumber \\
B_{O}  &=& B_0+ B_1 \cdot N_F, \nonumber \\
C_{O} &=& C_0+C_1 \cdot n_f +C_2 \cdot N_f\cdot N_F.
\label{Cparpar}
\end{eqnarray}
we can determine $\tilde{\mu}_{\rm II}$ for the region $(1-T)>0.33$
as shown in Eq.~\eqref{icfscale2}:
\begin{equation}
C_{\beta_0}=\left( \frac{C_1}{\frac{11}{3} C_A B_1-\frac{2}{3}B_0}\right)
\label{mu2}
\end{equation}
by identifying the $\beta_0$-term at NNLO. The LO and NLO
PMC$_\infty$ scales are shown in Fig.~\ref{Tscales}.
We notice that the two PMC$_\infty$ scales have similar behaviors
in the range $(1-T)<0.33$ and that the LO-PMC$_\infty$ scale
agrees with the PMC scale used in Ref.~\cite{Wang:2019ljl}. Small
numerical fluctuations are visible in the NLO-PMC$_\infty$ scale
as unphysical "kinks", due to the Monte-Carlo simulation results
of Ref.\cite{Weinzierl:2009ms}. In the range $(1-T)<0.33$, where
they are more evident, an interpolation has been performed.
However, for the case of thrust, these have a negligible effect on
the final distribution and do not depend on the method used for
setting the scale, e.g. the PMC$_\infty$, but on the method used
for performing the calculations (i.e. the Monte-Carlo
\cite{Weinzierl:2009ms}).

The PMC$_\infty$ method totally eliminates both the ambiguity in
the choice of the renormalization scale and the scheme dependence
to all orders in QCD.

\subsection{NNLO thrust distribution results}

We use here the results of Ref.~\cite{Weinzierl:2008iv,
Weinzierl:2009ms} and for the running coupling $\alpha_s(Q)$ we
use the RunDec program \cite{Chetyrkin:2000yt}. In order to
normalize the thrust distribution consistently, we expand the
denominator in $\alpha_0\equiv\alpha_s(\mu_0)$ while the numerator
has the couplings renormalized at different PMC$_\infty$ scales
$\alpha_I \equiv \alpha_s(\mu_{\rm I})$ and $\alpha_{II}\equiv
\alpha_s(\tilde{\mu}_{II})$. We point out here that the proper
normalization would be given by the integration of the total
cross-section after renormalization with the PMC$_\infty$ scales,
nonetheless since the PMC$_\infty$ prescription only involves
absorption of higher-order terms into the scales, the difference
would be within the accuracy of the calculations, i.e.\ $\sim
\mathcal{O}(\alpha_s^4(\mu_0))$.
Equation~\eqref{observable1-thrust} becomes:
\begin{equation}
\frac{1}{\sigma_{tot}} \, \frac{O d \sigma(\mu_{\rm
I},\tilde{\mu}_{\rm II},\mu_{0})}{d O}=
\left\{\overline{\sigma}_{\rm I}+\overline{\sigma}_{\rm
II}+\overline{\sigma}_{\rm III}+ {\cal O}(\alpha_{s}^4) \right\},
 \label{observable3}
\end{equation}
where the $\overline{\sigma}_{N}$ are normalized subsets that are
given by:
\begin{eqnarray}
\overline{\sigma}_{\rm I} &=& A_{\mathit{Conf}} \cdot x_{\rm I} \nonumber  \\
\overline{\sigma}_{\rm II} &=& \big( B_{\mathit{Conf}}+\eta
A_{\textrm{tot}} A_{\mathit{Conf}} \big)\cdot x_{\rm II}^2
- \eta A_{\textrm{tot}} A_{\mathit{Conf}} \cdot x_0^2
- A_{\textrm{tot}} A_{\mathit{Conf}}\cdot x_0 x_{\rm I} \nonumber \\
\overline{\sigma}_{\rm III} &=& \left( C_{\mathit{Conf}} -
A_{\textrm{tot}} B_{\mathit{Conf}}\!-\!(B_{\textrm
{tot}}-A_{\textrm{tot}}^{2}) A_{\mathit{Conf}}\right) \cdot x_0^3,
\label{normalizedcoeff}
\end{eqnarray}
$A_{\mathit{Conf}}, B_{\mathit{Conf}}, C_{\mathit{Conf}}$ are the
scale-invariant conformal coefficients (i.e.\ the coefficients of
each perturbative order not depending on the scale $\mu_r$) while
$x_{\rm I},x_{\rm II},x_0$ are the couplings determined at the
$\mu_{\rm I},\tilde{\mu}_{\rm II},\mu_0$ scales respectively.


Normalized subsets for the region $(1-T)>0.33$ can be achieved
simply by setting $A_{\mathit{Conf}}\equiv 0$ in the
Eq.~\eqref{normalizedcoeff}. Within the numerical precision of these
calculations there is no evidence of the presence of spurious
terms, such as any further UV-finite $N_F$ term up to
NNLO \cite{Gehrmann:2014uva}, besides the kinematic term at lowest
order in the multi-jet region. These terms, if there are any, must
remain rather small over the entire range of the thrust variable in
comparison with the $\beta$ term or even be compatible with
numerical fluctuations. Moreover, we notice a small rather
constant difference between the iCF-predicted and the calculated
coefficient for the $N_f^2$ color factor of
Ref.~\cite{Weinzierl:2008iv}, which might be due to an $n_f^2$
UV-finite coefficient or possibly to statistics. This small
difference must be included in the conformal coefficient but it
has a completely negligible impact on the total thrust
distribution.
\begin{figure}[htb]
\centering
\includegraphics[width=12cm]{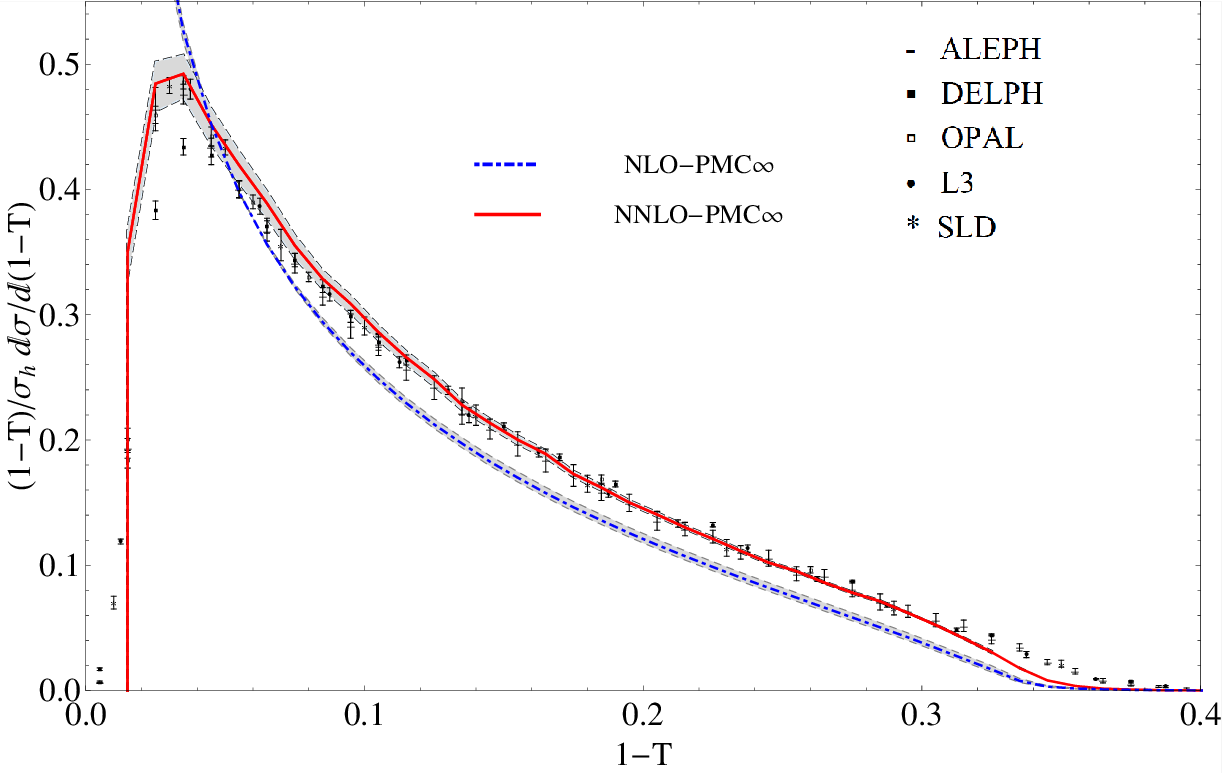}
\caption{The thrust distribution under the PMC$_\infty$ at NLO
(dot-dashed blue) and at NNLO (solid red)
\cite{DiGiustino:2020fbk}. The experimental data points are taken
from the ALEPH, DELPHI, OPAL, L3, SLD experiments
\cite{ALEPH:2003obs, DELPHI:2003yqh, OPAL:2004wof, L3:2004cdh,
SLD:1994idb}. The shaded area shows theoretical errors for the
PMC$_\infty$ predictions at NLO and at NNLO.} \label{thrust}
\end{figure}
In Fig.~\ref{thrust} we show the thrust distribution at NLO and at
NNLO with the use of the PMC$_\infty$ method. Theoretical errors
for the thrust distribution at NLO and at NNLO are also shown (the
shaded area). Conformal quantities are not affected by a change of
renormalization scale. Thus, the errors shown give an evaluation
of the level of conformality achieved up to the order of accuracy
and they have been calculated using standard criteria, i.e.\
varying the remaining initial scale value in the range $\sqrt{s}/2
\leq \mu_0 \leq 2 \sqrt{s}$.

We recall that the distributions are calculated by a Monte-Carlo
simulation, considering 50 equidistant bins in the range
$0<1-T<0.5$, thus the peak in the NNLO-PMC$_\infty$ distribution
appears more as a broken-straight line also due to the linear
interpolation used in the figure to join the points.

Using the same definition of the parameter $\bar{\delta}$ given in
Eq.~\eqref{delta}, we have in the interval $0<(1-T)<0.33$ an
average error of $\bar{\delta}\simeq 3.54\%$ and $1.77\%$ for the
thrust at NLO and at NNLO respectively. A greater improvement has
been obtained over the entire range of reliable results for thrust
distribution, i.e.\ $0<(1-T)<0.42$, from $\bar\delta\simeq7.36\%$
to $1.95\%$ from NLO to the NNLO accuracy with the PMC$_\infty$.

\begin{figure}[htb]
\centering
\includegraphics[width=12cm]{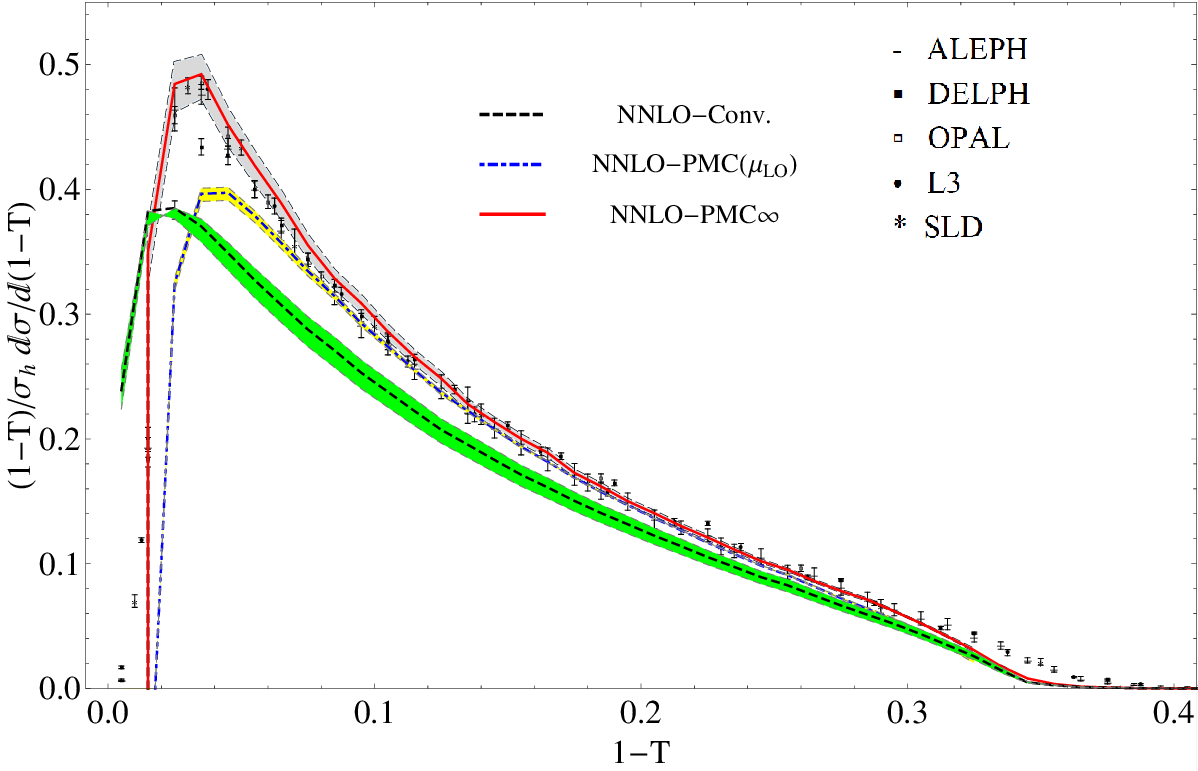}
\caption{The thrust distribution at NNLO under the Conventional
(dashed black), the PMC($\mu$\textsubscript{LO}) (dot-dashed blue)
and the PMC$_\infty$ (solid red) \cite{DiGiustino:2020fbk}. The
experimental data points are taken from the ALEPH, DELPHI, OPAL,
L3, SLD experiments \cite{ALEPH:2003obs, DELPHI:2003yqh,
OPAL:2004wof, L3:2004cdh, SLD:1994idb}. The shaded areas show
theoretical error predictions at NNLO, which have been calculated
varying the remaining initial scale value in the range $\sqrt{s}/2
\leq \mu_0 \leq 2 \sqrt{s}$.} \label{thrust2}
\end{figure}

In Fig.~\ref{thrust2} a direct comparison of the PMC$_\infty$ with
the CSS results (obtained in  \cite{Weinzierl:2008iv} and
 \cite{Gehrmann-DeRidder:2007nzq, GehrmannDeRidder:2007hr}) is
shown. In addition, we show the results of the first PMC approach
used in \cite{Wang:2019ljl}, which we indicate as
PMC($\mu$\textsubscript{LO}) extended to NNLO accuracy. In this
approach the last unknown PMC scale $\mu$\textsubscript{NLO} of
the NLO was set to the last known PMC scale
$\mu$\textsubscript{LO} of the LO, while the NNLO scale
$\mu$\textsubscript{NNLO}$\equiv\mu_0$ was left unset and varied
in the range $\sqrt{s}/2\leq\mu_0\leq2\sqrt{s}$.

Average errors calculated in different regions of the spectrum are
reported in Table~\ref{tab:1}. The PMC$(\mu_{LO})$ cannot be
defined in the range $0.33<1-T<0.42$ since $A_{\rm Conf}=0$; thus
the third and fifth row in Table~\ref{tab:1} are blank. In fact,
this further analysis was performed in order to show that the
procedure of setting the last unknown scale to the last known one
leads to stable and precise results and is consistent with the
proper PMC method over a wide range of accessible values of the
$(1-T)$ variable.

\begin{table}[h!]
\centering
 \begin{tabular}{||c|c|c|c||}
   \hline
 $\bar{\delta}[\%]$ & Conv. & PMC($\mu$\textsubscript{LO}) & PMC$_\infty$ \\
   \hline
   $0.10 < (1-T) < 0.33$ & 6.03 & 1.41 & 1.31 \\
   $0.21 < (1-T) < 0.33$ & 6.97 & 2.19 & 0.98 \\
   $0.33 < (1-T) < 0.42$ & 8.46 & ---  & 2.61 \\
   $0.00 < (1-T) < 0.33$ & 5.34 & 1.33 & 1.77\\
   $0.00 < (1-T) < 0.42$ & 6.00 & --- & 1.95 \\
   \hline
  \end{tabular}
 \caption{The average error $\bar{\delta}$ for the NNLO thrust distribution under Conventional, PMC($\mu$\textsubscript{LO}) and PMC$_\infty$
 scale settings calculated in different ranges of values of the $(1-T)$ variable.}
 \label{tab:1}
\end{table}
 From the comparison with the CSS, we
notice that the PMC$_\infty$ prescription significantly improves
the theoretical predictions. Moreover, results are in remarkable
agreement with the experimental data over a wider range of values
$(0.015 \leq 1-T \leq 0.33)$ and they show an improvement of the
PMC$(\mu_{LO})$ results when the two-jet and multi-jet
regions are approached, i.e.\ the region of the peak and the region
$(1-T)>0.33$ respectively. The use of the PMC$_\infty$ approach in
perturbative QCD thrust calculations restores the correct behavior
of the thrust distribution in the region $(1-T)>0.33$ and this is
a clear effect of the iCF property. Comparison with experimental
data has been improved over the entire spectrum and the introduction
of the $\rm N^3LO$ correction would improve this comparison
especially in the multi-jet $(1-T) > 0.33$ region. In the
PMC$_\infty$ method theoretical errors are given by the unknown
intrinsic conformal scale of the last order of accuracy. We expect
this scale not to be significantly different from that of the
previous orders. In this particular case, as shown in
Eq.~\eqref{normalizedcoeff}, we also have a dependence on the
initial scale $\alpha_s(\mu_0)$ left due to the normalization and
to the regularization terms. These errors represent 12.5\% and
1.5\% respectively of the full theoretical errors in the range
$0<(1-T)<0.42$ and they could be improved by means of a correct
normalization.

\subsection{NNLO $C$-parameter distribution results}

The same analysis applies straightforwardly to the $C$-parameter
distribution including the regularizing $\eta$ parameter, which
has been set to the same value~$3.51$. The same scales of
Eq.~\eqref{icfscale1} and Eq.~\eqref{icfscale2} apply to the
$C$-parameter distribution in the region $0<C<0.75$ and in the
region $0.75<C<1$. In fact, due to kinematic constraints that set
the $A_{\mathit{Conf}}=0$, we also have the same iCF effect for
the $C$-parameter.

\begin{figure}[htb]
\centering
\includegraphics[width=12cm]{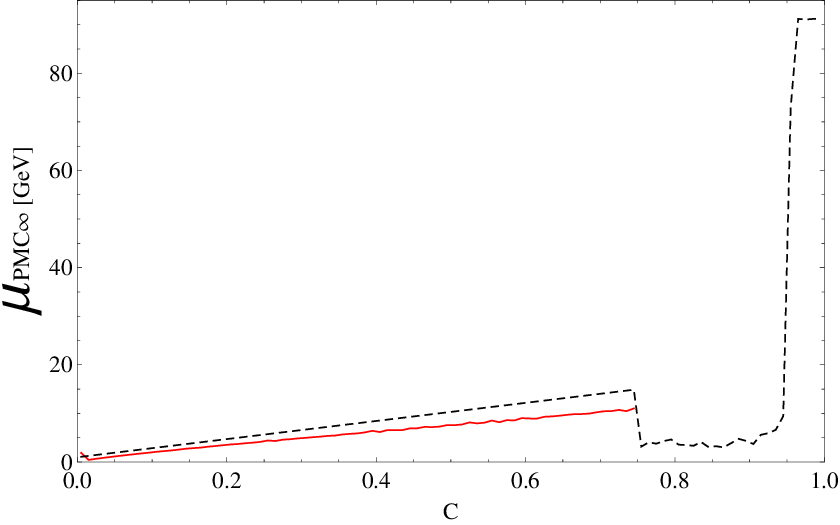}
\caption{The LO-PMC$_\infty$ (solid red) and the NLO-PMC$_\infty$
(dashed black) scales for the $C$-parameter
\cite{DiGiustino:2020fbk}.} \label{Cpar-scales}
\end{figure}

\begin{figure}[htb]
\centering
\includegraphics[width=12cm]{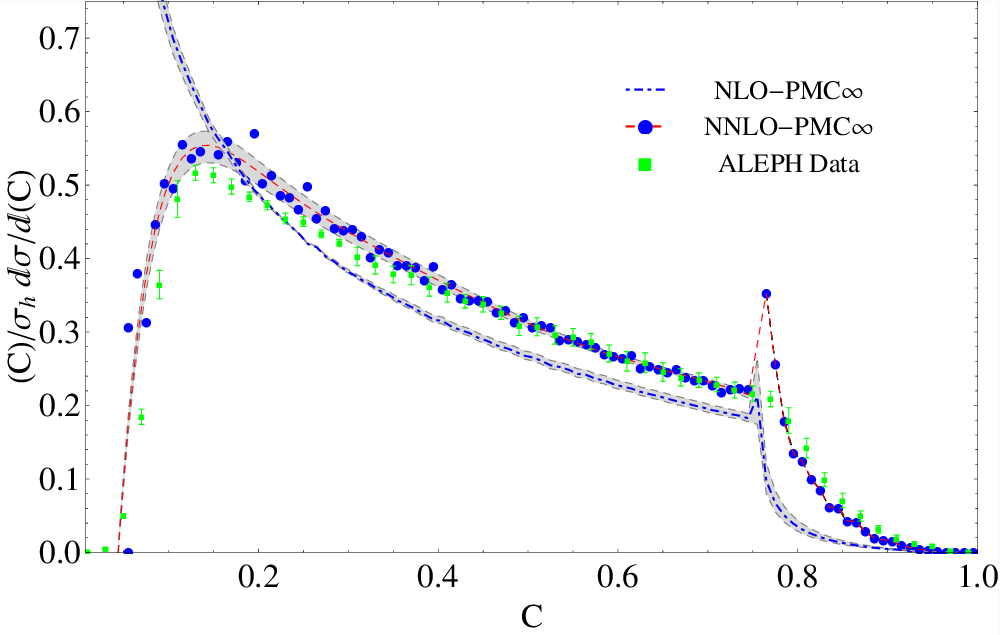}
\caption{The $C$-parameter distribution under the PMC$_\infty$ at
NLO (dot-dashed blue) and at NNLO (dashed red)
\cite{DiGiustino:2020fbk}. The blue points indicate the
NNLO-PMC$_\infty$ thrust distribution obtained with $\mu_{\rm
III}=\mu_0=M_Z$. The experimental data points (green) are taken
from the ALEPH experiment \cite{ALEPH:2003obs}. The dashed lines
of the NNLO distribution show fits of the theoretical calculations
with interpolating functions for the values of the remaining
initial scale $\mu_0 = 2 M_Z$ and~$M_Z/2$. The shaded area shows
theoretical errors for the PMC$_\infty$ predictions at NLO and at
NNLO calculated varying the remaining initial scale value in the
range $\sqrt{s}/2\leq \mu_0\leq2\sqrt{s}$.} \label{Cpar}
\end{figure}

Results for the $C$-parameter scales are shown in
Fig.~\ref{Cpar-scales}. We notice the effect of the iCF intrinsic
conformality on the LO-PMC$_\infty$ scale, which terminates at the
kinematic boundary $C=0.75$. In fact, at this boundary $A_{\rm
Conf}=0$ and thus sets the whole conformal subset $\sigma_{\rm
I}=0$. The NLO-PMC$_\infty$ scale has two distinct domains
separated by the kinematic constraint $C=0.75$. In the range
$0<C<0.75$, the NLO-PMC$_\infty$ scale has the same physical
behavior as the LO-PMC$_\infty$ scale, in the range $0.75<C<0.97$
is almost constant, due to a similar behavior of the numerator and
denominator in Eq.\ref{mu2} and at $C=0.97$ it goes through a
saturation effect given by the $C_1$ coefficient in Eq.\ref{mu2},
which becomes null. Results for the $C$-parameter distributions
are shown in Fig.~\ref{Cpar}.

Theoretical errors have been calculated, as in the previous case,
using standard criteria and results indicate an average error over
the entire spectrum $0<C<1$ of the $C$-parameter distribution at NLO
and at NNLO of $\bar{\delta}\simeq 7.26\%$ and $2.43\%$
respectively.
\begin{table}[htb]
\centering
 \begin{tabular}{||c|r|c|c||}
   \hline
 $\bar{\delta}\;$[\%] & Conv. & PMC($\mu$\textsubscript{LO}) & PMC$_\infty$ \\
   \hline
   $0.00 < (C) < 0.75$ & 4.77 & 0.85 & 2.43 \\
   $0.75 < (C) < 1.00$ & 11.51 & 3.68 & 2.42 \\
   $0.00 < (C) < 1.00$ & 6.47 & 1.55 & 2.43 \\
  \hline
 \end{tabular}
 \caption{The average error $\bar\delta$ for the NNLO $C$-parameter distribution under Conventional, PMC($\mu$\textsubscript{LO}) and PMC$_\infty$
 scale settings calculated in different ranges of values of the $(C)$ variable.}
 \label{tab:2}
\end{table}
A comparison of average errors according to the different methods
is displayed in Table~\ref{tab:2}. Results show that the PMC$_\infty$
improves the NNLO QCD predictions for the $C$-parameter distribution
over the entire spectrum.

\begin{figure}[htb]
\centering
\includegraphics[width=12cm]{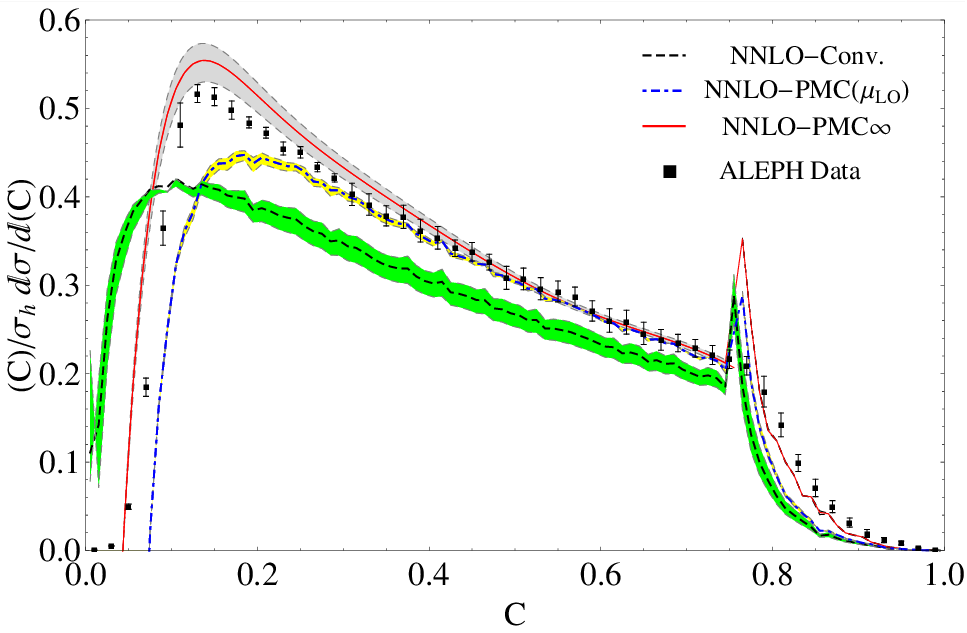}
\caption{The NNLO $C$-parameter distribution under CSS (dashed
black), the PMC($\mu$\textsubscript{LO}) (dot-dashed blue) and the
PMC$_\infty$ (solid red) \cite{DiGiustino:2020fbk}. The
experimental data points (Black) are taken from the ALEPH
experiment \cite{ALEPH:2003obs}. The shaded area shows theoretical
error predictions at NNLO calculated varying the remaining initial
scale value in the range $\sqrt{s}/2 \leq \mu_0 \leq 2 \sqrt{s}$.}
\label{Cpar2}
\end{figure}

A comparison of the distributions calculated with the CSS, the
PMC($\mu$\textsubscript{LO}) \cite{Wang:2019isi} and the
PMC$_\infty$ is shown in Fig.~\ref{Cpar2}. The results for the
PMC$_\infty$ display remarkable agreement with the experimental data
away from the regions $C<0.05$ and $C\simeq0.75$. The errors due
to the normalization and to the regularization terms
(Eq.~\eqref{normalizedcoeff}) are respectively $8.8\%$ and $0.7\%$
of the full theoretical errors.

The perturbative calculations could be further improved using a
correct normalization and also by introducing the resummation of
the large-logarithm technique in order to extend the perturbative
regime and to eliminate the unphysical spike at $C=0.75$, which is
due to enhanced logarithmic terms at the kinematic boundary.

\subsection{The thrust distribution in the QCD conformal
window and in QED\label{sec:confwin}}

For the first time, we employ the perturbative regime of the
quantum chromodynamics (pQCD) infrared conformal window as a
laboratory to investigate in a controllable manner (near)
conformal properties of physically relevant quantities, such as the
thrust distribution in electron--positron annihilation
processes \cite{DiGiustino:2021nep}.
The conformal window of pQCD has a long and noble history
conveniently summarized and generalized to arbitrary
representations in Ref.~\cite{Dietrich:2006cm}. Several lattice
gauge theory applications and results have been summarized in a
recent report on the subject in Ref.~\cite{Cacciapaglia:2020kgq}.

\subsubsection{The thrust distribution according to $N_f$}

It would be highly desirable to compare the PMC and CSS methods
along the entire renormalization group flow from the highest
energies down to zero energy. This is precluded in standard QCD
with a number of active flavors less than six because the theory
becomes strongly coupled at low energies. We therefore employ the
perturbative regime of the conformal window (Sec.~\ref{twoloops})
which allows us to arrive at arbitrary low energies and obtain the
corresponding results for the SU($3$) case at the cost of
increasing the number of active flavors. Here we are able to
deduce the full solution at NNLO in the strong coupling.
In this section we shall consider the region of flavors and
colors near the upper bound of the conformal window, i.e.\ $N_f\sim 11/2 N_c$, where the IR fixed point can be reliably
accessed in perturbation theory and we compare the two
renormalization scale setting methods, the CSS and the
PMC$_\infty$.

\begin{figure}[h]
 \centering
\hspace*{-0.5cm}
\includegraphics[width=12cm]{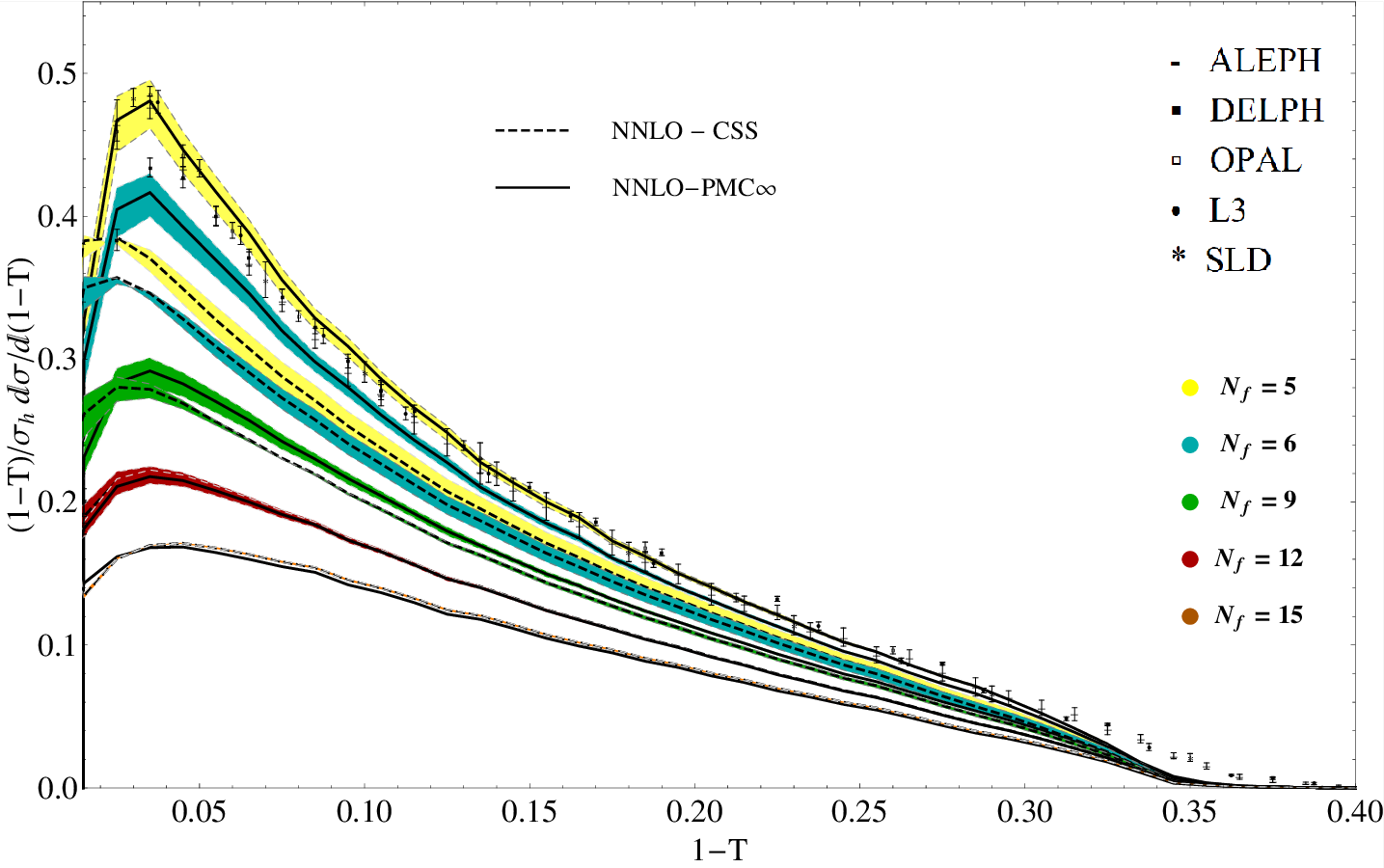}
\caption{Thrust distributions for different values of $N_f$, using
the PMC$_\infty$ (solid line) and the CSS (dashed line)
\cite{DiGiustino:2021nep}. Shaded colored areas show error bars
for each curve respectively. The experimental data points are
taken from the ALEPH, DELPHI, OPAL, L3, SLD experiments
\cite{ALEPH:2003obs, DELPHI:2003yqh, OPAL:2004wof, L3:2004cdh,
SLD:1994idb}.} \label{confthrust}
\end{figure}

Results for the thrust distribution calculated using the  NNLO
solution for the coupling $\alpha_s(\mu)$, at different values of
the number of flavors, $N_f$,  is shown in Fig.~\ref{confthrust}.
A direct comparison between PMC$_\infty$ (solid line) and CSS
(dashed line) is shown at different values of the number of
flavors. We notice that, despite the phase transition (i.e.\ the
transition from an infrared finite coupling to an infrared
divergent coupling), the curves given by the PMC$_\infty$ at
different $N_f$, preserve with continuity the same characteristics
of the conformal distribution setting $N_f$ outside the conformal
window of pQCD.

Technically, this is explained by the fact that the PMC$_\infty$
reabsorbs all the $N_f$ terms into the running coupling and the
PMC$_\infty$ scales are both above 2 GeV in almost the entire
range of the distribution; in particular, for values in the range
$0.015<1-T<0.42$, and thus the PMC$_\infty$ thrust distribution is
affected by the change of behavior of the coupling only in the
first two bins at $1-T\sim 0$. However, this region is a
multi-scale region and is not only affected by nonperturbative
effects, but also by the presence of large-logarithms deriving
from incomplete IR cancellation. On the other hand, the CSS
distribution is more sensitive to the decreasing of $N_f$.

\begin{figure}[htb]
\centering
\includegraphics[width=0.5\textwidth]{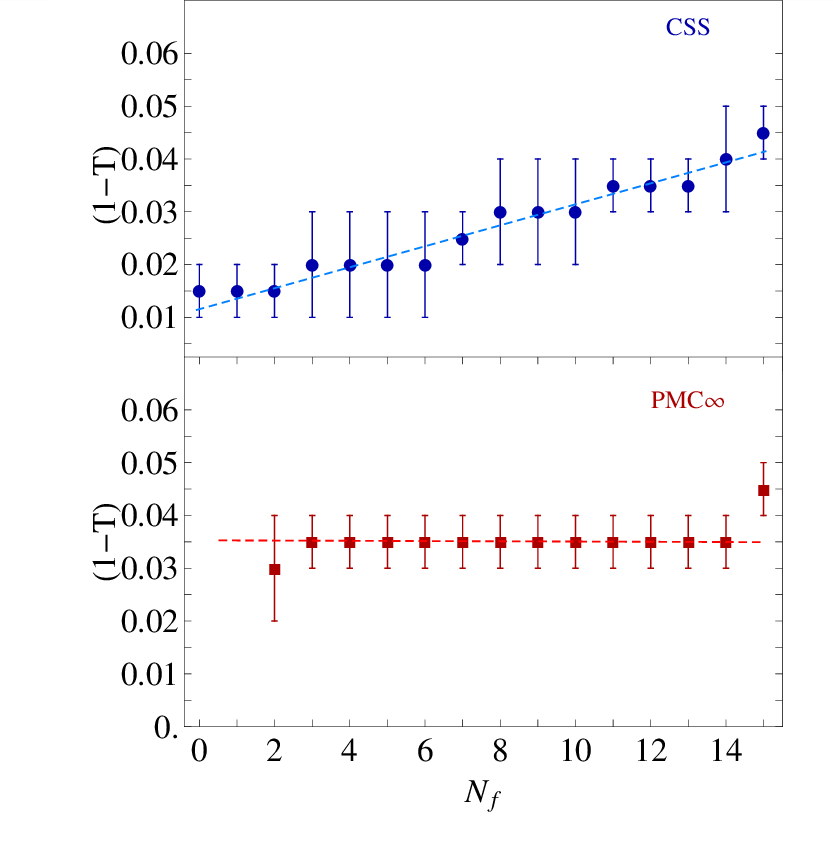}
\caption{Comparison of the position of the peak for the thrust
distribution using the CSS and the PMC$_\infty$ vs the number of
flavors, $N_f$. Dashed lines indicate the particular trend in each
graph~\cite{DiGiustino:2021nep}.} \label{peaks}
\end{figure}

In fact, the position of the thrust distribution peak is well
preserved varying $N_f$ in and outside the conformal window using
the PMC$_\infty$, while there is a constant shift towards lower
values using the CSS. These trends are shown in Fig.~\ref{peaks}.
We notice that in the central range, $2<N_f<15$, the position of
the peak is exactly preserved using the PMC$_\infty$ and overlaps
with the position of the peak shown by the experimental data.
According to our analysis for the case PMC$_\infty$, in the range
$N_f<2$, the number of bins is insufficient to resolve the peak,
although the behavior of the curve is consistent with the presence
of a peak in the same position, while for $N_f\rightarrow0$, the
peak is no longer visible. Theoretical uncertainties on the
position of the peak have been calculated using standard criteria,
i.e.\ by varying the remaining initial scale value in the range
$M_Z/2 \leq \mu_0 \leq 2 M_Z$ and considering the lowest
uncertainty given by the half the spacing between two adjacent
bins.

\begin{figure}[htb]
\centering
\includegraphics[width=0.5\textwidth]{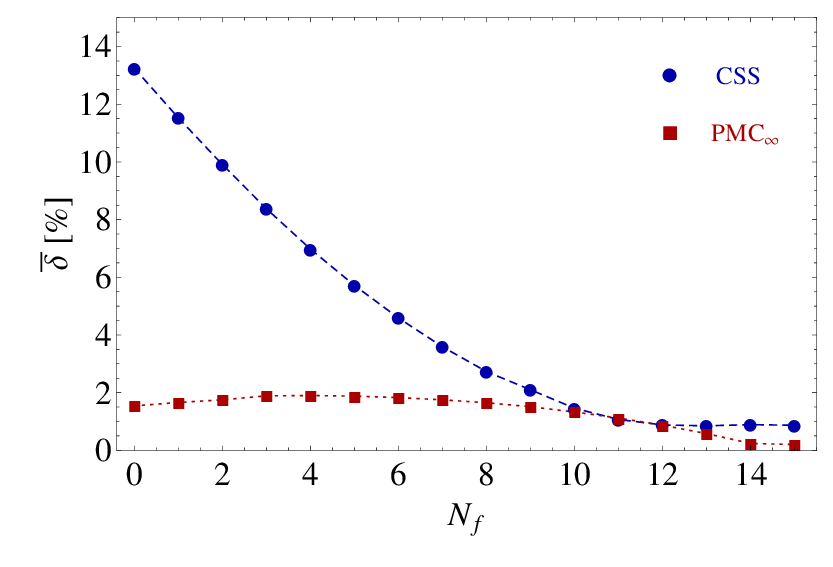}
\caption{Comparison of the average theoretical error,
$\bar{\delta}$, calculated using standard criteria in the range:
$0.005<(1-T)<0.4$, using the CSS and the PMC$_\infty$ for the
thrust distribution vs.\ the number of flavors, $N_f$
\cite{DiGiustino:2021nep}.} \label{err}
\end{figure}

Using the definition given in Eq.~\eqref{delta}, we have determined
the average error, $\bar{\delta}$, calculated in the interval
$0.005<(1-T)<0.4$ of thrust and results for CSS and PMC$_\infty$
are shown in Fig.~\ref{err}. We notice that the PMC$_\infty$ in
the perturbative and IR conformal window, i.e.\ $12<N_f<\bar{N}_f$,
which is the region where $\alpha_s(\mu)<1$ in the entire range of
the renormalization scale values, from $0$ up to $\infty$, the
average error given by PMC$_\infty$ tends to zero ($\sim
0.23-0.26\%$) while the error given by the CSS tends to remain
constant ($0.85-0.89\%$). Comparison of the two methods shows
that, outside the conformal window, $N_f<\frac{34 N_c^3}{13
N_c^2-3}$, the PMC$_\infty$ leads to higher precision.

From our analysis in section \ref{twoloops}, it follows that the
IR fixed point at $N_f=\frac{11}{2} N_c$ is not phenomenologically
accessible, assuming a measured value of the coupling at a certain
scale, since in the region $\bar{N}_f<N_f<\frac{11}{2}N_c$, the
coupling would no longer have the UV asymptotically free behavior.
Thus, in order to have a phenomenological application of the
Banks-Zaks variable, (i.e. $\Delta_f=33/2-N_f\simeq \epsilon$, as
shown in Refs.\cite{Ryttov:2016asb,Ryttov:2016hal}), one should be
able to reach any small value of the coupling above the Planck
scale in QCD, since the IR fixed point is interacting. Moreover, a
conformal result is just characterized by the absence of running
of the coupling and not necessarily by a null value. A straight
conformal solution of Eqs. \ref{xz1} and \ref{xz}, is given by:
$z=0$, $W=0$ with $ x(\mu)=x^\ast= x_0$ for any scale
$0<\mu<\infty$. Our application of the Banks-Zaks results shows
that the conformal limit of the thrust distribution is less
sensitive to the method adopted to set the renormalization scale,
CSS or PMC$_\infty$, once the same phenomenological value of the
coupling has been determined at a certain scale. We suggest that a
more suitable variable for the expansion would be given by
$\tilde{\Delta}=\bar{N}_f-N_f\simeq\epsilon$, where $\bar{N}_f$ is
the maximum allowed value of $N_f$ to be asymptotically free,
introduced by the phenomenological value of the coupling at an
initial scale $\mu_0$.

\subsubsection{The thrust distribution in the Abelian limit
$N_c\rightarrow 0$\label{sec:qedthrust}}

\begin{figure}[htb]
\centering
\includegraphics[width=12cm]{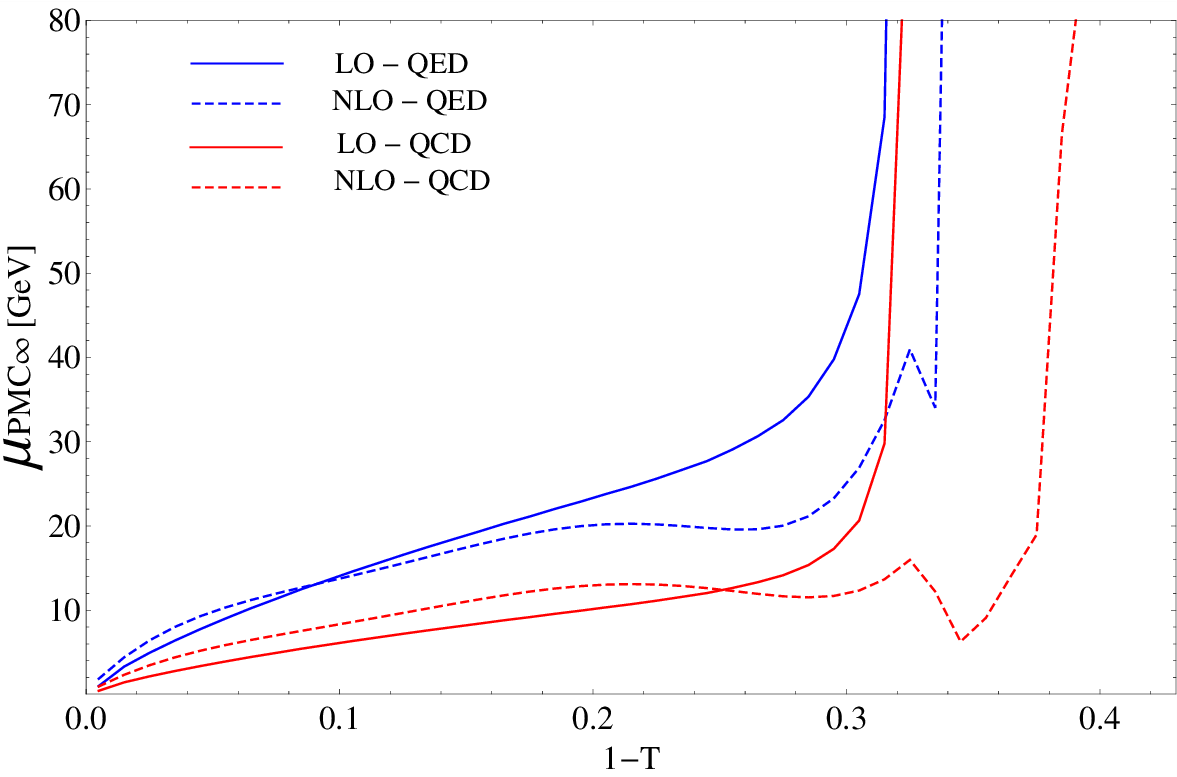}
\caption{PMC$_\infty$ scales for the thrust distribution: LO-QCD
scale (solid red); LO-QED scale (solid blue);NLO-QCD scale (dashed
red); NLO-QED scale (dashed blue) \cite{DiGiustino:2021nep}.}
\label{scales}
\end{figure}

We consider now the thrust distribution in U(1) Abelian QED, which
rather than being infrared interacting is infrared free. We obtain
the QED thrust distribution performing the $N_c\rightarrow 0$
limit of the QCD thrust at NNLO according
to \cite{Brodsky:1997jk, Kataev:2015yha}. In the zero number of
colors limit the gauge group color factors are fixed by $N_A=1,$
$C_F=1,$ $T_R=1,$ $C_A=0,$ $N_c=0,$ $N_f=N_l$, where $N_l$ is the
number of active leptons, while the $\beta$-terms and the coupling
rescale as $\beta_n/C_F^{n+1}$ and $\alpha_s \cdot C_F$
respectively. In particular, $\beta_0=-\frac{4}{3}N_l$ and
$\beta_1=-4 N_l$ using the normalization of Eq.~\eqref{betafun1}.
According to this rescaling of the color factors, we have
determined the QED thrust and the QED PMC$_\infty$ scales. For the
QED coupling, we have used the analytic formula for the effective
fine structure constant in the $\overline{\textrm{MS}}$ scheme:
\begin{equation}
{\alpha(Q)} = {\alpha \over  {\left(1 - \Re e
\Pi^{\overline{\textrm{MS}}} (Q^2)\right)}},
\end{equation}
with $\alpha^{-1}\equiv \alpha(0)^{-1}= 137.036$ and the vacuum
polarization function ($\Pi$) calculated perturbatively to two
loops, including contributions from leptons, quarks and the $W$ boson.
The QED PMC$_\infty$ scales have the same form of
Eqs.~\eqref{icfscale1} and \eqref{icfscale2} with the factor for the
$\overline{\textrm{MS}}$ scheme set to $f_{sc}\equiv 5/6$ and the
$\eta$ regularization parameter introduced to cancel singularities
in the NLO PMC$_\infty$ scale $\mu_{\rm II}$ in the $N_c
\rightarrow 0$ limit tends to the same QCD value, $\eta=3.51$. A
direct comparison between QED and QCD PMC$_\infty$ scales is shown
in Fig.~\ref{scales}.

\begin{figure}[htb]
\centering \vspace{1cm}
\includegraphics[width=12cm]{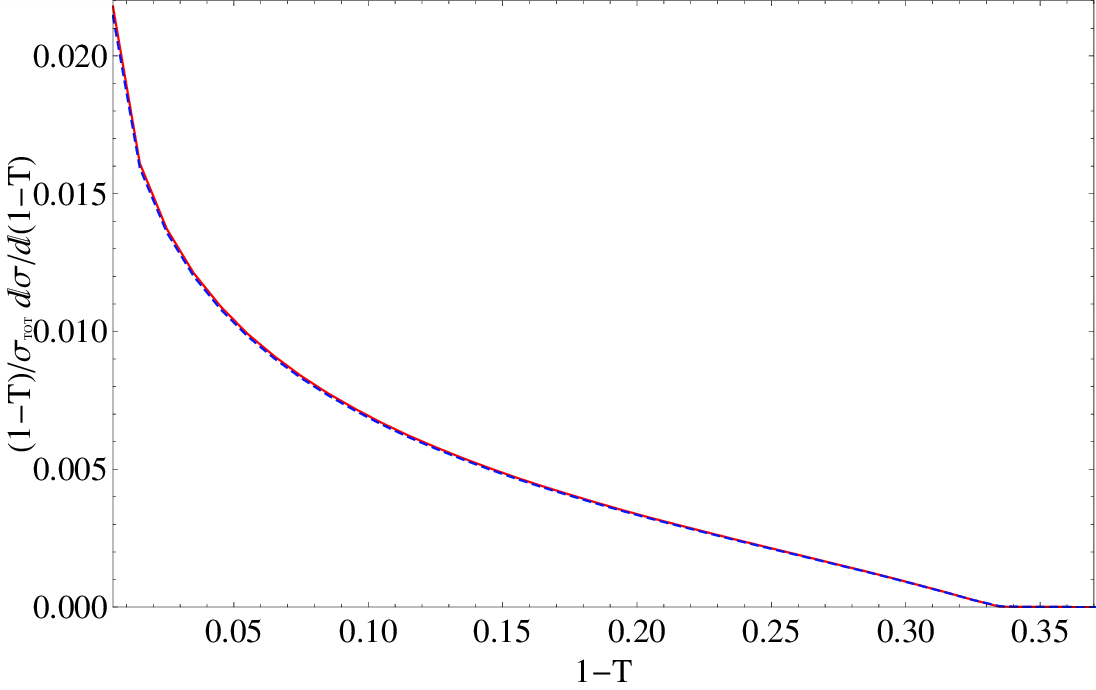}
\caption{Thrust distributions in the QED limit at NNLO using the
PMC$_\infty$ (solid red) and the CSS (dashed blue)
\cite{DiGiustino:2021nep}.} \label{qedthrust}
\end{figure}

We note that in the QED limit the PMC$_\infty$ scales have
analogous dynamical behavior to those calculated in QCD;
differences arise mainly owing to the $\overline{\textrm{MS}}$
scheme factor reabsorption, the effects of the $N_c$ number of
colors at NLO are very small. Thus, we notice that perfect
consistency is shown from QCD to QED using the PMC$_\infty$
method. The normalized QED thrust distribution is shown in
Fig.~\ref{qedthrust}. We note that the curve is peaked at the
origin, $T=1$, which suggests that the three-jet event in QED
occurs with a rather back-to-back symmetry. Results for the CSS
and the PMC$_\infty$ methods in QED are
$O(\alpha)$ and show very small differences, given the good
convergence of the theory.


\subsection{A novel method for the precise determination of the
strong coupling and its behavior\label{sec:novel}}

In this section we present a novel method for precisely
determining the running QCD coupling constant $\alpha_s(Q)$ over a
wide range of $Q$ from event-shape variables for
electron--positron annihilation measured at a single
center-of-mass energy $\sqrt{s}$, based on PMC scale setting. In
particular, we display the results obtained in
Refs.~\cite{Wang:2019isi, Wang:2019ljl} using the approach of a
single PMC scale at LO and NLO, i.e.\ the
PMC($\mu$\textsubscript{LO}) of the previous section.


The precise determination of the strong coupling $\alpha_s(Q)$ is
one of the crucial tests of QCD. The dependence of $\alpha_s(Q)$
on the renormalization scale $Q$ obtained from many different
physical processes shows consistency with QCD predictions and
asymptotic freedom. The Particle Data Group (PDG) currently gives
the world average:
$\alpha_s(M_Z)=0.1179\pm0.0009$ \cite{Workman:2022ynf} in the
$\overline{\rm{MS}}$ renormalization scheme.

Particularly suitable to the determination of the strong coupling
is the process $e^+e^-\rightarrow 3\,$jets since its leading order
is $\mathcal{O}(\alpha_s)$ \cite{Kluth:2006bw}. Currently,
theoretical calculations for event shapes are based on CSS. By
using conventional scale setting, only one value of $\alpha_s$ at
the scale $\sqrt{s}$ can be extracted and the main source of the
uncertainty is given by the choice of the renormalization scale.
Several values for the strong coupling have been extracted from
several processes, e.g.\ $\alpha_s(M_Z)=0.1224\pm 0.0035$
\cite{Dissertori:2009ik} is obtained by using perturbative
corrections and resummation of the large logarithms in the
NNLO+NLL accuracy predictions. Other evaluations improving the
resummation calculations up to N$^3$LL give a result of
$\alpha_s(M_Z)=0.1135\pm0.0011$ \cite{Abbate:2010xh} from thrust
and $\alpha_s(M_Z)=0.1123\pm0.0015$ \cite{Hoang:2015hka} from the
$C$-parameter. Nonperturbative corrections for hadronization
effects have also been included in Ref.~\cite{Dasgupta:2003iq},
but as pointed out in Ref.~\cite{Tanabashi:2018oca}, the
systematics of the theoretical uncertainties introduced by
hadronization effects are not well understood.

In this section we show that by using the PMC, it is possible to
eliminate the renormalization scale ambiguities and obtain
consistent results for the strong coupling using the precise
experimental data of event-shape variable distributions. We notice
that improved event-shape distributions have been obtained in
Refs.~\cite{Kramer:1990zt, Gehrmann:2014uva, Hoang:2014wka} using
BLM and soft and collinear effective theory (SCET).

\subsubsection{Running behavior}
Given that the PMC scale (PMC($\mu$\textsubscript{LO})) is not a
single-valued but rather a monotonically increasing function of
$\sqrt{s}$ and of the selected observable,(as shown in
Figs.~\ref{Tscales} and \ref{Cpar-scales}), it is possible to
determine the strong coupling at different scales from one
single experiment at one single center-of-mass energy. The
dependence of the scale on the observable reflects the dynamics of
the underlying gluon and quark subprocess. This dynamics also varies
the number of active flavors~$N_f$.
Considering that PMC scales for QCD and QED show the same behavior
and that their relation at LO is only given by a RS redefinition term,
$Q^2_{\rm QCD}/Q^2_{\rm QED}=e^{-5/3}$, this approach may also be
extended to QED.
\begin{figure}[htb]
\centering
\includegraphics[width=0.60\textwidth]{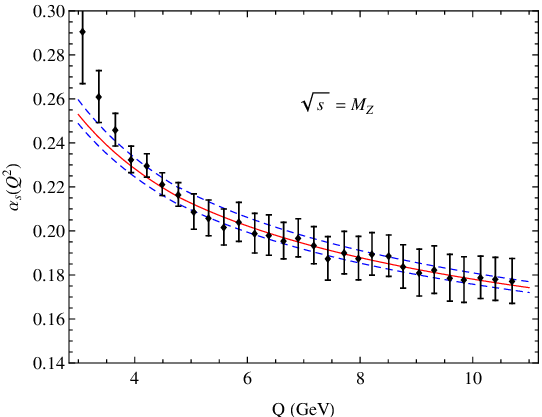}
\caption{The coupling constant $\alpha_s(Q)$ extracted by
comparing PMC predictions with the ALEPH data \cite{ALEPH:2003obs}
at a single energy of $\sqrt{s}=M_Z$ from the $C$-parameter
distributions in the $\overline{\rm MS}$ scheme (from
Ref.~\cite{Wang:2019isi}). The error bars are the squared averages
of the experimental and theoretical errors. The three lines are
the world average evaluated from $\alpha_s(M_Z)=0.1179\pm0.009$
\cite{Workman:2022ynf}.} \label{figasPMCC}
\end{figure}

\begin{figure}[htb]
\centering
\includegraphics[width=0.60\textwidth]{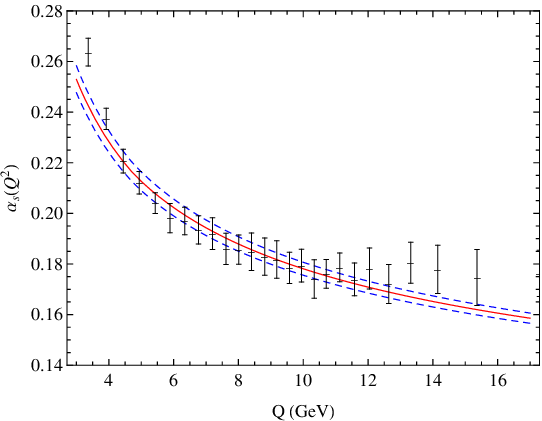}
\caption{The extracted $\alpha_s$ from the comparison of PMC
predictions with ALEPH data at $\sqrt{s}=M_Z$ (from
Ref.~\cite{Wang:2019ljl}). The error bars are from the
experimental data. The three lines are the world average evaluated
from $\alpha_s(M_Z)=0.1179\pm0.0009$ \cite{Workman:2022ynf}.}
\label{figasPMCT}
\end{figure}

We extract $\alpha_s$ at different scales bin-by-bin from the
comparison of PMC predictions for ($1-T$) and $C$ differential
distributions with measurements at $\sqrt{s}=M_Z$. The extracted
$\alpha_s$ values from the $C$-parameter distribution are shown in
Fig.~\ref{figasPMCC}. We note that the $\alpha_s$ values extracted in the
scale range of $3$\,GeV$\,<Q<11$\,GeV are in excellent agreement with
those evaluated from the world average
$\alpha_s(M_Z)$ \cite{Workman:2022ynf}. Given that the PMC scale
setting eliminates the scale uncertainties, the corresponding
extracted $\alpha_s$ values are not plagued by ambiguities in
the choice of~$\mu_r$. The extracted $\alpha_s$ values from the
thrust observable using PMC are shown in Fig.~\ref{figasPMCT}.
There is good agreement also in this case in the range
$3.5$\,GeV\,$<\mu_r<16$\,GeV (corresponding to the range
$0.05<(1-T)<0.29$). These extracted values of $\alpha_s$ are also in
good agreement with the world average value of the
PDG \cite{Workman:2022ynf}.
Thus, PMC scale setting provides a remarkable way to verify the
running of $\alpha_s(Q)$ from event shapes measured at a single
energy of $\sqrt{s}$. Analogously in QED, the running of the QED
coupling $\alpha(Q)$ can be measured at a single energy of
$\sqrt{s}$ (see e.g.\ \cite{OPAL:2005xqs}).

The differential distributions of event shapes are afflicted with
large logarithms especially in the two-jet region. Thus, the
comparison of QCD predictions with experimental data and thus the
extracted $\alpha_s$ values are restricted to the region where the
theory is able to describe the data well.
Choosing a different area of distributions leads to the different
values of~$\alpha_s$.

\subsubsection{$\alpha_s(M_Z)$ from a $\chi^2$ fit}

In order to obtain a reliable $\alpha_s$ at the scale of the $Z^0$
mass, we determine $\alpha_s(M_Z)$ from the fit of the PMC
predictions to measurements. In particular, we perform the fit by
minimizing the $\chi^2$ respect to the $\alpha_s(M_Z)$ parameter.
The variable $\chi^2$ is defined as:
$$\chi^2 =
\sum_{i}\left(\frac{\langle y\rangle^{\rm {expt}}_i - \langle
y\rangle^{\rm {th}}_i}{\sigma_i} \right)^2,$$
where $\langle
y\rangle^{\rm {expt}}_i$ is the value of the experimental data,
$\sigma_i$ is the corresponding experimental uncertainty and
$\langle y\rangle^{\rm {th}}_i$ is the theoretical prediction. The
fit for thrust and the $C$-parameter leads to the following results:
\begin{eqnarray}
\alpha_s(M_Z) &=& 0.1185\pm0.0011(\rm expt)\pm0.0005(\rm th) \nonumber \\
&=& 0.1185\pm0.0012,
\end{eqnarray}
with $\chi^2/$d.o.f.\,$=27.3/20$ for the thrust mean value and
\begin{eqnarray}
\alpha_s(M_Z) &=&
0.1193^{+0.0009}_{-0.0010}(\rm expt)\,^{+0.0019}_{-0.0016}(\rm th) \nonumber \\
&=& 0.1193^{+0.0021}_{-0.0019},
\end{eqnarray}
with $\chi^2/$d.o.f.\,$=43.9/20$ for the $C$-parameter mean value,
where the first error is the experimental uncertainty and the
second the theoretical uncertainty. Both results are consistent
with the world average $\alpha_s(M_Z)=0.1179\pm0.0009$
\cite{Workman:2022ynf}.

The precision of the extracted $\alpha_s$ has been greatly
improved by using the PMC: the dominant $\mu_r$ scale
uncertainties are eliminated and the convergence of pQCD series is
greatly improved. In particular, a strikingly much faster pQCD
convergence is obtained for the mean thrust
value \cite{Wang:2019ljl}, theoretical uncertainties are even
smaller than the experimental uncertainties. We remark that these
results for $\alpha_s(M_Z)$ are among the most precise determinations
of the strong coupling at the $Z^0$ mass from event-shape variables.

 \newpage

\section{Comparison of the PMCm, PMCs and PMC$\infty$ for fully integrated fundamental quantities\label{sec:comparison}}

In this section, we show predictions for three important
quantities: $R_{e^+e^-}$, $R_\tau$ and $\Gamma(H\to b\bar{b})$,
which have been calculated up to four-loop QCD
corrections using alternative PMC scale-setting procedures.
Numerical results for the conventional, PMCm, PMCs and
PMC$_\infty$ approaches will be presented. For self-consistency,
the same loop $\alpha_s$-running behavior will be adopted for
calculating the same loop perturbative series. The QCD asymptotic
scale ($\Lambda_{\rm QCD}$) is then fixed by using
$\alpha_s(M_Z)=0.1179$ \cite{Workman:2022ynf} and for a four-loop
prediction, we obtain $\Lambda_{\rm{QCD}}^{n_f=4}=291.7$\,MeV and
$\Lambda_{\rm{QCD}}^{n_f=5}=207.2$\,MeV in a conventional
$\overline{\rm MS}$ renormalization scheme.

\subsection{The pQCD predictions for $R_{e^+e^-}$, $R_\tau$ and
$\Gamma(H\to b\bar{b})$}

The annihilation of electrons and positrons into hadrons provides
one of the most important platforms for determining the running
behavior of the QCD coupling. The $R$ ratio is defined as
\begin{eqnarray}
R_{e^+ e^-}(Q)&=&\frac{\sigma\left(e^+e^-\rightarrow {\rm hadrons} \right)}{\sigma\left(e^+e^-\rightarrow \mu^+\mu^-\right)}\nonumber\\
&=& 3\sum_q e_q^2\left[1+R(Q)\right], \label{Re+e-}
\end{eqnarray}
where $Q=\sqrt s$, corresponding to the electron--positron
collision energy in the center-of-mass frame. The pQCD series for
$R(Q)$, up to $(n{+}1)$-loop QCD corrections, is
\begin{displaymath}
R_n(Q)=\sum_{i=0}^{n} {\cal C}_{i}(Q,\mu_r) a_s^{i+1}(\mu_r).
\end{displaymath}
The perturbative coefficients ${\cal C}_{i}(Q,\mu_r)$ in the
$\overline{\rm MS}$ scheme up to four-loop level have been
calculated in
Refs.~\cite{Baikov:2008jh, Baikov:2010je, Baikov:2012zm, Baikov:2012zn}.
As a reference point, at $\sqrt{s}=31.6$\,GeV, we have
$\frac{3}{11}R_{e^+ e^-}^{\rm expt}=1.0527\pm0.0050$ \cite{Marshall:1988ri}.

Another useful ratio is $R_\tau$ for $\tau$-lepton decays into
hadrons, defined as
\begin{eqnarray}
R_\tau(M_\tau)&=&\frac{\Gamma\left(\tau\rightarrow \nu_\tau+{\rm hadrons} \right)}{\Gamma\left(\tau\rightarrow \nu_\tau+ \bar{\nu}_{l}+l\right)}\nonumber\\
&=& 3|V_{ud}|^2S_{\rm EW}\left[1+\hat{R}(M_\tau)+\delta_{\rm
EW}^{'}+\delta_2+\delta_{\rm NP}\right], \label{Rtau}
\end{eqnarray}
where $V_{ud}=0.97373\pm0.00031$ \cite{Workman:2022ynf} is the
relevant Cabibbo--Kobayashi--Maskawa matrix element, $S_{\rm
EW}=1.0198\pm0.0006$ and $\delta_{\rm EW}^{'}=0.001$ for the
electroweak corrections, $\delta_2=(-4.4\pm2.0)\times 10^{-4}$ for
light-quark mass effects\footnote{An improved determination of the
$m_s$ and $|V_{\rm us}|$ from the $\tau$-decay has been recently
shown in Ref. \cite{Ananthanarayan:2022ufx}.}, $\delta_{\rm
NP}=(-4.8\pm1.7)\times 10^{-3}$ for the nonperturbative effects
and $M_\tau=1.777$\,GeV \cite{Davier:2005xq, ALEPH:2005qgp,
Baikov:2008jh}. The pQCD series for $\hat{R}(M_\tau)$ up to
$(n{+}1)$-loop QCD corrections is
\begin{displaymath}
\hat{R}_{n}(M_\tau)=\sum_{i=0}^{n} \hat{\cal
C}_i(M_\tau,\mu_r) a_s^{i+1}(\mu_r),
\end{displaymath}
where the perturbative coefficients $\hat{\cal
C}_i(M_\tau,\mu_r)$ up to four-loop QCD corrections can be
derived by using the relation between $R_\tau(M_\tau)$ and
$R_{e^+ e^-}(\sqrt{s})$ \cite{Lam:1977cu}.

The decay width for Higgs boson decay into a bottom and
anti-bottom pair, $H\to b\bar{b}$, can be written as
\begin{eqnarray}
\Gamma(H\to b\bar{b})=\frac{3G_{F} M_{H} m_{b}^{2}(M_{H})}
{4\sqrt{2}\pi} [1+\tilde{R}(M_{H})],
\end{eqnarray}
where the Fermi constant
$G_{F}=1.16638\times10^{-5}\,\rm{GeV}^{-2}$, the Higgs mass $M_{H}
=125.1$\,GeV and the $b$-quark $\overline{\rm{MS}}$ running mass
is $m_b(M_H)=2.78$\,GeV \cite{Wang:2013bla}. The pQCD series for
$\tilde{R}(M_{H})$ up to $(n{+}1)$-loop QCD corrections is
\begin{displaymath}
\tilde{R}_n(M_{H})=\sum_{i=0}^{n} \tilde{\cal C}_i(M_{H},\mu_r)
a_s^{i+1}(\mu_r).
\end{displaymath}
The perturbative coefficients $\tilde{\cal C}_i(M_{H},\mu_r)$ up
to four-loop QCD corrections have been calculated in
Ref.~\cite{Baikov:2005rw}. In the following, we give the properties
for the pQCD series of $R_{n}(Q{=}31.6$\,GeV),
$\hat{R}_{n}(M_\tau)$ and $\tilde{R}_{n}(M_H)$ using each
scale-setting approach. As for the leading-order ratios with
$n=0$, we have no information to set the renormalization scale for
all the scale-setting approaches;
and for convenience, we directly set it to be $Q$, $M_\tau$, or $M_H$,
respectively, which gives $R_0=0.04428$, $\hat{R}_0=0.0891$ and $\tilde{R}_0=0.2034$.
We point out that all four-loop calculations for these observables derive from analytic properties (of the Adler function)
rather than from a direct multi-loop calculation of the perturbative series.

\subsection{Properties using the conventional scale-setting approach}

\begin{table*}[htb]
\centering
\begin{tabular}{ccccc}
\hline
& ~~~${\cal C}_1(\mu_r)$~~~ & ~~~${\cal C}_2(\mu_r)$~~~ & ~~~${\cal C}_3(\mu_r)$~~~ & ~~~${\cal C}_4(\mu_r)$~~~ \\
 \hline
 $R(Q)$ & $4$ & $22.55^{+42.51}_{-42.51}$ & $-819.50^{+1145.54}_{-266.26}$ & $-20591^{+31244.8}_{-7812.4}$ \\
 \hline
 $\hat{R}(M_\tau)$ & $4$ & $83.24^{+49.91}_{-41.39}$ & $1687.42^{+3054.60}_{-1588.77}$ & $32532.1^{+139210.0}_{-37674.5}$ \\
 \hline
 $\tilde{R}(M_H)$ & $22.6667$ & $466.347^{+240.907}_{-240.907}$ & $2672.49^{+13688.40}_{-8567.51}$ & $-211391^{+358424}_{-27651.9}$ \\
 \hline
\end{tabular}
\caption{The scale-dependent coefficients ${\cal C}_i(\mu_r)$ of
the conventional series for $R_{n}(Q{=}31.6$\,GeV),
$\hat{R}_{n}(M_\tau)$ and $\tilde{R}_{n}(M_H)$, respectively. The
central values are for $\mu_r=Q$, $M_\tau$, or $M_H$,
respectively. The errors are evaluated by taking $\mu_r\in [Q/2, 2Q]$
for $R_{n}(Q{=}31.6$\,GeV), $\mu_r\in [1\,{\rm GeV},
2M_\tau]$ for $\hat{R}_{n}(M_\tau)$ and $\mu_r\in [M_H/2, 2
M_H]$ for $\tilde{R}_{n}(M_H)$, respectively.} \label{cijn}
\end{table*}

We present the perturbative coefficients ${\cal C}_i(\mu_r)$ in
Table~\ref{cijn}, where the errors are evaluated by taking $\mu_r\in
[Q/2,2Q]$ for $R(Q{=}31.6$\,GeV), $\mu_r\in [1\,{\rm GeV},
2M_\tau]$ for $\hat{R}(M_\tau)$ and $\mu_r\in [M_H/2, 2
M_H]$ for $\tilde{R}(M_H)$, respectively. Table~\ref{cijn} shows
that those coefficients are highly scale dependent.

\begin{table*}[htb]
\centering
\begin{tabular}{clllrlr}
\hline
& \multicolumn{1}{c}{$n=1$}
& \multicolumn{1}{c}{$n=2$}
& \multicolumn{1}{c}{$n=3$}
& \multicolumn{1}{c}{$\kappa_1$}
& \multicolumn{1}{c}{$\kappa_2$}
& \multicolumn{1}{c}{$\kappa_3$} \\
 \hline
 $R_n|_{\rm Conv.}$ & $0.04753^{+0.00044}_{-0.00138}$ & $0.04638^{+0.00012}_{-0.00070}$ & $0.04608^{+0.00015}_{-0.00009}$ & $7.3^{-2.9}_{+9.2}\%$ & $\hphantom02.4^{+0.7}_{-1.4}\%$ & $0.6^{-0.0}_{+0.1}\%$ \\
 \hline
 $\hat{R}_n|_{\rm Conv.}$ & $0.1522^{+0.0482}_{-0.0295}$ & $0.1826^{+0.0360}_{-0.0268}$ & $0.1980^{+0.0170}_{-0.0194}$ & $70.8^{+2.6}_{+2.3}\%$ & $20.0^{-10.9}_{+7.0}\%$ & $8.4^{-6.8}_{+6.2}\%$ \\
 \hline
 $\tilde{R}_n|_{\rm Conv.}$ & $0.2404^{+0.0074}_{-0.0075}$ & $0.2423^{+0.0002}_{-0.0007}$ & $0.2409^{+0.0015}_{-0.0007}$ & $18.2^{-8.0}_{+6.7}\%$ & $\hphantom00.8^{+1.3}_{+2.9}\%$ & $0.6^{-0.6}_{+0.0}\%$ \\
 \hline
\end{tabular}
\caption{Results for $R_{n}(Q{=}31.6$\,GeV),
$\hat{R}_{n}(M_\tau)$, $\tilde{R}_{n}(M_H)$ up to four-loop QCD
corrections using the conventional scale-setting approach. The
central values are obtained by setting $\mu_r$ as $Q$, $M_\tau$,
or $M_H$, respectively. The errors are evaluated by taking $\mu_r\in
[Q/2, 2 Q]$ for $R_{n}(Q{=}31.6$\,GeV), $\mu_r\in [1\,{\rm GeV}, 2M_\tau]$ for
$\hat{R}_{n}(M_\tau)$ and $\mu_r\in [M_H/2, 2 M_H]$ for $\tilde{R}_{n}(M_H)$.}
\label{conv}

\end{table*}

We present the results of $R_{n}(Q{=}31.6$\,GeV),
$\hat{R}_{n}(M_\tau)$ and $\tilde{R}_{n}(M_H)$ up to four-loop
QCD corrections using the conventional scale-setting approach in
Table~\ref{conv}, where the errors are evaluated by taking $\mu_r\in
[Q/2, 2 Q]$ for $R_3(Q{=}31.6$\,GeV), $\mu_r\in [1\,{\rm GeV}, 2M_\tau]$ for $\hat{R}_3(M_\tau)$ and $\mu_r\in [M_H/2, 2 M_H]$ for $\tilde{R}_3(M_H)$, respectively. For
self-consistency, we adopt the $(n{+}1)$th-loop
$\alpha_s$-running behavior in deriving $(n{+}1)$th-loop
prediction for $R_{n}(Q{=}31.6$\,GeV), $\hat{R}_{n}(M_\tau)$
and $\tilde{R}_{n}(M_H)$. We define the ratio
\begin{displaymath}
\kappa_n = \left|\frac{{\cal R}_n-{\cal R}_{n-1}}{{\cal
R}_{n-1}}\right|,
\end{displaymath}
where ${\cal R}$ stands for $R$, $\hat{R}$ and $\tilde{R}$,
respectively. It shows how the ``known'' prediction ${\cal
R}_{n-1}$ is affected by the one-order-higher terms.
Table~\ref{conv} shows that generally we have $\kappa_1>\kappa_2>\kappa_3$
for all those quantities, consistently with the
perturbative nature of the series and indicates that one can
obtain more precise predictions by including more loop terms.
To show the perturbative nature more explicitly, we present the
magnitudes of each loop term for the four-loop approximants
$R_3(Q=31.6\,\mathrm{GeV})$, $\hat{R}_3(M_\tau)$ and
$\tilde{R}_3(M_H)$ in Table~\ref{convorder}, which displays the
relative importance among the LO, NLO, N$^2$LO and N$^3$LO terms, which for
those approximants are
\begin{eqnarray}
&& 1 : +0.063^{+0.099}_{-0.100} : -0.026^{+0.034}_{-0.016} : -0.007^{+0.001}_{+0.009}, \label{conv41} \\
&& 1 : +0.531^{-0.108}_{+0.101} : +0.275^{-0.205}_{+0.152} : +0.135^{-0.186}_{+0.159}, \label{conv42} \\
&& 1 : +0.184^{+0.071}_{-0.085} : +0.009^{+0.039}_{-0.035} :
-0.007^{+0.010}_{-0.002}, \label{conv43}
\end{eqnarray}
where the central values are for $\mu_r=Q$, $\mu_r=M_\tau$ and
$\mu_r=M_H$; and the errors are for $\mu_r\in[Q/2, 2Q]$,
$\mu_r\in[1\,{\rm GeV}, 2M_\tau]$ and $\mu_r\in[M_H/2, 2M_H]$,
respectively. Consistently with Table~\ref{conv}, the scale
dependence for each loop term is large, but due to the
cancellation of scale dependence among different orders, the net
scale dependence is small, e.g.\ $\left(^{+0.3\%}_{-0.2\%}\right)$,
$\left(^{+8.6\%}_{-9.8\%}\right)$ and
$\left(^{+0.6\%}_{-0.3\%}\right)$ for $R_3(Q{=}31.6$\,GeV),
$\hat{R}_3(M_\tau)$ and $\tilde{R}_3(M_H)$, respectively.
Note that due to the usual renormalon divergence and a larger
$\alpha_s$ value at a smaller scale $M_\tau$, i.e.\ $\alpha_s(M_\tau) \sim 0.33$, the net scale dependence of the
four-loop prediction $\hat{R}_3(M_\tau)$ is still sizable.

\begin{table}[htb]
\begin{center}
\begin{tabular}{clllll}
\hline
 & \multicolumn{1}{c}{$\rm LO$}
 & \multicolumn{1}{c}{$\rm NLO$}
 & \multicolumn{1}{c}{$\rm N^2LO$}
 & \multicolumn{1}{c}{$\rm N^3LO$}
 & \multicolumn{1}{c}{$\rm Total$} \\
 \hline
 $R_3|_{\rm Conv.}$ & $0.04473^{-0.00499}_{+0.00512}$ & $0.00282^{+0.00360}_{-0.00468}$ & $-0.00115^{+0.00147}_{-0.00095}$ & $-0.00032^{+0.00007}_{+0.00042}$ & $0.04608^{+0.00015}_{-0.00009}$ \\
 \hline
 $\hat{R}_3|_{\rm Conv.}$ & $0.1020^{+0.0471}_{-0.0261}$ & $0.0542^{+0.0089}_{-0.0062}$ & $\hphantom-0.0280^{-0.0176}_{+0.0044}$ & $\hphantom-0.0138^{-0.0214}_{+0.0085}$ & $0.1980^{+0.0170}_{-0.0194}$ \\
 \hline
 $\tilde{R}_3|_{\rm Conv.}$ & $0.2030^{-0.0175}_{+0.0226}$ & $0.0374^{+0.0100}_{-0.0151}$ & $\hphantom-0.0019^{+0.0070}_{-0.0077}$ & $-0.0014^{+0.0020}_{-0.0005}$ & $0.2409^{+0.0015}_{-0.0007}$ \\
 \hline
\end{tabular}
\caption{The value of each loop term (LO, NLO, N$^2$LO and
N$^3$LO) for the four-loop QCD predictions $R_3(Q{=}31.6$\,GeV), $\hat{R}_3(M_\tau)$ and $\tilde{R}_3(M_H)$ using the
conventional scale-setting approach. The errors are evaluated by
taking $\mu_r\in [Q/2, 2 Q]$ for $R_3(Q{=}31.6$\,GeV),
$\mu_r\in [1\,{\rm GeV}, 2M_\tau]$ for $\hat{R}_3(M_\tau)$
and $\mu_r\in [M_H/2, 2 M_H]$ for $\tilde{R}_3(M_H)$.}
\label{convorder}
\end{center}
\end{table}

In the pQCD calculation, it is helpful to give a reliable
prediction of the uncalculated higher-order terms. The Pad\'{e}
approximant approach
(PAA) \cite{Basdevant:1972fe, Samuel:1992qg, Samuel:1995jc}, shown
in Eq.~\eqref{padeapp}, provides an effective method to estimate the
$(n{+}1)$th-order coefficient from a given $n$-th-order
series.\footnote{Another method, which uses the scale-invariant
conformal series together with the Bayesian
model \cite{Cacciari:2011ze, Bonvini:2020xeo, Duhr:2021mfd} to
provide probabilistic estimates of the unknown higher-orders terms
has been proposed \cite{Shen:2022nyr}} In practice, it has been
found that the PAA becomes more effective when more loop
terms are known. For a pQCD approximant, $\rho(Q)=c_1 a_s+c_2
a_s^2+c_3 a_s^3+c_4 a_s^4+\cdots$, the predicted N$^3$LO and
the N$^4$LO terms are
\begin{eqnarray}
\rho^{\rm N^3LO}_{[1/1]}&=&\frac{c_3^2} {c_2} a_s^4, \\
\rho^{\rm N^3LO}_{[0/2]}&=&\frac{2c_1 c_2 c_3-c_2^3} {c_1^2} a_s^4, \\
\rho^{\rm N^4LO}_{[1/2]}&=&\frac{2c_2 c_3 c_4-c_3^3-c_1 c_4^2} {c_2^2-c_1 c_3} a_s^5, \\
\rho^{\rm N^4LO}_{[2/1]}&=&\frac{c_4^2} {c_3} a_s^5, \\
\rho^{\rm N^4LO}_{[0/3]}&=&\frac{c_2^4-3c_1 c_2^2 c_3+2c_1^2 c_2
c_4+c_1^2 c_3^2} {c_1^3} a_s^5.
\end{eqnarray}
\begin{table}[htb]
\begin{center}
\begin{tabular}{cll}
\hline
 & \multicolumn{1}{c}{$\rm N^3LO$}
 & \multicolumn{1}{c}{$\rm N^4LO$} \\
\hline
 \raisebox {0ex}[0pt]{$R_3|_{\rm Conv.}$}
 & $[1/1]$: $0.00047^{+0.00000}_{-0.00284}$ & $[1/2]$: $-0.00002^{+0.00011}_{-0.00010}$ \\
 & \multicolumn{1}{c}{---} & $[2/1]$: $-0.00009^{+0.00028}_{-0.00000}$ \\
\hline
 \raisebox {0ex}[0pt]{$\hat{R}_3|_{\rm Conv.}$}
 & $[1/1]$: $0.0145^{+0.0074}_{-0.0128}$
 & $[1/2]$: $+0.0061^{+0.0093}_{-0.0142}$ \\
 & \multicolumn{1}{c}{---} & $[2/1]$: $+0.0068^{+0.0085}_{-0.0011}$ \\
\hline
 \raisebox {0ex}[0pt]{$\tilde{R}_3|_{\rm Conv.}$}
 & $[1/1]$: $0.0001^{+0.0016}_{-0.0000}$
 & $[1/2]$: $-0.0006^{+0.0005}_{-0.0000}$ \\
 &\multicolumn{1}{c}{---} & $[2/1]$: $+0.0010^{+0.0000}_{-0.0016}$ \\
\hline
\end{tabular}
\caption{The preferable diagonal-type PAA predictions of the $\rm
N^3LO$ and $\rm N^4LO$ terms of $R_3(Q{=}31.6$\,GeV),
$\hat{R}_3(M_\tau)$ and $\tilde{R}_3(M_H)$ using the
conventional scale-setting approach. The uncertainties correspond
to taking $\mu_r\in [Q/2, 2 Q]$, $\mu_r\in [1\,{\rm GeV},
2M_\tau]$ and $\mu_r\in [M_H/2, 2 M_H]$, respectively.}
\label{convpaa1}
\end{center}
\end{table}
Table~\ref{convpaa1} displays the preferable diagonal-type PAA
predictions \cite{Gardi:1996iq} of the $\rm N^3LO$ and $\rm N^4LO$
terms of $R_3(Q{=}31.6$\,GeV), $\hat{R}_3(M_\tau)$ and
$\tilde{R}_3(M_H)$ using the conventional scale-setting
approach. Owing to the large scale dependence of each
loop term, the PAA predictions show large scale dependence. The
two allowable diagonal-type PAA predictions for $\rm N^4LO$ terms
are consistent with each order within errors. Comparing Table
\ref{convpaa1} with Table \ref{convorder}, we note that the
values of the predicted $\rm N^3LO$ terms agree with their exact
values within errors. Thus, by employing a perturbative series
with enough higher-order terms, the PAA prediction can be made
reliable.

\subsection{Properties using the PMCm approach}

Following the standard PMCm procedures, the nonconformal
$\{\beta_i\}$-terms are eliminated by using the RGE recursively,
determining the effective $\alpha_s$ at each perturbative order and resulting in the renormalon-free and scheme-independent
conformal series \eqref{PMCmseries}.
\begin{table}[htb]
\begin{center}
\begin{tabular}{cccrr}
\hline
& ${\hat r}_{1,0}$ & ${\hat r}_{2,0}$
& \multicolumn{1}{c}{${\hat r}_{3,0}$}
& \multicolumn{1}{c}{${\hat r}_{4,0}$} \\
 \hline
 $R(Q)$ & $4$ & $29.44$ & $-64.25$ & $-2812.74$ \\
 \hline
 $\hat{R}(M_\tau)$ & $4$ & $34.33$ & $219.89$ & $1741.15$ \\
 \hline
 $\tilde{R}(M_H)$ & $22.6667$ & $216.356$ & $-8708.09$ & $-110597$ \\
 \hline
\end{tabular}
\caption{Conformal coefficients ${\hat r}_{i,0}$ for
$R_{n}(Q{=}31.6$\,GeV), $\hat{R}_{n}(M_\tau)$ and
$\tilde{R}_{n}(M_H)$, respectively.} \label{rijn}
\end{center}
\end{table}
We present the conformal coefficients ${\hat r}_{i,0}$ in
Table~\ref{rijn}. The PMC scales are of a perturbative nature, which
leads to the \emph{first kind of residual scale dependence} for
PMCm predictions. If the pQCD approximants are known up to
four-loop QCD corrections, three PMC scales ($Q_1$, $Q_2$ and
$Q_3$) can be determined up to $\rm N^2LL$, $\rm NLL$ and $\rm LL$
order, which are $\{$41.19, 36.85, 168.68$\}$\,GeV for
$R_{n}(Q{=}31.6$\,GeV), $\{$1.26, 0.98, 0.36$\}$\,GeV for
$\hat{R}_{n}(M_\tau)$ and $\{$62.03, 40.76, 52.76$\}$\,GeV for
$\tilde{R}_{n}(M_H)$, accordingly. There is no $\{\beta_i\}$-term
to set the scale $Q_4$, the PMCm prediction has the \emph{second
kind of residual scale dependence}. As mentioned in
Sec.~\ref{sec:pmcm}, there is the \emph{second kind of residual
scale dependence} for PMCm series and, for convenience, we set
$Q_4=Q_3$ as the default choice of~$Q_4$. A discussion of the
magnitude of the \emph{second kind of residual scale dependence}
by taking some other typical choices of $Q_4$ will be presented
at the end of this subsection. For the scales $\gg\Lambda_{\rm
QCD}$, we adopt the usual approximate four-loop analytic solution
of the RGE to derive the value of~$\alpha_s$. Due to the sizable
difference between the approximate analytic solution and the
exact numerical solution of the RGE at scales below a few
GeV \cite{Wu:2019mky, Workman:2022ynf}, we adopt the exact
numerical solution of the RGE to evaluate $R_\tau(M_\tau)$ at
1.26\,GeV and 0.98\,GeV. For the scales close to $\Lambda_{\rm
QCD}$, various low-energy models have been suggested in the
literature; a detailed comparison of various low-energy models can
be found in Ref.~\cite{Zhang:2014qqa}. For definiteness, we shall
adopt the Massive Perturbation Theory (MPT)
model \cite{Shirkov:2012ux} to evaluate $R_\tau(M_\tau)$ at
$Q_3=0.36$\,GeV, which gives
$\alpha_s|_{\rm MPT}^{\xi=10\pm2}(0.36)=0.559^{+0.042}_{-0.032}$,
where $\xi$ is the parameter in the MPT model.
\begin{table}[htb]
\begin{center}
\begin{tabular}{clllrrr}
\hline
& \multicolumn{1}{c}{$n=1$}
& \multicolumn{1}{c}{$n=2$}
& \multicolumn{1}{c}{$n=3$}
& \multicolumn{1}{c}{$\kappa_1$}
& \multicolumn{1}{c}{$\kappa_2$}
& \multicolumn{1}{c}{$\kappa_3$} \\
 \hline
 $R_n|_{\rm PMCm}$ & $0.04735$ & $0.04640$ & $0.04610$ & $6.9\%$ & $2.0\%$ & $0.6\%$ \\
 \hline
 $\hat{R}_n|_{\rm PMCm}$ & $0.2133$ & $0.1991$ & $0.2087$ & $139.4\%$ & $6.7\%$ & $4.8\%$ \\
 \hline
 $\tilde{R}_n|_{\rm PMCm}$ & $0.2481$ & $0.2402$ & $0.2400$ & $22.0\%$ & $3.2\%$ & $0.1\%$ \\
 \hline
\end{tabular}
\caption{Results for $R_{n}(Q{=}31.6$\,GeV),
$\hat{R}_{n}(M_\tau)$, $\tilde{R}_{n}(M_H)$ up to four-loop QCD
corrections using the PMCm scale-setting approach. The
renormalization scale is set as $Q$, $M_\tau$ and $M_H$,
respectively.} \label{pmcm}
\end{center}
\end{table}

We present the results of $R_{n}(Q{=}31.6$\,GeV),
$\hat{R}_{n}(M_\tau)$ and $\tilde{R}_{n}(M_H)$ up to four-loop
QCD corrections using the PMCm scale-setting approach in
Table~\ref{pmcm}. Table~\ref{pmcm} shows that the PMCm predictions
generally have behavior close to the central predictions under
conventional scale-setting procedures, especially when more loop
terms are known. This is due to the fact that when the renormalization
scale of the conventional series is set as the one to eliminate
the large logarithms, the divergent renormalon terms may also be
simultaneously removed, since the $\{\beta_i\}$-terms are always
accompanied by the logarithm terms.
\begin{table*}[htb]
\begin{center}
\begin{tabular}{clllll}
\hline
 & \multicolumn{1}{c}{$\rm LO$}
 & \multicolumn{1}{c}{$\rm NLO$}
 & \multicolumn{1}{c}{$\rm N^2LO$}
 & \multicolumn{1}{c}{$\rm N^3LO$}
 & \multicolumn{1}{c}{$\rm Total$} \\
 \hline
 $R_3|_{\rm PMCm}$ & $0.04267^{+0.00003}_{-0.00001}$ & $0.00349^{+0.00003}_{-0.00004}$ & $-0.00004$ & $-0.00002$ & $0.04610^{+0.00006}_{-0.00005}$ \\
 \hline
 $\hat{R}_3|_{\rm PMCm}$ & $0.1272^{+0.0062}_{-0.0090}$ & $0.0553^{+0.0147}_{-0.0178}$ & $\hphantom-0.0194$ & $\hphantom-0.0068$ & $0.2087^{+0.0209}_{-0.0268}$ \\
 \hline
 $\tilde{R}_3|_{\rm PMCm}$ & $0.2258$ & $0.0247^{+0.0001}_{-0.0001}$ & $-0.0093$ & $-0.0012$ & $0.2400^{+0.0001}_{-0.0001}$ \\
 \hline
\end{tabular}
\caption{The value of each loop term (LO, NLO, N$^2$LO and
N$^3$LO) for the four-loop QCD predictions $R_3(Q{=}31.6$\,GeV), $\hat{R}_3(M_\tau)$ and $\tilde{R}_3(M_H)$ using the
PMCm scale-setting approach. The uncertainties correspond to
taking $\mu_r\in [Q/2, 2 Q]$ for $R_3(Q{=}31.6$\,GeV),
$\mu_r\in [1\,{\rm GeV}, 2M_\tau]$ for $\hat{R}_3(M_\tau)$ and $\mu_r\in [M_H/2, 2 M_H]$ for $\tilde{R}_3(M_H)$.}
\label{pmcmorder}
\end{center}
\end{table*}

We present the value of each loop term for the four-loop
predictions $R_3(Q{=}31.6$\,GeV), $\hat{R}_3(M_\tau)$ and $\tilde{R}_3(M_H)$ using the PMCm scale-setting approach in
Table~\ref{pmcmorder}.\footnote{By using the RGE recursively, one
can obtain correct $\alpha_s$ value and achieve good matching of
$\alpha_s$ to its coefficients at the same perturbative order.
However, such treatment is a sort of $\{\beta_i\}$-resummation and
the resultant PMC series is no longer the usual fixed-order series.
Thus, the function of Table~\ref{pmcmorder} is to show its own
perturbative behavior.} The relative importance among the
LO terms, the NLO terms, the N$^2$LO terms and the N$^3$LO terms
for those approximants are
\begin{eqnarray}
&& 1:+0.0818: -0.0009: -0.0005 \quad (\mu_r{=}Q), \\
&& 1:+0.4347: +0.1525: +0.0535 \quad (\mu_r{=}M_\tau), \\
&& 1:+0.1094: -0.0412: -0.0053 \quad (\mu_r{=}M_H).
\end{eqnarray}
Table \ref{pmcmorder} shows that there are residual scale
dependences of $R_3(Q{=}31.6$\,GeV) and $\tilde{R}_3(M_H)$
for the LO and NLO terms that are quite small (e.g.\ the errors
are only about $\pm 0.1\%$ of the LO terms and $\pm 0.04\%$ of the
NLO terms, respectively), which are smaller than the corresponding
ones under the conventional scale-setting approach. However, the
residual scale dependence of $\hat{R}_3(M_\tau)$ is sizable --
approximately $\pm10\%$ -- and is comparable to the conventional
scale dependence. Here the large residual scale dependence of
$\hat{R}_3(M_\tau)$ is reasonable, caused by the poor pQCD
convergence for the PMC scales at higher orders and the
uncertainties of the $\alpha_s$-running behavior in the low-energy
region.
\begin{table*}[htb]
\begin{center}
\begin{tabular}{clllll}
\hline
 & \multicolumn{1}{c}{$Q_4=2Q_3$}
 & \multicolumn{1}{c}{$Q_4=\frac{1}{2}Q_3$}
 & \multicolumn{1}{c}{$Q_4=Q_1$}
 & \multicolumn{1}{c}{$Q_4=Q_2$}
 & \multicolumn{1}{c}{$Q_4={(Q_1+Q_2+Q_3)}/{3}$}
 \\ \hline
 $R_3|_{\rm PMCm}$ & $0.04611$ & $0.04610$ & $0.04608$ & $0.04608$
 &\hspace{2em} $0.04610$ \\
 \hline
 $\hat{R}_3|_{\rm PMCm}$ & $0.2054^{+0.0054}_{-0.0036}$ & $0.2105^{+0.0088}_{-0.0053}$ & $0.2033^{+0.0049}_{-0.0032}$ & $0.2041^{+0.0050}_{-0.0033}$
 &\hspace{2em} $0.2046^{+0.0051}_{-0.0034}$ \\
 \hline
 $\tilde{R}_3|_{\rm PMCm}$ & $0.2404$ & $0.2392$ & $0.2401$ & $0.2398$
 &\hspace{2em} $0.2400$ \\
 \hline
\end{tabular}
\caption{The four-loop pQCD predictions of $R_3(Q{=}31.6$\,GeV), $\hat{R}_3(M_\tau)$ and $\tilde{R}_3(M_H)$ under
some other choices of the undetermined $Q_4$ as an estimate of
the \emph{second kind of residual scale dependence} under the
PMCm scale-setting approach. The uncertainty for
$\hat{R}_3(M_\tau)$ is obtained by changing the MPT-model
parameter $\xi=10\pm2$.} \label{Q4effect}
\end{center}
\end{table*}

As a final remark, we discuss the possible magnitudes of the
\emph{second kind of residual scale dependence} under the PMCm
scale-setting approach by taking some other typical choices of
$Q_4$, e.g.\ $2 Q_3$, $Q_3/2$, $Q_1$, $Q_2$ and
$(Q_1+Q_2+Q_3)/3$, which also ensures the scheme independence of
the PMCm series. Table \ref{Q4effect} shows that the
\emph{second kind of residual scale dependence} of $R_3$,
$\hat{R}_3$ and $\tilde{R}_3$ are $(^{+0.00001}_{-0.00002})$,
$(^{+0.0018}_{-0.0054})$ and $(^{+0.0004}_{-0.0008})$,
respectively. It shows that those choices will change the
magnitudes of $R_3$, $\hat{R}_3$ and $\tilde{R}_3$ at the default
choice of $Q_4=Q_3$ by about $\pm 0.04\%$, $\pm 2.6\%$ and $\pm
0.3\%$, respectively. For $\hat{R}_3(M_\tau)$, another
uncertainty caused by the MPT model parameter $\xi=10\pm2$ is
$\hat{R}_3|_{\rm PMCm}=0.2087^{+0.0072}_{-0.0046}$ in the choice
of $Q_4=Q_3$. We also give the numerical results under some other
choices of the undetermined $Q_4$ in Table \ref{Q4effect}. It
shows that the MPT model parameter $\xi=10\pm2$ will lead to
$\sim3\%$ uncertainties. These uncertainties caused by the small
scale $Q_3$ and undetermined scale $Q_4$ indicate that we still
need a more appropriate scale-setting approach to suppress the
theoretical uncertainties.

\subsection{Properties using the PMCs approach}

The PMCs approach provides a method to suppress the residual scale
dependence. Applying the standard PMCs scale-setting procedures,
we obtain an overall effective $\alpha_s$ and, accordingly, an
overall effective scale ($Q_*$) for $R_{n}(Q{=}31.6$\,GeV),
$\hat{R}_{n}(M_\tau)$ and $\tilde{R}_{n}(M_H)$, respectively.
If they are known up to two-loop, three-loop and four-loop
levels, the PMC scale $Q_*$ can be determined up to ${\rm LL}$,
${\rm NLL}$ and ${\rm N^2LL}$ accuracies, respectively. That is,
for $n=1,2,3$, we have
\begin{eqnarray}
Q_{*}|_{e^+e^-} &=& \{35.36, 39.49, 40.12\} \, {\rm GeV}, \\
Q_{*}|_\tau \quad &=& \{\hphantom00.90, \hphantom01.06, \hphantom01.07 \} \,
  {\rm GeV}, \\
Q_{*}|_{H\to b\bar{b}} &=& \{60.94, 56.51, 58.80\} \, {\rm GeV}.
\end{eqnarray}
Their magnitudes become more precise as one includes more loop
terms and the difference between the two nearby values becomes
smaller and smaller when more loop terms are included, e.g.\ the
${\rm N^2LL}$ scales only shift by about $2\%-4\%$ to the ${\rm NLL}$
ones. Since these PMCs scales are numerically sizable, one avoids
confronting the possibly small scale problem at certain
perturbative orders of the multi-scale-setting approaches, such as
PMCm, PMC$_\infty$.
\begin{table}[htb]
\begin{center}
\begin{tabular}{clllrrr}
\hline
& \multicolumn{1}{c}{$n=1$}
& \multicolumn{1}{c}{$n=2$}
& \multicolumn{1}{c}{$n=3$}
& \multicolumn{1}{c}{$\kappa_1$}
& \multicolumn{1}{c}{$\kappa_2$}
& \multicolumn{1}{c}{$\kappa_3$} \\
 \hline
 $R_n|_{\rm PMCs}$ & $0.04735$ & $ 0.04629$ & $ 0.04613$ & $6.9\%$ & $2.2\%$ & $0.3\%$ \\
 \hline
 $\hat{R}_n|_{\rm PMCs}$ & $0.2136$ & $ 0.1997$ & $ 0.2064$ & $139.7\%$ & $6.5\%$ & $3.4\%$ \\
 \hline
 $\tilde{R}_n|_{\rm PMCs}$ & $0.2481$ & $ 0.2424$ & $ 0.2398$ & $22.0\%$ & $2.3\%$ & $1.1\%$ \\
 \hline
\end{tabular}
\caption{Results for $R_{n}(Q{=}31.6$\,GeV),
$\hat{R}_{n}(M_\tau)$, $\tilde{R}_{n}(M_H)$ up to four-loop QCD
corrections using the PMCs scale-setting approach, which are
independent of any choice of renormalization scale.} \label{pmcs}
\end{center}
\end{table}

\begin{table}[htb]
\begin{center}
\begin{tabular}{clllll}
\hline
 & \multicolumn{1}{c}{$\rm LO$}
 & \multicolumn{1}{c}{$\rm NLO$}
 & \multicolumn{1}{c}{$\rm N^2LO$}
 & \multicolumn{1}{c}{$\rm N^3LO$}
 & \multicolumn{1}{c}{$\rm Total$} \\
 \hline
 $R_3|_{\rm PMCs}$
 & $0.04287$ & $0.00338$ & $-0.00008$ & $-0.00004$ & $0.04613$ \\
 \hline
 $\hat{R}_3|_{\rm PMCs}$ & $0.1465$ & $0.0460$
 & $\hphantom-0.0108$ & $\hphantom-0.0031$ & $0.2064$ \\
 \hline
 $\tilde{R}_3|_{\rm PMCs}$ & $0.2278$ & $0.0219$ & $-0.0088$ & $-0.0011$ & $0.2398$ \\
 \hline
\end{tabular}
\caption{The value of each loop term (LO, NLO, N$^2$LO or
N$^3$LO) for the four-loop predictions $R_3(Q{=}31.6$\,GeV),
$\hat{R}_3(M_\tau)$, $\tilde{R}_3(M_H)$ using the PMCs
scale-setting approach.} \label{pmcsorder}
\end{center}
\end{table}

\begin{figure}[htb]
\begin{center}
\includegraphics[width=0.5\textwidth]{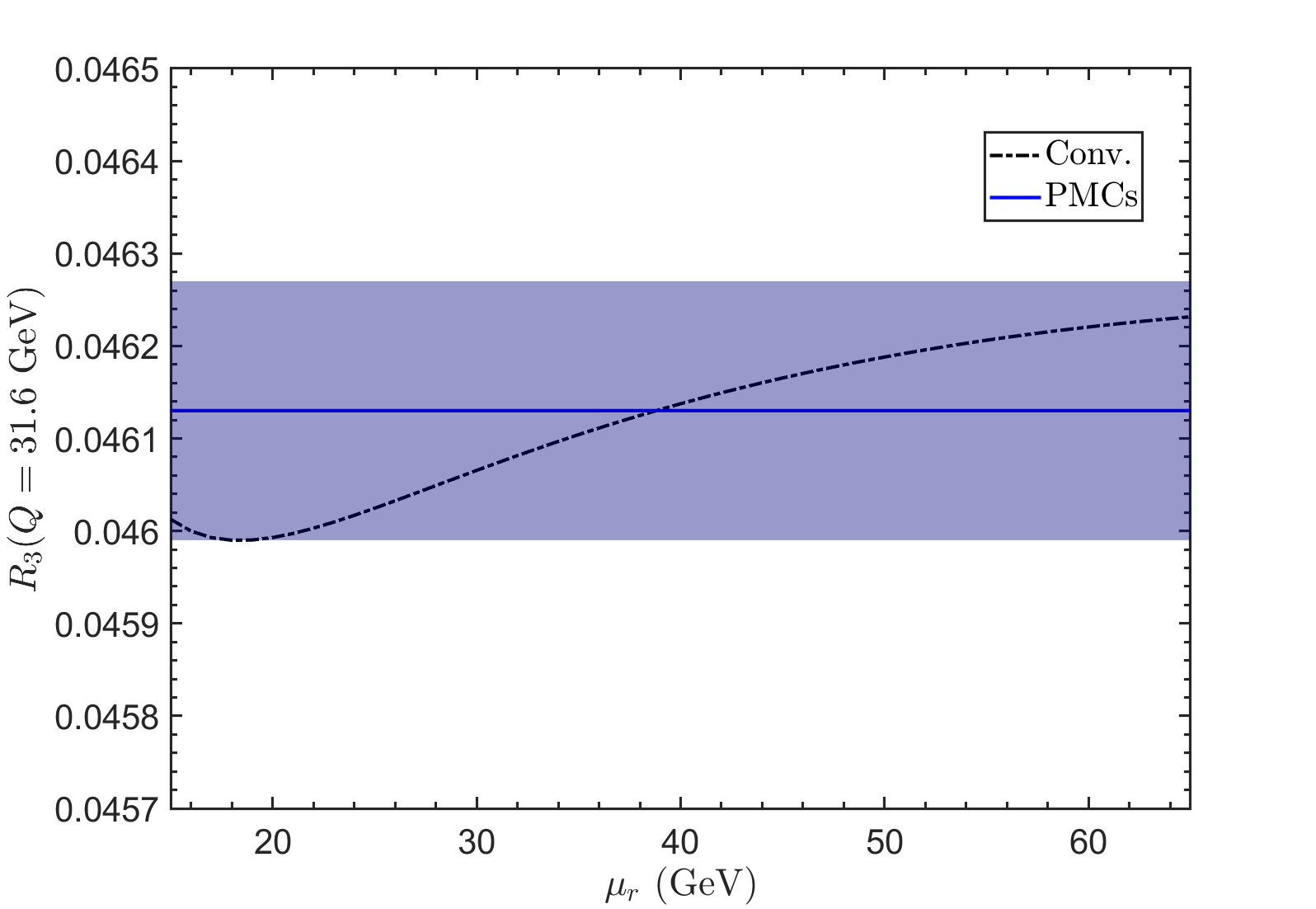}
\caption{The renormalization scale dependence of the four-loop
prediction $R_3(Q{=}31.6$\,GeV) using the conventional and
PMCs scale-setting procedures (from Ref.~\cite{Huang:2021hzr}).
The band represents a conservative estimate
\eqref{PMCsfirstres1} of the \emph{first kind of residual scale
dependence} under the PMCs.}  \label{urdepen11}
\end{center} \end{figure}

\begin{figure}[htb]
\begin{center}
\includegraphics[width=0.5\textwidth]{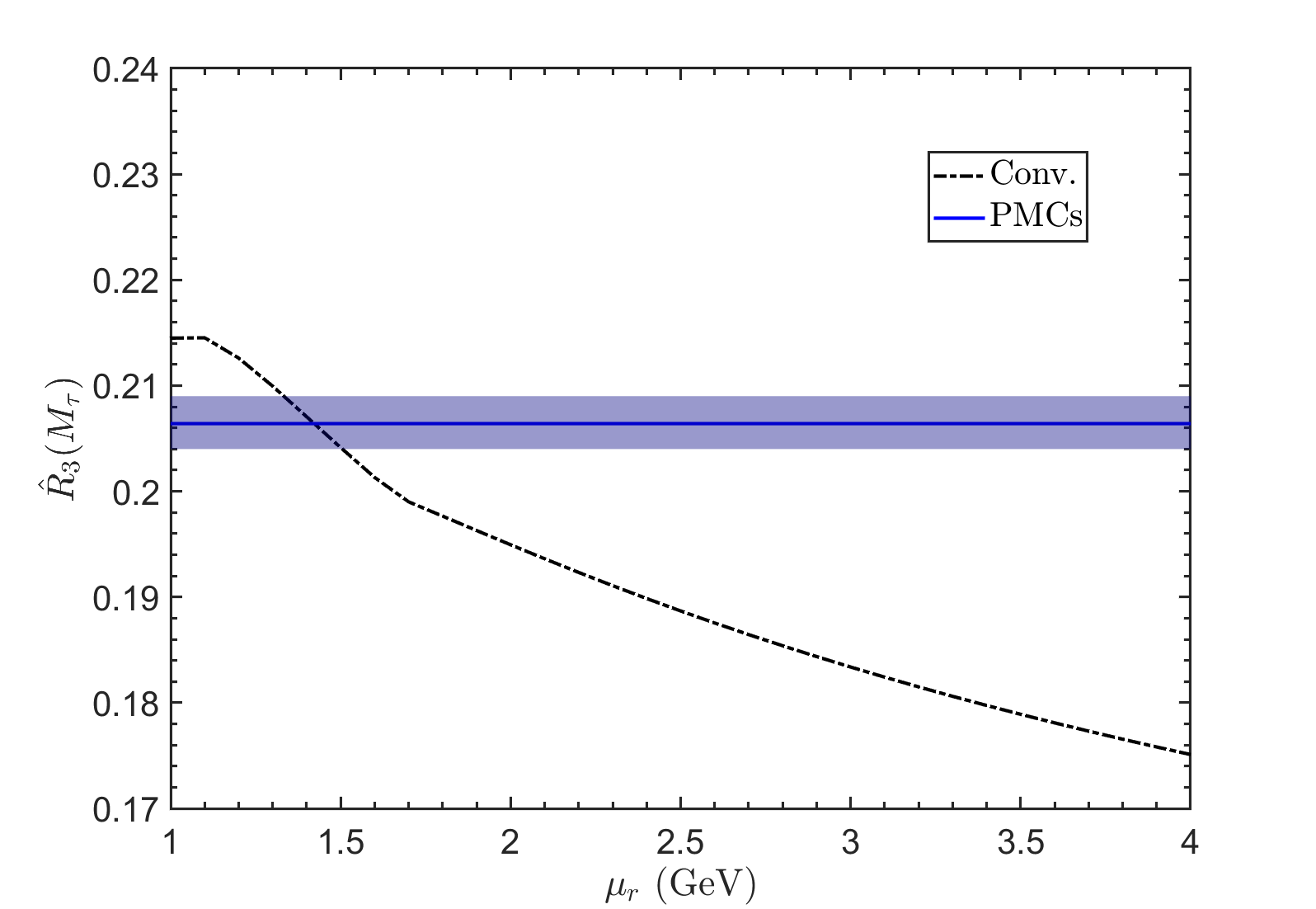}
\caption{The renormalization scale dependence of the four-loop
prediction $\hat{R}_3(M_\tau)$ using the conventional and PMCs
scale-setting procedures (from Ref.~\cite{Huang:2021hzr}). The
band represents a conservative estimate \eqref{PMCsfirstres2} of
the \emph{first kind of residual scale dependence} under the
PMCs.} \label{urdepen21}
\end{center} \end{figure}

\begin{figure}[htb]
\begin{center}
\includegraphics[width=0.5\textwidth]{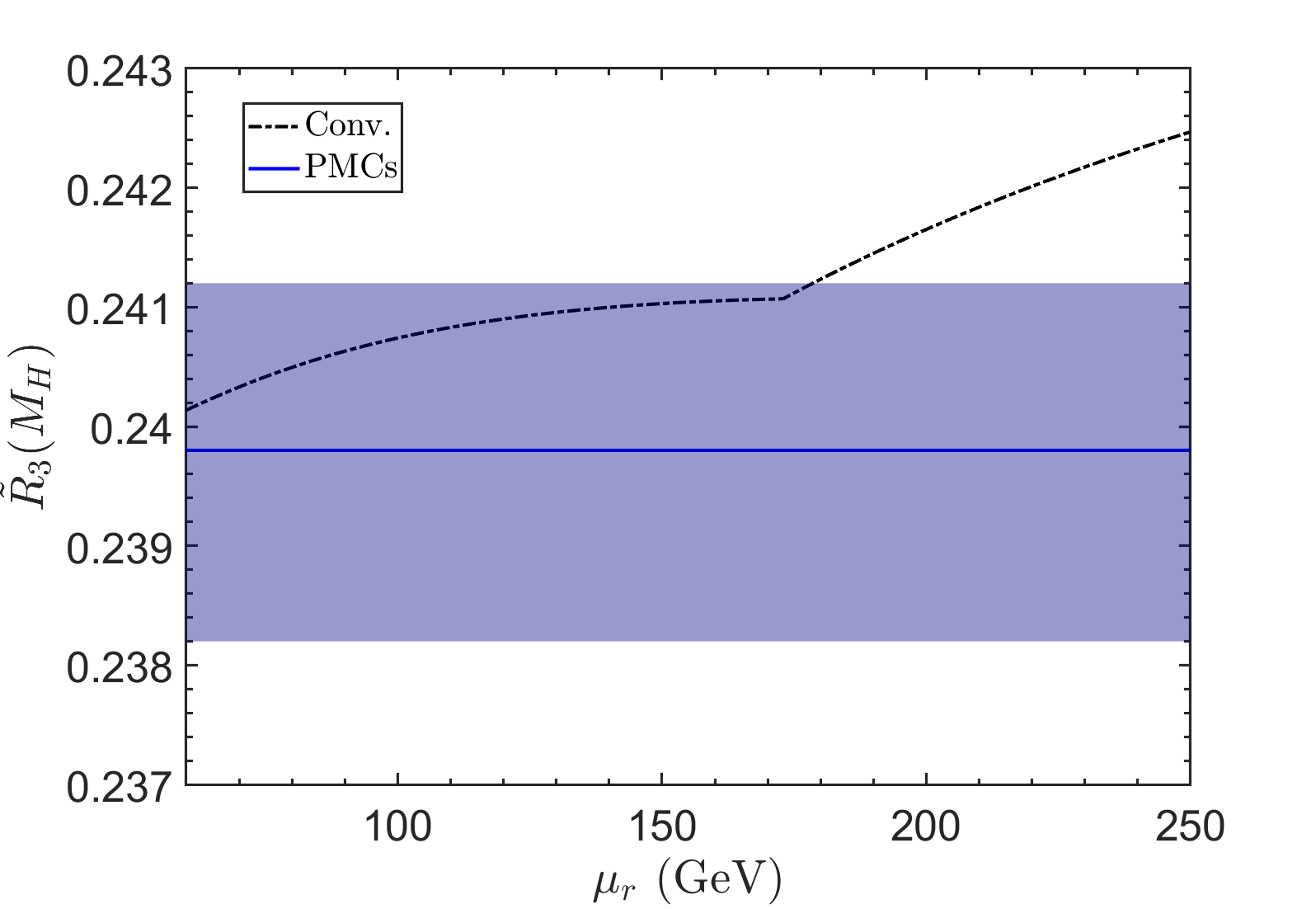}
\caption{The renormalization scale dependence of the four-loop
prediction $\tilde{R}_3(M_H)$ using the conventional and PMCs
scale-setting procedures (from Ref.~\cite{Huang:2021hzr}). The
band represents a conservative estimate \eqref{PMCsfirstres3} of
the \emph{first kind of residual scale dependence} under the
PMCs.} \label{urdepen31}
\end{center}
\end{figure}

We present the results of $R_{n}(Q{=}31.6$\,GeV),
$\hat{R}_{n}(M_\tau)$ and $\tilde{R}_{n}(M_H)$ up to four-loop
QCD corrections using the PMCs scale-setting approach in
Table~\ref{pmcs}. We also present the value of each loop term for
the four-loop approximants $R_3(Q{=}31.6$\,GeV),
$\hat{R}_3(M_\tau)$ and $\tilde{R}_3(M_H)$ using the PMCs
scale-setting approach in Table~\ref{pmcsorder}. The relative
importance among the LO terms, the NLO terms, the N$^2$LO terms
and the N$^3$LO terms for those approximants are
\begin{eqnarray}
&& 1: +0.0788: -0.0019: -0.0009, \\
&& 1: +0.3140: +0.0737: +0.0212, \\
&& 1: +0.0961: -0.0386: -0.0048.
\end{eqnarray}
These are comparable to the convergent behavior of the PMCm
series and are more convergent than conventional predictions.
Moreover, the sizable residual scale dependence of
$\hat{R}_3(M_\tau)$ appearing in Table~\ref{pmcmorder} has been
eliminated by using the PMCs procedure. Thus, the PMCs approach,
which requires a much simpler analysis, can be adopted as a
reliable substitute for the basic PMCm approach to setting the
renormalization scales for high-energy processes, with small
residual scale dependence. As a conservative estimate of the
\emph{first kind of residual scale dependence}, we take the
magnitude of its last known term as the unknown N$^3$LL term, e.g.\ $\pm( |S_2 a^2_s(Q_*)|)$ for $R_3(Q{=}31.6$\,GeV), ${\hat
R}_3(M_\tau)$ and ${\tilde R}_3(M_H)$. We then obtain
\begin{eqnarray}
R_3(Q{=}31.6\,{\rm GeV}) &=&0.04613\pm 0.00014, \label{PMCsfirstres1} \\
{\hat R}_3(M_\tau) &=&0.2064^{+0.0026}_{-0.0024}, \label{PMCsfirstres2} \\
{\tilde R}_3(M_H) &=&0.2398^{+0.0014}_{-0.0016}, \label{PMCsfirstres3}
\end{eqnarray}
which show that the \emph{first kind of residual scale dependence}
is about $\pm0.3\%$, $\pm1.3\%$ and $\pm0.7\%$.\footnote{Since
the N$^2$LL accuracy $\ln Q^2_*/Q^2$-series of these pQCD
approximants already show good perturbative behavior, it is found
that by using the PAA predicted N$^3$LL term, e.g.\ $\pm
|{S^2_2}/{S_1} a^3_s(Q_*)|$, to perform the estimates, one can obtain
smaller \emph{first kind of residual scale dependence} than the
ones listed in Eqs.~\eqref{PMCsfirstres1}, \ref{PMCsfirstres2},
\ref{PMCsfirstres3}), which are $\pm 0.00002$,
$\left(^{+0.0002}_{-0.0003}\right)$ and
$\left(^{+0.0007}_{-0.0009}\right)$, respectively.} More
explicitly, we show the conservative estimate of the
\emph{first kind of residual scale dependence} under the PMCs in
Figs.~\ref{urdepen11}, \eqref{urdepen21} and \eqref{urdepen31},
respectively.

The conformal PMCs series is scheme and scale independent; it thus
provides a reliable basis for estimating the effect of unknown
higher-order contributions. At present, there is no way to use a
series with different effective $\alpha_s(Q_i)$ at different
orders; and if there were any, its effectiveness would also be
greatly affected by the possibly large residual scale dependence.
Thus, we shall not use PMCm, PMC$_\infty$ series to estimate the
contribution of the unknown terms. As for the PMCs series, with an
overall effective $\alpha_s(Q_*)$, we may directly use the
PAA \cite{Du:2018dma}.
\begin{table}[htb]
\begin{center}
\begin{tabular}{lll}
\hline
& ~~~~~${\hat r}_{4,0}$~~~ & ~~~~~${\hat r}_{5,0}$~~~  \\
 \hline
 $R(Q)$ & [0/2]: $-2541.35$ & [0/3]: $-181893$ \\
 \hline
 $\hat{R}(M_\tau)$ & [0/2]: $\hphantom-1245.3$ & [0/3]: $\hphantom-15088.6$ \\
 \hline
 $\tilde{R}(M_H)$ & [0/2]: $-185951$ & [0/3]: $\hphantom-3802450$ \\
 \hline
\end{tabular}
\caption{The preferable [$0/n{-}1$]-type PAA predictions of the
conformal coefficients ${\hat r}_{4,0}$ and ${\hat r}_{5,0}$ for
$R_{n}(Q{=}31.6$\,GeV), $\hat{R}_{n}(M_\tau)$ and
$\tilde{R}_{n}(M_H)$, respectively.} \label{rijn2}
\end{center}
\end{table}
\begin{table}[htb]
\begin{center}
\begin{tabular}{lll}
\hline
 & ~~~~~~$\rm N^3LO$~~~~~ & ~~~~~~$\rm N^4LO$~~~~~ \\
\hline
 \raisebox {0ex}[0pt]{$R_3|_{\rm PMCs}$}
 & [0/2]: $-0.00003$ & [0/3]: $-0.000003$ \\
\hline
 \raisebox {0ex}[0pt]{$\hat{R}_3|_{\rm PMCs}$}
 & [0/2]: $+0.0022$ & [0/3]: $+0.0010$ \\
\hline
 \raisebox {0ex}[0pt]{$\tilde{R}_3|_{\rm PMCs}$}
 & [0/2]: $-0.0019$ & [0/3]: $+0.0004$ \\
\hline
\end{tabular}
\caption{The preferable [$0/n{-}1$]-type PAA predictions of the $\rm
N^3LO$ and $\rm N^4LO$ terms of $R_3(Q{=}31.6$\,GeV),
$\hat{R}_3(M_\tau)$ and $\tilde{R}_3(M_H)$ using the PMCs
scale-setting approach.} \label{pmcpaa1}
\end{center}
\end{table}

We present the preferable [$0/n{-}1$]-type PAA predictions for the
PMCs series of $R_3(Q{=}31.6$\,GeV), $\hat{R}_3(M_\tau)$ and
$\tilde{R}_3(M_H)$ in Table~\ref{rijn2} and \ref{pmcpaa1}.
Table~\ref{rijn2} shows the predicted $\rm N^3LO$ and $\rm N^4LO$
conformal coefficients ${\hat r}_{4,0}$ and ${\hat r}_{5,0}$. Note
that the predicted ${\hat r}_{4,0}$ values are close to the exact
values shown in Table~\ref{rijn} and those known conformal
coefficients do not change when more loop terms are known. To
obtain the final numerical result, we need to combine the
coefficients ${\hat r}_{4,0}$ and ${\hat r}_{5,0}$ with the
effective $\alpha_s(Q_*)$ at corresponding orders.
Table~\ref{pmcpaa1} displays the numerical results, these values
will be very slightly changed for a more accurate $Q_*$, since the
${\rm N^2LL}$ accuracy $Q_*$ is already changed from that at ${\rm
NLL}$ by less than~$\sim5\%$.

\subsection{Properties using the PMC$_\infty$ approach}

Given the unique form of intrinsic conformality iCF, any
other attempt (such as PMCa \cite{Chawdhry:2019uuv}) to write the
perturbative expansion in a scale-invariant form would lead to the
iCF (as shown in Ref.~\cite{Huang:2021hzr}).

Following the standard PMC$_\infty$ procedures, we calculate
$R_{n}(Q{=}31.6$\,GeV), $\hat{R}_{n}(M_\tau)$ and
$\tilde{R}_{n}(M_H)$ up to four-loop QCD corrections. The
perturbative coefficients ($\mathcal{C}_{i=(1,\cdots,4),\rm{IC}}$)
are exactly the same as those of the PMCm and PMCs conformal
coefficients (${\hat r}_{i=(1,\cdots,4),0}$). As shown by
Eqs.~\eqref{Bconf}, \eqref{Cconf}, \eqref{Dconf}, the PMC$_\infty$
scales are definite and have no perturbative nature, they are
free of renormalization scale ambiguities and do not have the
\emph{first kind of residual scale dependence}. Using the
four-loop QCD corrections, we can determine their first three
scales, i.e.
\begin{eqnarray}
\{\mu_{\rm I},\mu_{\rm II}, \mu_{\rm III}\}|_{e^+e^-} &=& \{35.36, 71.11, 0.003\} \,{\rm GeV}, \\
\{\mu_{\rm I},\mu_{\rm II}, \mu_{\rm III}\}|_\tau &=& \{0.90, 1.16, 1.82\} \,{\rm GeV}, \\
\{\mu_{\rm I},\mu_{\rm II}, \mu_{\rm III}\}|_{H\to b\bar{b}} &=&
\{60.94, 41.24, 46.44\} \,{\rm GeV}.
\end{eqnarray}
For the case $R_3(Q{=}31.6$\,GeV), its third scale
$\mu_{\rm III}=0.003$\,GeV is quite small and we adopt the above-mentioned
MPT model to estimate its contribution, which gives
$\alpha_s|_{\rm MPT}(0.003)=0.606$. As mentioned in
Sec.~\ref{pmcinfty}, the fourth scale $\mu_{\rm IV}$ is fixed to the
initial scale $\mu_r$, i.e.\ the kinematic scale of the process, and
varied in the range $\mu_{\rm IV}\in [\mu_r/2, 2\mu_r]$ to
ascertain the level of conformality achieved by the series. In fact,
the last PMC scale is entangled with the missing higher-order
contributions. This is referred as \emph{second kind of residual
scale dependence}.
\begin{table}[htb]
\begin{center}
\begin{tabular}{clllrrr}
\hline
& \multicolumn{1}{c}{$n=1$}
& \multicolumn{1}{c}{$n=2$}
& \multicolumn{1}{c}{$n=3$}
& \multicolumn{1}{c}{$\kappa_1$}
& \multicolumn{1}{c}{$\kappa_2$}
& \multicolumn{1}{c}{$\kappa_3$} \\
 \hline
 $R_n|_{\rm PMC_\infty}$ & $0.04750$ & $0.04652$ & $0.03937$ & $7.3\%$ & $2.1\%$ & $15.4\%$ \\
 \hline
 $\hat{R}_n|_{\rm PMC_\infty}$ & $0.1805$ & $0.2112$ & $0.2199$ & $102.6\%$ & $17.0\%$ & $4.1\%$ \\
 \hline
 $\tilde{R}_n|_{\rm PMC_\infty}$ & $0.2438$ & $0.2448$ & $0.2405$ & $19.9\%$ & $0.4\%$ & $1.8\%$ \\
 \hline
\end{tabular}
\caption{Results for $R_{n}(Q{=}31.6$\,GeV),
$\hat{R}_{n}(M_\tau)$, $\tilde{R}_{n}(M_H)$ up to four-loop QCD
corrections using the PMC$_\infty$ scale-setting approach. For
each case, the undetermined PMC$_\infty$ scale of the highest
order terms is set as $Q$, $M_\tau$ and $M_H$, respectively.}
\label{pmcinf}
\end{center}
\end{table}
\begin{table*}[htb]
\begin{center}
\begin{tabular}{clllll}
\hline
   & \multicolumn{1}{c}{$\rm LO$}
   & \multicolumn{1}{c}{$\rm NLO$}
   & \multicolumn{1}{c}{$\rm N^2LO$}
   & \multicolumn{1}{c}{$\rm N^3LO$}
   & \multicolumn{1}{c}{$\rm Total$} \\
 \hline
 $R_3|_{\rm PMC_\infty}$ & $0.04383$ & $0.00280$ & $-0.00722$ & $-0.00004^{+0.00001}_{-0.00004}$ & $0.03937^{+0.00001}_{-0.00004}$ \\
 \hline
 $\hat{R}_3|_{\rm PMC_\infty}$ & $0.1761$ & $0.0396$ & $\hphantom-0.0035$ & $\hphantom-0.0007^{+0.0034}_{-0.0005}$ & $0.2199^{+0.0034}_{-0.0005}$ \\
 \hline
 $\tilde{R}_3|_{\rm PMC_\infty}$ & $0.2265$ & $0.0246$ & $-0.0099$ & $-0.0007^{+0.0002}_{-0.0004}$ & $0.2405^{+0.0002}_{-0.0004}$ \\
 \hline
\end{tabular}
\caption{The value of each loop term (LO, NLO, N$^2$LO or
N$^3$LO) for the four-loop predictions $R_3(Q{=}31.6$\,GeV),
$\hat{R}_3(M_\tau)$ and $\tilde{R}_3(M_H)$ using the
PMC$_\infty$ scale-setting approach. The errors, representing the
\emph{second kind of residual scale dependence}, are estimated
by varying the undetermined PMC$_\infty$ scale $\mu_{\rm IV}$
within the range $[Q/2, 2Q]$ for $R_3(Q{=}31.6$\,GeV),
$[1\,{\rm GeV}, 2M_\tau]$ for $\hat{R}_3(M_\tau)$ and
$[M_H/2, 2M_H]$ for $\tilde{R}_3(M_H)$.} \label{pmcinforder}
\end{center}
\end{table*}

We present the results of $R_{n}(Q{=}31.6$\,GeV),
$\hat{R}_{n}(M_\tau)$ and $\tilde{R}_{n}(M_H)$ up to four-loop
QCD corrections using the PMC$_\infty$ scale-setting approach in
Table~\ref{pmcinf}. For the cases $R_{n}(Q{=}31.6$\,GeV) and
$\tilde{R}_{n}(M_H)$, we have $\kappa_2 < \kappa_3$, indicating
that the \emph{second kind of residual scale dependence} is sizable
for these two quantities, which largely affects the magnitude of
the lower-order series. When one has enough higher-order terms,
the residual scale dependence is highly suppressed owing to the more
convergent renormalon-free series. For example, we present the
value of each loop term (LO, NLO, N$^2$LO or N$^3$LO) for the
four-loop predictions $R_3(Q{=}31.6$\,GeV),
$\hat{R}_3(M_\tau)$ and $\tilde{R}_3(M_H)$ in
Table~\ref{pmcinforder}. At the four-loop level, the PMC$_\infty$
series already exhibits convergent behavior. As shown in
Table~\ref{pmcinforder}, the relative importance among the
LO terms, the NLO terms, the N$^2$LO terms and the N$^3$LO terms
for those approximants are
\begin{eqnarray}
&& 1:+0.0639: -0.1647: -0.0009 \quad (\mu_r{=}Q), \\
&& 1:+0.2249: +0.0199: +0.0040 \quad (\mu_r{=}M_\tau), \\
&& 1:+0.1086: -0.0437: -0.0031 \quad (\mu_r{=}M_H).
\end{eqnarray}
This perturbative behavior is similar to the predictions of PMCm
and PMCs, except for $R_3(Q{=}31.6$\,GeV), which due to a
much smaller scale $\mu_{\rm III}$ leads to quite large
N$^2$LO terms.

\subsection{A comparison of the renormalization scale dependence of the various PMC approaches}

\begin{figure}[htb]
\begin{center}
\includegraphics[width=0.5\textwidth]{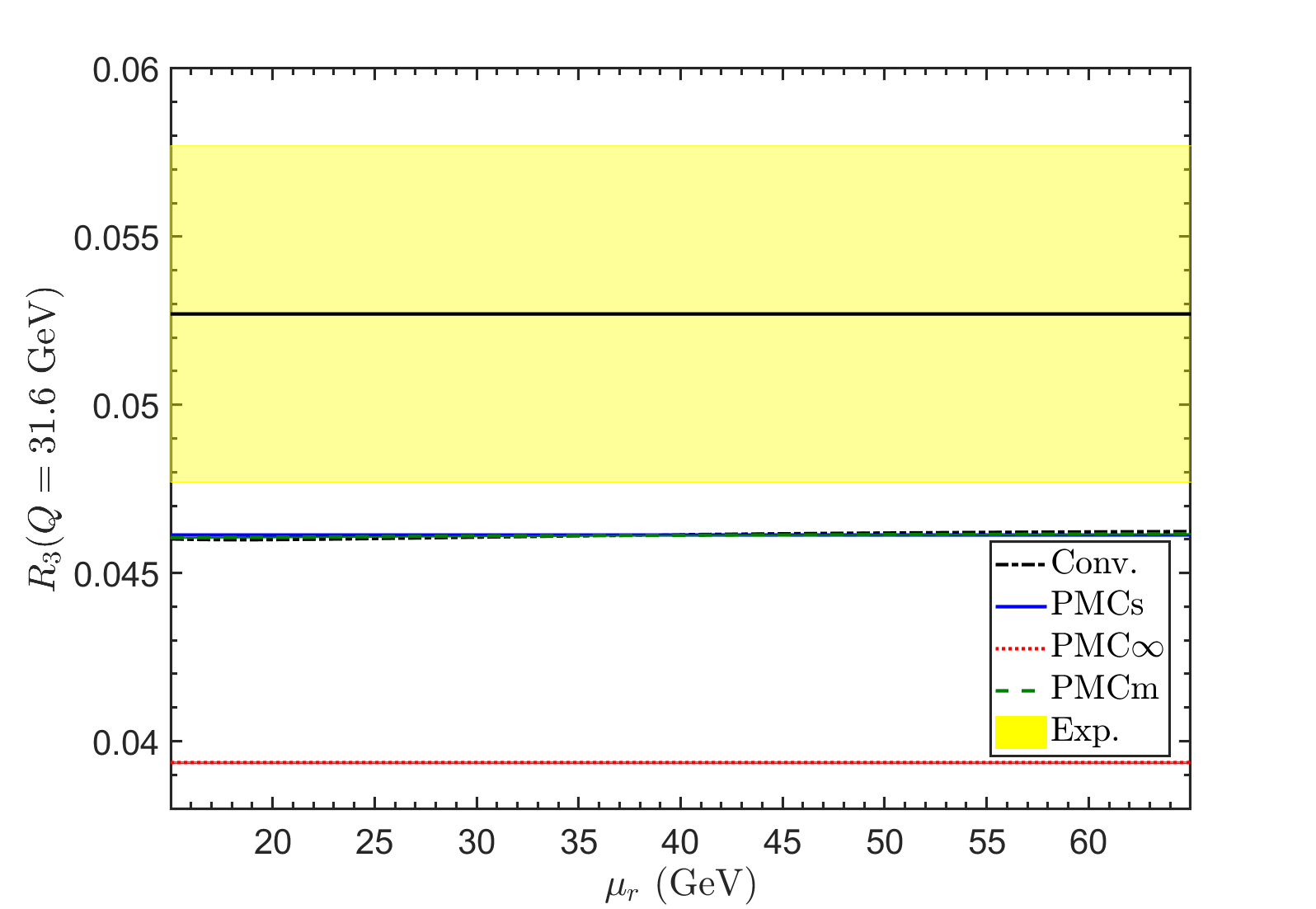}
\caption{The renormalization scale dependence of the four-loop
prediction $R_3(Q{=}31.6$\,GeV) using the conventional PMCm,
PMCs and PMC$_\infty$ scale-setting procedures (from
Ref.~\cite{Huang:2021hzr}). The green/lighter band represents the
\emph{second kind of residual scale dependence} under the PMCm,
which is obtained by changing the undetermined $Q_4$ to be $2
Q_3$, $Q_3/2$, $Q_1$, $Q_2$ and $(Q_1+Q_2+Q_3)/3$,
respectively. The red/darker band represents the \emph{second
kind of residual scale dependence} under the PMC$_\infty$, which
is obtained by varying the undetermined PMC$_\infty$ scale
$\mu_{\rm IV}$ within the range $[Q/2,2Q]$. The experimental
result $R^{\rm expt}(Q=31.6\,{\rm GeV})=0.0527\pm 0.0050$ is
extracted from $\frac{3}{11}R_{e^+ e^-}^{\rm expt}=1.0527
\pm0.0050$ \cite{Marshall:1988ri}. } \label{urdepen1}
\end{center}
\end{figure}

\begin{figure}[htb]
\begin{center}
\includegraphics[width=0.5\textwidth]{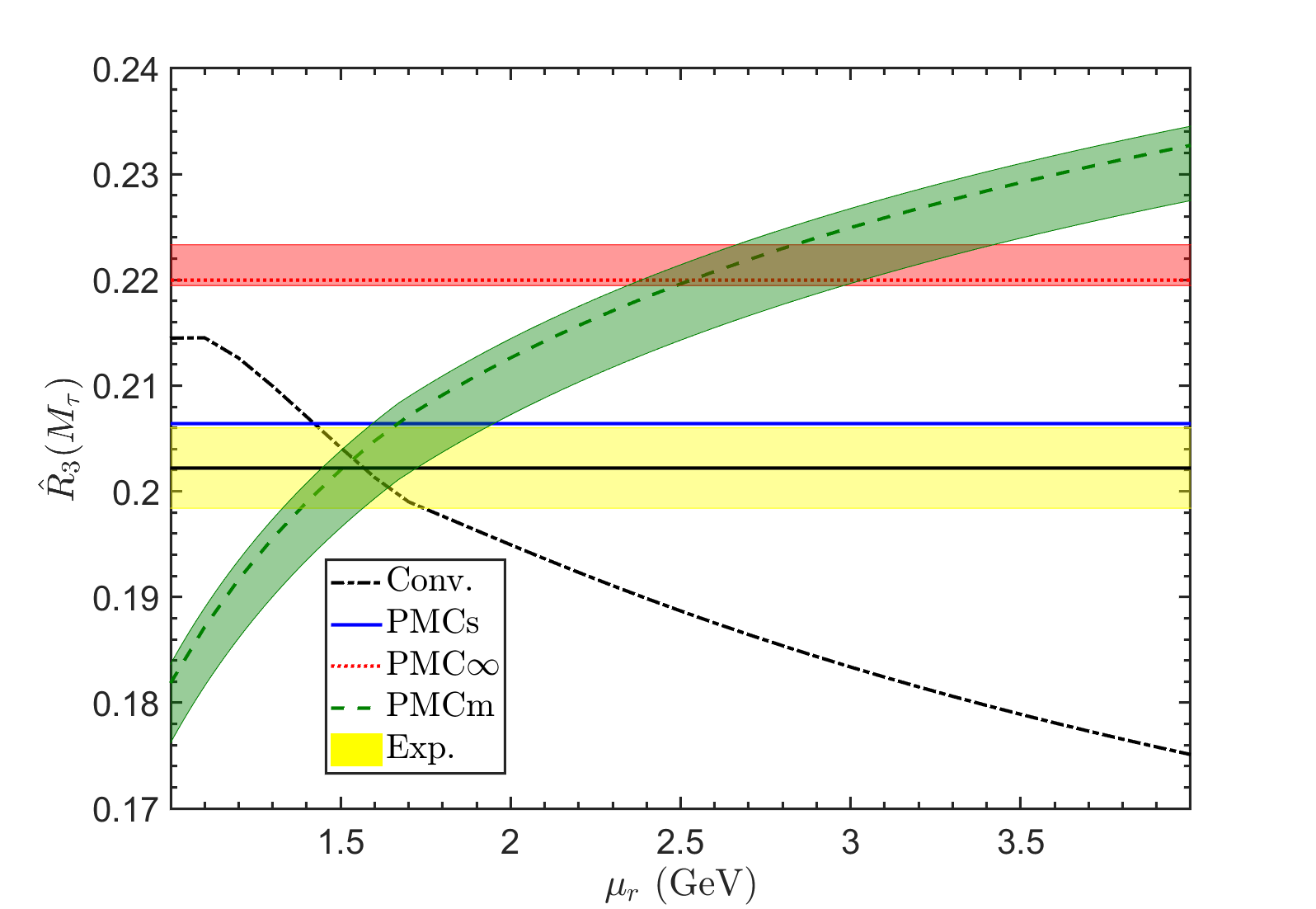}
\caption{The renormalization scale dependence of the four-loop
prediction $\hat{R}_3(M_\tau)$ using the conventional, PMCm,
PMCs and PMC$_\infty$ scale-setting procedures (from
Ref.~\cite{Huang:2021hzr}). The green/lighter band represents the
\emph{second kind of residual scale dependence} under the PMCm,
which is obtained by changing the undetermined $Q_4$ to be $2
Q_3$, $Q_3/2$, $Q_1$, $Q_2$ and $(Q_1+Q_2+Q_3)/3$,
respectively. The red/darker band represents the \emph{second
kind of residual scale dependence} under the PMC$_\infty$, which
is obtained by varying the undetermined PMC$_\infty$ scale
$\mu_{\rm IV}$ within the region of $[1\,{\rm GeV},2M_\tau]$. The
experimental result $\hat{R}^{\rm
expt}(M_\tau)=0.2022^{+0.0038}_{-0.0038}$ is extracted from
$R_\tau^{\rm expt}(M_\tau)=3.475
\pm0.011$ \cite{Davier:2013sfa}. } \label{urdepen2}
\end{center}
\end{figure}

\begin{figure}[htb]
\begin{center}
\includegraphics[width=0.5\textwidth]{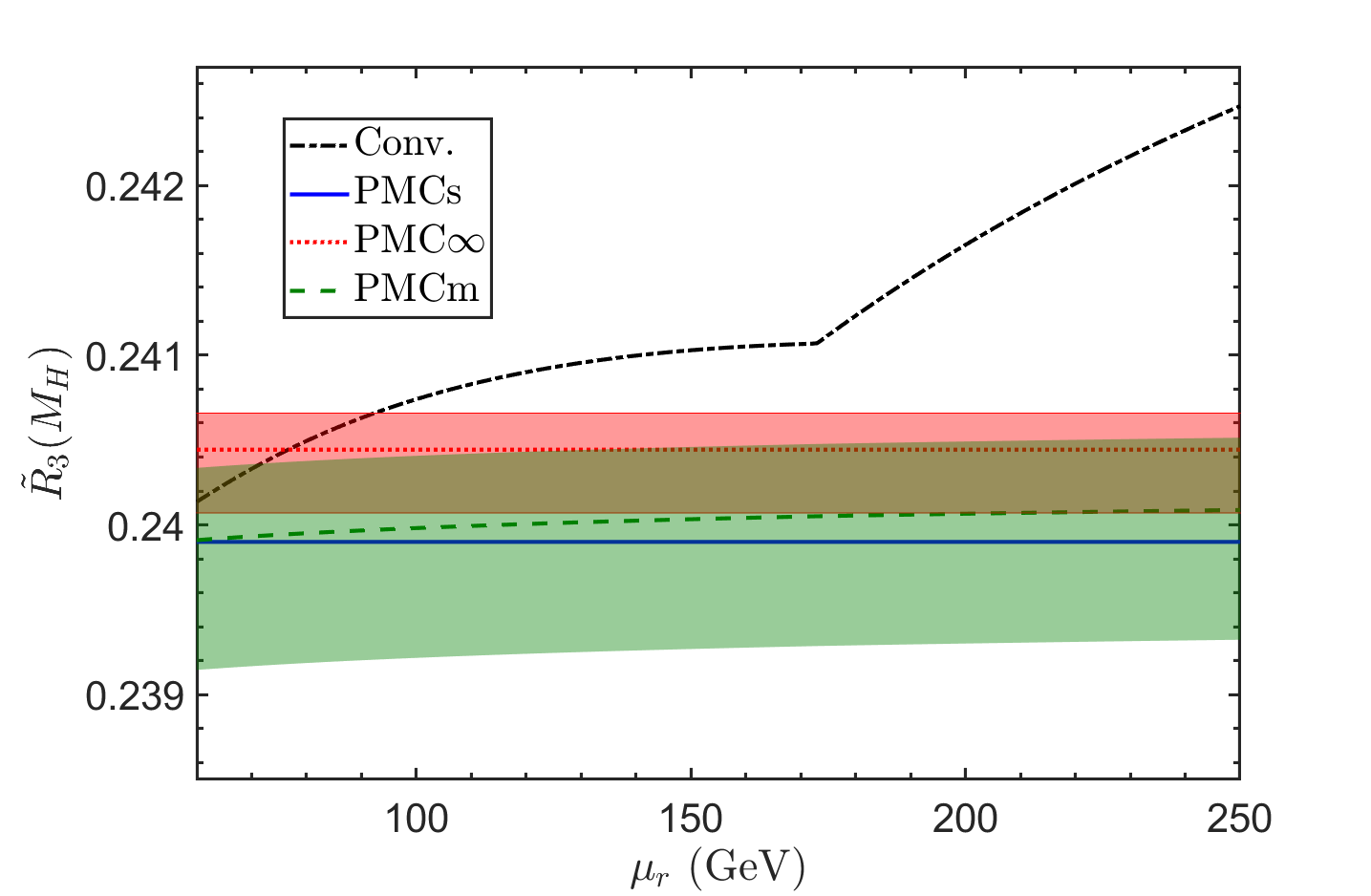}
\caption{The renormalization scale dependence of the four-loop
prediction $\tilde{R}_3(M_H)$ using the conventional, PMCm, PMCs
and PMC$_\infty$ scale-setting procedures (from
Ref.~\cite{Huang:2021hzr}). The green/lighter band represents the
\emph{second kind of residual scale dependence} under the PMCm,
which is obtained by changing the undetermined $Q_4$ to be $2
Q_3$, $Q_3/2$, $Q_1$, $Q_2$ and $(Q_1+Q_2+Q_3)/3$,
respectively. The red/darker band represents the \emph{second
kind of residual scale dependence} under the PMC$_\infty$, which
is obtained by varying the undetermined PMC$_\infty$ scale
$\mu_{\rm IV}$ within the region of $[M_H/2, 2M_H]$.}
\label{urdepen3}
\end{center}
\end{figure}

We present the renormalization scale ($\mu_r$) dependence of the
four-loop predictions $R_3(Q{=}31.6$\,GeV),
$\hat{R}_3(M_\tau)$ and $\tilde{R}_3(M_H)$ using the
conventional, PMCm, PMCs and PMC$_\infty$ scale-setting
procedures in Figs.~\ref{urdepen1}, \ref{urdepen2} and
\ref{urdepen3}, respectively. In these figures, we show the
\emph{second kind of residual scale dependence} of
$R_3(Q{=}31.6$\,GeV), $\hat{R}_3(M_\tau)$ under the PMCm
and PMC$_\infty$ scale-setting procedures with the shaded bands.
The green/lighter bands are obtained by changing the undetermined
$Q_4$ to $2 Q_3$, $Q_3/2$, $Q_1$, $Q_2$ and $(Q_1+Q_2+Q_3)/3$.
And the red/darker bands are obtained by varying the undetermined
PMC$_\infty$ scale $\mu_{\rm IV}$ within the range $[Q/2,2Q]$
for $R_3(Q{=}31.6$\,GeV), $[1\,{\rm GeV}, 2M_\tau]$ for
$\hat{R}_3(M_\tau)$ and $[M_H/2, 2M_H]$ for
$\tilde{R}_3(M_H)$.
Figure~\ref{urdepen1} shows that the
theoretical predictions are smaller than the experimental result.
This is reasonable since we have adopted the world average
$\alpha_s(M_Z)=0.1179$ \cite{Workman:2022ynf} to set $\Lambda_{\rm
QCD}$ for all these observables and, if we adopt a strong
coupling $\alpha_s(M_Z)$ fixed by using the $e^+e^-$
annihilation data alone, we obtain consistent predictions in
agreement with the data. For example, using
$\alpha_s(M_Z)=0.1224$ \cite{Dissertori:2009ik}, fixed by
using the hadronic event shapes in $e^+e^-$ annihilation to set
$\Lambda_{\rm QCD}$, we obtain a larger $R_3(Q{=}31.6$\,GeV),
e.g.\ $R_3(Q=31.6\,{\rm GeV})=0.04826$ for the PMCs
approach, which is consistent with
$R^{\rm expt}(Q=31.6\,{\rm GeV})=0.0527\pm 0.0050$ within errors.
It has been noticed that the
\emph{second kind of residual scale dependence} of
$R_3(Q{=}31.6$\,GeV) under the PMCm and PMC$_\infty$
scale-setting procedure are both very small, since the order
$\alpha_s^4$ correction is highly suppressed in
$R_3(Q{=}31.6$\,GeV). These figures show that by including
enough higher-order terms, the following hold.
\begin{itemize}
\item
The renormalization scale dependence of the conventional
prediction depends strongly on the convergence of the perturbative
series and the cancellation of scale dependence among different
orders. For a numerically strongly convergent series, such as
$R_3(Q{=}31.6$\,GeV) and $\tilde{R}_3(M_H)$, the net scale
dependence is only parts per thousand for a wide range of scale
choices. For a less convergent series, such as
$\hat{R}_3(M_\tau)$, the net renormalization scale uncertainty
is sizable, which is up to $\sim18\%$ for
$\mu_r\in[1\,{\rm GeV},2M_\tau]$, $\sim 24\%$ for
$\mu_r\in[1\,{\rm GeV},3M_\tau]$ and
$\sim 28\%$ for $\mu_r\in [1\,{\rm GeV}, 5M_\tau]$;
\item
The PMCm predictions have two kinds of residual scale
dependence due to unknown terms. The second residual scale
dependence can be greatly suppressed by the extra requirement of
conformal invariance; the first property then dominates the net
residual scale dependence. For numerically convergent series, such
as $R_3(Q{=}31.6$\,GeV) and $\tilde{R}_3(M_H)$, the
residual scale dependences are small, i.e.\ less than four parts
per thousand.\footnote{As a comparison, the conventional scale
dependence of $\tilde{R}_3(M_H)$ is about nine parts per
thousand for $\mu_r\in[M_H/2, 2M_H]$} For a less convergent
series, such as $\hat{R}_3(M_\tau)$, due to the large residual
scale dependence of the NLO terms, its net residual scale
dependence is sizable, it is $\sim 23\%$ for $\mu_r\in [1\,{\rm GeV}, 2M_\tau]$, $\sim 28\%$ for $\mu_r\in [1\,{\rm GeV},
3M_\tau]$ and $\sim 32\%$ for $\mu_r\in [1\,{\rm GeV},
5M_\tau]$. Although in some special cases, such as
$\hat{R}_3(M_\tau)$, the residual scale dependence may be
comparable to the conventional prediction, the PMCm series has no
renormalon divergence and it generally has a better pQCD
convergence. For the case of $R_3(Q{=}31.6$\,GeV) and
$\tilde{R}_3(M_H)$, the PMCm predictions show weaker dependence
on $\mu_r$ and its prediction can be more accurate than
conventional pQCD predictions;
\item
The PMC$_\infty$ predictions only have the \emph{second
kind of residual scale dependence}, which are suppressed for the
present four-loop predictions. The magnitude of the residual scale
dependence depends on the convergence of the resultant series and
for the present processes, the \emph{second kind of residual
scale dependence} are only about parts per thousand to a few
percent. Due to the application of \,``intrinsic conformality'' or
equivalently the requirement of scale invariance at each order,
the PMC$_\infty$ scales determined are not of a perturbative
nature, but they can be very small in certain cases. For the case
$R_3(Q{=}31.6$\,GeV), we obtain a much smaller scale
$\mu_{\rm III}=0.003$\,GeV, which is unreasonable and indicates that the
PMC$_\infty$ approach may not be applicable for this process. To
obtain a numerical estimate, we have adopted the MPT model to
calculate the magnitude of $\alpha_s$ at such a small scale;
Fig.~\ref{urdepen1} shows that the MPT prediction deviates from other
approaches by about $15\%$. By including the uncertainty from the
MPT model parameter $\xi=10\pm2$, the ${\rm PMC}_{\infty}$
prediction still deviates from other approaches by about~$11\%$;
\item
The PMCs predictions for the dependence of observables on
the renormalization scale are flat lines. The \emph{first kind
of residual scale dependence} of the PMCs predictions only affects
the precision of the magnitude of effective $\alpha_s$ and the
PMCs predictions are exactly independent of the choice of $\mu_r$
at any fixed order.
\end{itemize}

 \newpage

\section{Summary}\label{sec:sum}

The Principle of Maximum Conformality (PMC) provides a rigorous
first-principles method to eliminate conventional renormalization
scheme and scale ambiguities for high-momentum-transfer processes.
Its predictions have a solid theoretical foundation, satisfying
renormalization group invariance and all other self-consistency
conditions derived from the renormalization group. The PMC has now
been successfully applied to many high-energy processes.

In this review, we have presented a new scale-setting procedure,
namely PMC$_\infty$, which stems from the general PMC and
preserves a particular property that we have defined as
\emph{intrinsic conformality} (iCF). The iCF is a particular
parametrization of the perturbative series that exactly preserves
the scale invariance of an observable perturbatively. We point out
that this is a unique property of the perturbative expansion, any
other attempt (such as PMCa \cite{Chawdhry:2019uuv}) to write the
perturbative expansion in a scale-invariant form would lead to the
iCF (as shown in Ref. \cite{Huang:2021hzr}).

The PMC$_\infty$ solves the conventional renormalization scale
ambiguity in QCD, it preserves not only the iCF but also all the
features of the PMC approach and leads to a final conformal series
at any order of the perturbative calculation. In fact, the final
series is given by perturbative conformal coefficients with the
couplings determined at conformal renormalization scales. The
PMC$_\infty$ scale setting agrees with the Gell-Man--Low scheme
and can be considered the non-Abelian analog of
Serber--Uehling \cite{Serber:1935ui, Uehling:1935uj} scale setting,
which is essential in precision tests of QED and atomic physics.

Given the iCF form, a new ``How-To'' method for identifying
conformal coefficients and scale has been developed and can be
applied to either numerical or analytical calculations.
The PMC$_\infty$ has been applied to the NNLO thrust and $C$-parameter
distributions and the results show perfect agreement with the
experimental data.

The evaluation of theoretical errors using standard criteria demonstrate
that the PMC$_\infty$ significantly improves the theoretical
predictions over the entire spectra of the shape variables
order-by-order and both the IR conformal and QED limits of
thrust respect the theoretical consistency requirements. Moreover,
the position of the thrust peak is in perfect agreement with
experiment and is preserved on varying~$N_f$. Hence, even though
for the thrust distribution the peak stems directly from resummation
(or partial resummation) of the large logarithms in the
low-momentum region, its correct position is fixed by the PMC
scale and can be considered a ``conformal'' property, given its
independence from the $N_f$ or $\beta_i$ terms.

Unlike the previous BLM/PMC
approaches, the PMC$_\infty$ scales are not perturbatively
calculated but are conformal functions of the physical
scale(s) of the process and any other unintegrated
momentum or variable, e.g.\ the event-shape variable $(1-T)$ or~$C$.
The PMC$_\infty$ is totally independent of the initial scale
$\mu_0$ used for renormalization in perturbative calculations and
it preserves the scale invariance at all stages of calculation,
independently of the kinematic boundary conditions,
of the starting order of the observable or of the order of the
truncated expansion. Moreover, this property leads to the
possibility of determining the entire coupling from a single
experiment at a single center-of-mass energy (this new method is
in progress and will soon appear.)

The iCF improves the general BLM/PMC procedure and point~``3'' of
Section~\ref{sec:blm}. In the same section, we suggested that an
improvement and simplification of the perturbatively calculated
BLM/PMC scales, would be achieved by setting the renormalization
scale $\mu_r$ directly to the physical scale $Q$ of the process,
before applying the BLM/PMC procedure. This would remove the
initial scale $\mu_0$ dependence from the perturbatively
calculated BLM/PMC scales.

We stress that, in contrast with the other PMCm and PMCs
approaches, the PMC$_\infty$ preserves the iCF; scales are thus set
straightforwardly in kinematic regions where constraints cancel
the effects of the lower-order conformal coefficients. These
effects are particularly visible in the case of event-shape
variables in the multi-jet region. For this case we have shown
only fixed-order calculation results and other effects due to
factorization, such as large logarithms coming from soft and
collinear configurations, have not been included.
The iCF effects in these kinematic regions are neglected by the PMCm and
PMCs approach, unless an \emph{ad hoc} prescription is introduced.

Another application of the PMC$_\infty$ is presented in
Ref.~\cite{Gao:2021wjn} and shows an improvement of the results on
$\Gamma(H \to gg)$ with respect to the CSS also in this process.
From the detailed comparison shown in Sec.~\ref{sec:comparison},
it follows that, though the application of the PMC$_\infty$
improves the theoretical predictions also for the $R_{e^+e^-}$,
$R_\tau$, $\Gamma(H \to b \bar{b})$ with respect to the CSS, the
PMCs leads to more stable results for these quantities.

In general:
\begin{itemize}
\item[$\circ$]
The PMCs approach determines an overall effective
$\alpha_s$ by eliminating all the RG-dependent nonconformal
$\{\beta_i\}$-terms; this results in a single effective scale
which effectively replaces the individual PMC scales of PMCm
approach in the sense of a mean-value theorem.
The PMCs prediction is renormalization scale-and-scheme
independent up to any fixed order. The \emph{first kind of
residual scale dependence} is highly suppressed, since the PMC
scale at all known orders is determined at the same highest-order
accuracy. There is no \emph{second kind of residual scale
dependence}. The PMCs prediction also avoids the small-scale
problem, which sometimes emerges in multi-scale approaches.
\item[$\circ$]
The PMC$_\infty$ approach fixes the PMC scales at
each order by using the property of intrinsic conformality, which
ensures scale invariance of the pQCD series at each order. The
resulting PMC scales have no ambiguities, are not of a
perturbative nature and thus avoid the \emph{first kind of
residual scale dependence}. Since the last effective scale of the
highest-order perturbative term is set to the kinematic scale or
physical scale of the process, the PMC$_\infty$ prediction still
has a reduced \emph{second kind of residual scale dependence}.
However, when more loop terms are included and scales are not in
the nonperturbative regime, all PMC's lead to similar results.
\end{itemize}
The PMCs approach is close to the PMCm approach in achieving the
goals of the PMC by inheriting most of the features of the PMCm
approach. It works remarkably well with fully integrated
quantities, but has some difficulties in application to
differential distributions, besides the fact that this approach
may have an effect of averaging the differences of the
PMC$_\infty$ scales arising at each order, which might be
significant to achieve a given precision at a certain level of
accuracy.
However, given the small differences that we have found in the
first two consecutive PMC$_\infty$ scales for thrust and
$C$-parameter, i.e.\ $\mu_I,\mu_{II}$, in the LO allowed kinematic
region, i.e.\ $0<(1-T)<1/3$ and in $0<C<0.75$, we may argue that in
the same accessible kinematic domain two consecutive PMC$_\infty$
scales have such small differences that a single-scale approach,
such as the PMCs, would be justified, leading to analogously
precise predictions.

We recall that only the $n_f$ terms related to the UV-divergent
diagrams (i.e.\ the $N_f$ terms) must be reabsorbed into the
PMC$_\infty$ scales. Thus, PMC$_\infty$ perfectly agrees with the
PMCm when an observable has a manifestly iCF form. We remark that
the iCF underlies scale invariance perturbatively, i.e.\ the
ordered scale invariance. We also remark that PMC$_\infty$ agrees
with the single-scale approach PMCs in the case of an observable
with a particular iCF form with all scales equal, i.e.\ $\mu_{\rm
I}=\mu_{\rm II}=\mu_{\rm III}=\cdots=\mu_{\rm N}$. In this sense,
the PMCm and PMCs may be considered more as ``optimization
procedures'' that follow the purpose of the maximal conformal
series by transforming the original perturbative series into an
iCF-like final series by using the PMC scales. In contrast, the
PMC$_\infty$ does not indicate any particular value of the
renormalization scale to be used, but indicates the final limit
obtained by each conformal subset and then by the perturbative
expansion, once all the terms related to each conformal subset are
resummed.

The PMC$_\infty$ is RG invariant at each order of accuracy, which
means we may perform a change of scale at any stage and reobtain
the initial perturbative quantity. In this sense PMC$_\infty$ is
not to be understood as an ``optimization procedure'', but as an
explicit RG-invariant form to parametrize a perturbative quantity
that leads to the conformal limit faster.
By setting the renormalization scale of each subset to the
corresponding PMC$_\infty$ scale, one simply cancels the infinite
series of $\beta$ terms, leading to the same conformal result as
the original series. Given that both scales and coefficients are
conformal in the PMC$_\infty$, the scheme and scale
dependence is also completely removed in the perturbative series up to
infinity.

It was pointed out in Sec.~\ref{sec:comparison} that the
PMC$_\infty$ scale might become quite small at a certain order for
the case of fully integrated quantities, whose calculations were
carried out using the analyticity property of the Adler function
(this seems not to occur in direct multi-loop calculations, e.g.\ for the shape variables); and that the PMC$_\infty$ retains the second
kind of scale dependence. We stress that the last scale in the
PMC$_\infty$ controls the level of convergence and the
conformality of the perturbative series and is thus entangled with
the theoretical error of a given prediction. According to the
PMC$_\infty$ procedure the last scale must be set to the
invariant physical scale of the process, given by
$\sqrt{s},M_H,\dots$. In this review we have shown that the usual PMC
practice of setting the last scale equal to the last unknown scale
is also consistent for the PMC$_\infty$ and leads to precise and
stable results. Improvements to these points are currently
under investigation.

We finally remark that the evaluation of the theoretical errors using
standard criteria shows that the PMC$_\infty$ significantly
improves the precision of pQCD calculations and eliminates the
scheme and scale ambiguities. An improved and more reliable
analysis of theoretical errors might be obtained by using a
statistical approach for evaluating the contributions of the
uncalculated higher-order terms, as suggested in
Refs.~\cite{Cacciari:2011ze, Bonvini:2020xeo, Duhr:2021mfd} and
recently applied with the PMCs in
Refs.~\cite{Wang:2023ttk, Shen:2022nyr}. This implementation would
lead to a more rigorous method to evaluate errors, also giving
indications on the possible range of values for the last unknown
PMC$_\infty$ scale.

\section*{Acknowledgements}
We thank Francesco Sannino, Andr\'e Hoang for useful discussions.
XGW is supported in part by the Natural Science Foundation of
China under Grant No.12175025 and No.12147102. SQW is supported in
part by the Natural Science Foundation of China under Grant
No.12265011. SJB is supported in part by the Department of Energy
Contract No. DE-AC02-76SF00515. SLAC-PUB-17737.

\end{document}